\documentclass[12pt]{report}
\usepackage{epsfig}

\usepackage{epsfig}
\usepackage{amsmath}
\usepackage{amssymb}
\usepackage{graphicx}
\def\permil{\%\raise.10ex\hbox{$_{\scriptstyle 0}$}}
\def\be{\begin{equation}}
\def\ee{\end{equation}}
\def\bea{\begin{eqnarray}}
\def\eea{\end{eqnarray}}
\def\bsp{\begin{split}}
\def\esp{\end{split}}

\def\cdott{\!\cdot\!}

\def\delll{\delta^{b_1'b_2'}}
\def\dell{\tilde\delta^{a_1',a_2'}}

\def\cVv{{\cal V}_{LON_c}^{vl\{a'\}}}
\def\cVvb{{\cal V}_{LON_c}^{vl\{b'\}}}
\def\B{{\cal C}}
\def\cVr{{\cal V}_{LON_c}^{r\{a'\}}}
\def\cVrr{{\cal V}_{LON_c}^{r\{b'\}}}
\def\cVrrr{{\cal V}_{subN_c}^{r\{b'\}}}
\def\csVv{{\cal V}_{subN_c}^{v\{a'\}}}
\def\csVr{{\cal V}_{subN_c}^{r\{a'\}}}

\def\imp{\phi}

\def\impf{\imp_{\{b'\}}\otimes}
\def\impfa{\imp_{\{a'\}}\otimes}
\def\verta{{\cal V}^{a'_1a'_2;a_1a_2a_3a_4}}

\def\ak{|k|}
\def\ak_1{|k_1|}
\def\ak_2{|k_2|}
\def\ak_3{|k_3|}
\def\ak_4{|k_4|}
\def\aw{|w|}
\def\aw_1{|w_1|}
\def\aw_2{|w_2|}
\def\aw_3{|w_3|}
\def\aw_4{|w_4|}

\def\gev{GeV}
\def\k{\textbf{k}}
\def\l{\textbf{l}}
\def\q{\textbf{q}}
\def\m{\textbf{m}}
\def\r{\textbf{r}}
\def\w{\textbf{w}}
\def\p{\textbf{p}}
\def\x{\textbf{x}}
\def\b{\textbf{b}}
\def\kpp{\mbox{\boldmath{$\kappa$}}}

\newcommand\epjc[3]{{\it Eur. Phys. J.}{\bf C #1} (#2) #3}
\newcommand\plb[3]{{\it Phys. Lett. }{\bf B #1} (#2) #3}
\newcommand{\hep}[1]{{\tt hep-ph/#1}}
\newcommand\appb[3]{{\it Acta Phys. Polon. }{\bf B #1} (#2) #3}

\newcommand\npb[3]{{\it Nucl. Phys. }{\bf B #1} (#2) #3}
\newcommand\npa[3]{{\it Nucl. Phys. }{\bf A #1} (#2) #3}
\newcommand\jhep[3]{{\it JHEP }{\bf #1} (#2) #3}

\newcommand\prd[3]{{\it Phys. Rev. }{\bf D #1} (#2) #3}
\newcommand\prl[3]{{\it Phys. Rev. Lett. }{\bf  #1} (#2) #3}

\newcommand\zpc[3]{{\it Z. Phys. }{\bf C #1}(#2) #3}
\newcommand\app[3]{{\it Acta Phys. Polon.}{\bf B #1}(#2) #3}
\begin{document}
 \begin{titlepage}
\rightline{DESY-THESIS-2006-034}
    \vspace*{2cm}
    \begin{center}
      \LARGE \bfseries Studies of the Triple Pomeron Vertex in perturbative QCD
                 and its applications in phenomenology
    \end{center}
    \vspace{2cm}
    \begin{center}\large \bfseries
      Dissertation\\[2mm]
      zur Erlangung des Doktorgrades\\[2mm]
      des Departments Physik\\[2mm]
      der Universit\"{a}t Hamburg
    \end{center}
    \vspace{2.5cm}
    \begin{center}\large
      vorgelegt von\\[2mm]
      Krzysztof Kutak\\[2mm]
      aus Nowy S\c acz
    \end{center}
    \vspace{\fill}
    \begin{center}
      Hamburg\\[2mm]
      2006
    \end{center}
  \end{titlepage}
\newpage
\thispagestyle{empty}
\vspace*{\fill}
  \begin{tabular}{ll}
    Gutachter des Dissertation: & Prof.~Dr.~J.~Bartels\\[1mm]
                                & Prof.~Dr.~B.~Kniehl\\[1mm]
    Gutachter der Disputation: &  Prof.~Dr.~J.~Bartels\\[1mm]
                               &  Prof.~Dr.~J.~Louis\\[1mm]
    Datum der Disputation:     & 19.~12.~2006\\[1mm]
    Vorsitzender des Pr\"{u}fungsausschusses: & Dr.~K.~Petermann\\[1mm]
    Dekan der Fakult\"{a}t Mathematik,\\
    Informatik und Naturwissenschaften: & Prof.~Dr.~G.~Huber
  \end{tabular}
  \pagebreak
\newpage
\thispagestyle{empty}
  \bigskip
  \begin{center}
    {\large \bfseries\sffamily Abstract}
   
    \vspace{1cm}
    \begin{minipage}{\textwidth}
 We study the properties of the Triple Pomeron Vertex in the perturbative QCD 
 using the twist expansion method.
 Such analysis allows us to find the momenta configurations preferred by the vertex. 
 When the momentum transfer is zero, the dominant contribution in the 
 limit when $N_c\rightarrow\infty$ comes from anticollinear pole.
 This is in agreement with result obtained without expanding, but by direct averaging of the Triple
 Pomeron Vertex over angles. Resulting theta functions show that the anticollinear configuration is optimal for 
 the vertex.
 In the finite $N_c$ case the collinear term also contributes. 
 Using the Triple Pomeron Vertex we construct a pomeron loop and we also consider four gluon 
 propagation between two Triple Pomeron Vertices.
 We apply the Triple Pomeron Vertex to construct the Hamiltonian from which we derive the Balitsky-
 Kovchegov equation for an unintegrated gluon density.\\
 In order to apply this equation to phenomenology,
 we apply the Kwieci\'nski-Martin-Sta\'sto model for higher order corrections to a linear part of the Balitsky-Kovchegov
 equation. We introduce the definition of the saturation scale which reflects properties of this 
 equation. Finally, we use it for computation of observables, such as the $F_2$ 
 structure function and diffractive Higgs boson production cross section. The impact of screening
 corrections on $F_2$ is negligible, but those effects turn out to be significant for diffractive
 Higgs boson production at LHC.   
 
    \end{minipage}
  \end{center}
\vspace{0.5cm}
\newpage
\thispagestyle{empty}
\mbox{}
\newpage
\thispagestyle{empty}
\bigskip
  \begin{center}
    {\large \bfseries\sffamily Zusammenfassung}
  
    \vspace{1cm}
    \begin{minipage}{\textwidth}
Wir untersuchen die Eigenschaften der Triple-Pomeron-Vertex in der st\"orungstheoretischen 
QCD mittels einer Twistentwicklung. 
Eine solche Analyse erm\"oglicht  die Bestimmung der von der Vertex bevorzugten Impulskonfiguration. Es ergibt sich, dass f\"ur $N_c \to \infty$ und  verschwindenden 
Impulstransfer der dominierende Beitrag von dem antikollinearen Pol stammt. Dies stimmt mit dem Ergebnis \"uberein, das man erh\"alt, indem man auf eine Twistentwicklung verzichtet, 
jedoch \"uber die Winkel der Vertex mittelt. Die resultierende Theta-Funktion zeigt dann, dass die antikollineare Konfiguration bevorzugt ist. Dies \"andert sich f\"ur endliches $N_c$, 
da in diesem Fall auch die 
kollinearen Terme beitragen. Mittels der Vertex k\"onnen wir einen Pomeron-Loop konstruieren. Dar\"uber hinaus betrachten wir den Fall in dem vier Gluonen zwischen den Vertices propagieren. 
Wir benutzen die Triple-Pomeron-Vertex um die Balitsky-Kovchegov-Gleichung f\"ur die unintegrierte Gluonendichte abzuleiten.\\
Um diese Gleichung in der Ph\"anomenolgie anwenden zu k\"onnen,
benutzen wir das KMS-Modell um Korrekturen h\"oherer Ordnung in den
linearen Teil der Balitsky-Kovchegov-Gleichung einzuschlie\ss en. Wir
f\"uhren eine f\"ur diese Gleichung nat\"urliche Definition einer
Saturierungsskala ein. Wir benutzen diese schlie\ss lich, um
Observable wie die Strukturfunktion $F_2$ und den Wirkungsquerschnitt
f\"ur diffraktive Higgsbosonproduktion zu berechnen. Es stellt sich heraus, dass die Bedeutung der
Abschirmeffekte f\"ur $F_2$ vernachl\"ssigbar ist, w\"ahrend sie f\"ur die diffraktive Higgsproduktion am LHC bedeutsame Beitr\"age liefern. 

    \end{minipage}
  \end{center}
  \newpage
\thispagestyle{empty}
\mbox{}
\setcounter{page}{0}
\tableofcontents
\newpage


\chapter{Introduction}
The strong interactions in a field theory language are described by Quantum
Chromodynamics (QCD). Quark and gluon fields are its degrees of freedom. Its
structure and its predictions for phenomenology can be investigated in a 
perturbative or nonperturbative way, depending on the question we ask. In
this dissertation, we are going to address the questions which are tractable within
the perturbative approach, but which also shed light in the regime when nonperturbative
physics  starts to become important. In particular, we are interested in the high
energy limit, or Regge limit of QCD, which is defined as the limit where the energy of colliding
particles is much higher then other scales.
To be explicit let us consider two virtual photons scattering.  
The leading order diagram that contributes is just two gluon (in
singlet state) exchange in the $t$-channel. To have a realistic description one should, however, 
calculate higher order corrections in order to check the validity of the Born
term (just two gluons). 
The higher order corrections, which can be separated into a real emission part
(gluon in the $s$-channel) and a virtual emission part (gluon in the $t$-channel), turn out to be proportional      
to the logarithm of the energy multiplied by a strong coupling constant. Since the energy can be large 
the smallness of the coupling constant may be compensated and the whole series 
should be resumed.
The summation of the large logarithms of the energy accompanied by the strong coupling constant 
is performed by the Bethe-Salpeter like equation i.e. BFKL \cite{BFKL} (Balitsky,
Fadin, Kuraev, Lipatov) integro-differential equation. This equation is the
basic mathematical tool which enables us to investigate the Regge physics
within the QCD and it is believed that one should go along the line of BFKL equation to
understand its high energy behavior. In the color octet channel,
BFKL describes the propagation of a reggeized gluon. It is the gluon which
represents collective excitation of the gluonic field and which turns out to
be the basic degree of freedom in the high energy limit of QCD. On the other hand,
in the color singlet channel, the BFKL equation defines so called BFKL pomeron
which is a bound state of two reggeized gluons, and which posses vacuum quantum
numbers. Studies show \cite{KMS} that LO BFKL is not enough and 
higher order contribution is necessary in order to make BFKL applicable to 
phenomenology. What is even more 
important is the fact that at very high energies BFKL amplitude violates the unitarity
bound \cite{Froissart} which states that the amplitude of hadronic process
should not grow faster than the energy logarithm. One cannot saturate this bound
using the linear equation (even including subleading corrections). The total cross section
of a scattering process, as obtained from the BFKL, grows like a power of the
center of mass energy squared $s$. The corresponding exponent equals at leading order 
$\omega_{BFKL}=0.5$. The violation of unitarity is related to the
fact that at very high energies we apply the BFKL equation to physical system 
which is very dense (for example Deep Inelastic Scattering at Bjorken 
$x=10^{-6}$) and where the probability of recombination, i.e. probability for 
fusion of gluons, is high and this is not included in the BFKL equation. 
Such processes reduce the growth of
the partonic density and lead to the effect called perturbative saturation.

There are two main streams of activity to construct unitary theory of high energy
scattering. One of them is persuaded in momentum space where reggeized gluons are
its basic degrees of freedom. Here one can distinguish two approaches: one based on Lipatov's effective
action \cite{Lev}, and the other based on investigations of Bartels \cite{BB}. 
The latter approach is of particular interest for us. Bartels \cite{BB} generalized 
the BFKL equation to situation where more then two reggeized gluons 
(in $t$-channel) are exchanged between colliding objects
\cite{BB,BW,BE}. The physical process that one considers here is again scattering of two 
virtual photons. 
The integral equations which generalize the BFKL are written for
an amplitude which is a coupling of gluons to virtual photon via the quark loop.
The solutions of the considered equations are used to construct imaginary part of
amplitude for photon-photon scattering. 
This approach leads to the emergence of the field theory like
structure where transitions from $n$ to $m$ reggeized gluons are possible
(in contrast to BFKL Green's function which is just the propagation of two gluons or 
Bartels, Kwieci\'nski, Praszalowicz (BKP) \cite{BKP1,BKP2} which describes propagation 
of constant number of $n$ gluons).
The integral equations for $n$ gluon amplitudes form an open set
of integral equations. The equation for the three gluon amplitude can be solved. 
The equation for the four gluon amplitude can be
partially solved. It can be rewritten as an equation with
reducible and irreducible parts. The basic ingredient of the irreducible
part is a two-to-four reggeized gluons transition vertex. 
This vertex plays a central role in the physics of saturation because it allows
gluons to recombine and therefore to tame the strong power like growth of gluon 
density and therefore the cross section. 
There is a hope that the approach outlined above can be mapped to the conformal field
theory because of the fact that M\"obius symmetry is encoded in the BFKL equation and also in the two-to-four
transition vertex \cite{BLW}. If it would be possible one could use the methods of conformal
field theory to study higher order vertices and perhaps to solve this
theory.\\ 
Another approach to high density systems is based on the dipole degrees of
freedom of the high
energy limit of QCD. The basic equation here is again BFKL, but
formulated in the coordinate space \cite{Mueller94}. That equation is written for an amplitude
for scattering of an color dipole off a target (which may be another dipole). 
The evolution of a dipole is defined by summation over a
cascade of dipoles scattering off the target. There is, however, assumption that
every child dipole scatters independently of others. When the density of
dipoles is high one should allow for simultaneous scattering of child dipoles
what leads to set of coupled integro-differential equations known as
Balitsky hierarchy \cite{Bal}. The equation for scattering of a one dipole contains contribution
from two dipoles. In turn, the equation for two dipoles, contains contribution
from three dipoles and so on. 
That chain can be broken under assumption of a large target and large $N_c$. In
particular, if we are interested in the first equations from that 
hierarchy, we can assume that the amplitude for two dipoles scattering is proportional to the
product of one dipole
scattering. Under this assumption nonlinear integro-differential evolution 
equation known as
the Balitsky-Kovchegov equation \cite{Bal,Kov} can be obtained. 
The linear part, where creation of gluons is
described by the BFKL term, is accompanied by the nonlinear term
which is given by the Triple Pomeron Vertex (the same as in the $t$-channel
approach) \cite{BW} and which allows for recombination of gluons.
Balitsky-Kovchegov (BK) equation leads to the unitarisation of the dipole
amplitude at fixed impact parameter. It is the best and relatively simple equation which can be applied to
phenomenology.\\
This equation is, however, theoretically satisfying.
The proper evolution equation which could describe saturation physics
should incorporate pomeron loops. Pomeron loops, or, quantum fluctuations of, pomeron are required to
ensure projectile target symmetry which is violated by BK. They are expected
to be important at low parton densities and may be important for understanding of double 
diffractive processes  which probe their structure.
Another limitation of BK comes from the fact that the BFKL kernel and the Triple Pomeron Vertex
are considered at leading order in $\ln1/x$. 
From studies of Kwieci\'nski, Martin and Sta\'sto \cite{KMS} it is known that subleading corrections (which are modeled in 
consistency with exact NLOln1/x \cite{FL} calculation) to
linear part reduce the value of the BFKL intercept
and lower the normalization of its solution. Therefore, in order to have equation applicable
in phenomenology, one should incorporate those corrections at least in linear part of BK.
The NLO ln1/x corrections to the Triple Pomeron Vertex are still not known. 
It is not clear how to model those
corrections in a consistent way.\\ 
Some of the problems of high energy QCD listed above have been already attacked.
There are promising approaches \cite{Iancu, KovLub} to include pomeron
loops in evolution equation. There NLOln1/x corrections to BFKL are also
known\cite{FL} so there is hope that NLOln1/x corrections to Triple Pomeron Vertex can be calculated
in a similar manner. What is missing and what we find important from
the theoretical point of view is the better understanding of properties of the
Triple Pomeron Vertex  which is the basic element of already mentioned pomeron
loop and of the BK equation. By knowing the high energy QCD problems at high
energy we are going to improve the current state of understanding.\\ Understanding of the
properties of the Triple Pomeron Vertex is the main subject of this thesis. We
focus in particular on dependence on momentum variables in case of strong
ordering which is crucial for understanding of the region of phase space where
the pomeron loops become important. This vertex also appears in perturbative
approach to diffractive processes. Diffraction is still not really understood
and we hope that our analysis will be useful in understanding it. Another
motivation to study momentum dependence of the TPV is to clarify relations
between different nonlinear evolution equations that appear in the literature
(GLR, GLR-MQ, BK). To study properties of TPV we apply the
method of twist expansion which allows us to perform the analysis via expanding considered 
amplitude in terms of a largest momentum.
We also find it important to derive analogous equations to Balitsky equations directly in
momentum space. Here the interpretation is much easier and follows directly from Feynman diagrams.
This is somewhat obscure in the original formulation. Later on, assuming mean field approximation,
we derive the BK equation for the unintegrated gluon density which is in agreement with the one obtained
from transformation from coordinate space to momentum space.\\
As we have already mentioned the BK equation misses important subleading corrections and therefore can not
be directly applied for phenomenology. In order to use that equation to describe the data we propose 
to  model the subleading corrections in $\ln 1/x$ following 
\cite{KMS}. This approach turned out to be very successful in describing $F_2$ HERA data and is 
consistent with NLOln1/x corrections to the BFKL kernel. 
The next important point is the fact that, due to nonlinearity, the gluon density in saturation 
region strongly depends on the profile function of the proton. In order to have realistic 
description of  $F_2$, we study the properties of the gluon density in two important cases: 
the cylinder like profile function and the Gaussian-like. We argue that the later is  more accurate.
Finally, equipped with the fitted nonlinear equation for the gluon density, we ask ourselves
whether saturation effects, which result in hard rescattering corrections, may have any impact on
diffractive Higgs production at LHC. In existing calculations such effects were not considered.
In order to estimate such corrections we modify the standard approach introducing rescattering
corrections via the Triple Pomeron Vertex and the gluon density provided by the BK equation.   
Our estimation is rough, but can be considered as a first step towards more accurate description.\\
The thesis is organized as follows:
\begin{itemize}
\item chapter two - in this overview chapter we discuss linear evolution equations of QCD and give arguments why
saturation effects are needed.   
\item chapter three - we construct amplitude for elastic scattering of hadronic object which exchange pomeron
loop and also BKP state. We perform the collinear analysis of the Triple Pomeron Vertex.
\item chapter four - is devoted to derivation of the BK equation for the unintegrated gluon density
from an effective
Hamiltonian consisting of the BFKL kernel and the Triple Pomeron Vertex. We also discuss the relation 
of the BK equation to other nonlinear equations.
\item in chapter five we apply the BK equation with subleading corrections in $\ln 1/x$ to
describe the $F_2$ structure
function and we calculate a saturation line. We also calculate hard rescattering corrections
to diffractive Higgs production at LHC energies. 
\end{itemize}
We end this thesis with summary and conclusions.
\chapter{Evolution equations of QCD}
\section {Linear evolution equations of QCD}
In this overview chapter we are going to set up conventions that will be used 
later on in this thesis. We start with the Deep Inelastic
Scattering of an electron off a proton which is a basic scattering process in
QCD.  Then we introduce
DGLAP evolution equation which is a realization of the renormalization group in QCD.
In the next step we introduce the BFKL evolution equation which is the first step in 
perturbative QCD towards understanding of the physics of dense partonic systems. 
Finally we give arguments why rescattering corrections, where the Triple Pomeron Vertex plays 
a central role, are needed, and
we give examples of formalisms within which those corrections are discussed.
\subsection{DIS kinematics and variables}

The high energy limit of QCD can be explored thanks to the asymptotic freedom, in a
perturbative way. To introduce perturbative parameters and quantities with which we will 
work let
us consider the scattering of an electron off a proton at high momentum
transfer
\be
e(p)+N(P)\rightarrow e'(p')+ X(P_X),
\label{eq:DIS}
\ee
where $X$ is an undetected hadronic system.
The quantities that are measured are the energy and the scattering angle of the outgoing
electron.
This process can be described using standard variables:
\be
q^2=(p-p')^2=-Q^2<0,
\ee
the four momentum squared of the exchanged photon,
\be
x=\frac{Q^2}{2 p\cdot
q}=\frac{Q^2}{Q^2+W^2-m_N^2},
\ee
the Bjorken scaling variable and
\be
y=\frac{W^2+Q^2-m_N^2}{s-m_N^2},
\ee
the inelasticity.\\
In the formulas above $W^2=(q+P)^2$ is the energy squared of 
the photon-nucleon system and $s=(p+P)^2$
is the total energy squared of the electron-nucleon system.
The nucleon mass is denoted by $m_N$. When $Q^2$ involved in the scattering considered
above is much larger than $m_N^2$ we speak about Deep Inelastic Scattering.\\
\begin{figure}[t!]
\centerline{\epsfig{file=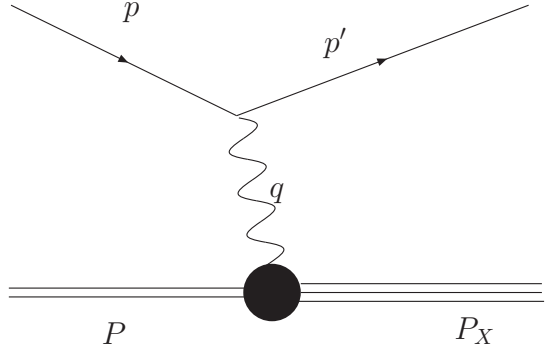,height=5cm}}
\caption{\em Deep inelastic scattering process.}
\label{fig:dis}
\end{figure}
In perturbative QCD (pQCD) the cross section for process depicted in Fig. \ref{eq:DIS} is expressed 
via kinematical variables and dimensionless quantities that contain the information about the structure 
of the nucleon and which are called structure functions $F_1(x,Q^2)$, $F_2(x,Q^2)$:
\bea
\frac{d^2\sigma}{dxdQ^2}=\frac{4\pi\alpha_{em}^2}{xQ^4}\big[\frac{1}{2}y^2(2xF_1(x,Q^2)-
F_2(x,Q^2))+\frac{1}{2}[1+(1-y)^2]F_2(x,Q^2)\big].
\eea
In pQCD structure functions can be calculated in two approaches:
\begin{itemize}
\item collinear factorization where the basic evolution equation in the 
momentum variable $Q^2$ is the DGLAP equation
\item high energy factorization where the basic evolution equation in the energy 
variable $s$ is the BFKL equation
\end{itemize}
In the two following sections we are going to give an overview of those approaches.
\subsection{DGLAP evolution}
In the collinear factorization approach the structure function is written as a convolution in
longitudinal momentum fraction:
\be
F_2(x,Q^2)=\sum_{i,q,\overline q,g}\int_x^1\!\!dy\,\,\frac{x}{y}f_i\left(\frac{x}{y},Q^2\right)C_{i}(y,Q^2)
\ee
where $C_i(y)$ (which is calculable in a perturbative way) is a coefficient
function which  represents the hard subprocess cross section for an incoming
parton. At leading order (LO) in $\alpha_s$, $C_q=e_q^2\delta(1-z)$,
$C_{\overline q}=e_{\overline q}^2\delta(1-z)$, $C_{g}=0$. The other quantity
$xf_i(x,Q^2)$ is the parton density of type $i$ and represents the probability
to pick a parton of type $i$ with longitudinal momentum fraction $x$ and
virtuality smaller than $Q^2$ out of the hadron. To relate parton densities at
different scales one uses the DGLAP evolution equation which is a  realization
of the renormalization group in QCD.
To introduce that framework let us consider scattering of a highly virtual photon with
a virtuality $Q^2$ on a proton which is characterized by a scale
$Q_0^2\!\!<\!\!\!<\!Q^2$. According to the naive parton model when the
structure of the proton is resolved,  the probe interacts with point like
partons inside the proton. The parton density turns out to be  independent on
the momentum scale and depends only on $x$. This is known as Bjorken scaling. 
However, if we introduce QCD corrections, which allow for parton interaction,
the situation changes. If we change (increase) the virtuality of the probe we
realize that the parton itself is  surrounded by a cloud of partons, so
consequently the parton density grows effectively towards larger $Q^2$.
The dependence of the parton density on the momentum scale can be written as integro-differential 
evolution equation:
\be
\frac{\partial xf_i(x,Q^2)}{\partial
\ln Q^2/Q_0^2}=\frac{\alpha_s(Q^2)}{2\pi}\sum_{j}\int_x^1dy\frac{x}{y}f_j\left(\frac{x}{y},Q^2\right)P_{ij}
(y)
\label{eq:DGLAP}
\ee
\begin{figure}[t!]
\centerline{\epsfig{file=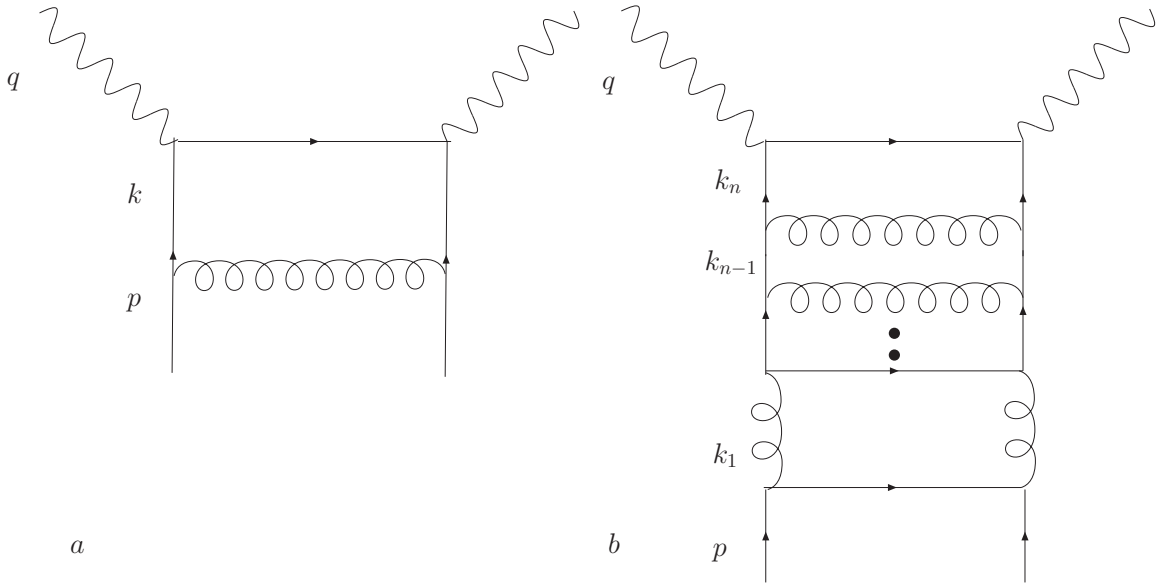,height=8cm}}
\caption{\em Ladder diagrams contributing to DGLAP evolution.}
\label{fig:dglap}
\end{figure}
The $P_{ij}$ terms are the  so called splitting functions which describe the probability of finding
a parton with momentum fraction $x$ within a parton of momentum fraction $y$.
The scale $Q_0^2$ acts as a cutoff, above which we may use perturbative methods.
Equation (\ref{eq:DGLAP}) can be interpreted as summation of ladder diagrams
(in a physical gauge) (\ref{fig:dglap}) of whose rungs are strongly ordered in
the transverse momentum. The natural evolution parameter in (\ref{eq:DGLAP})
is the resolution  $\ln Q^2/\ln Q_0^2$ (we shall put $Q_0^2= 1GeV^2$ from now
on). The full set of DGLAP equations describing evolution of quarks and gluons
takes the following form: 
\be
\frac{\partial xg(x,Q^2)}{\partial
\ln Q^2}=\frac{\alpha_s(Q^2)}{2\pi}\int_x^1dy\left[
P_{gg}(y)\frac{x}{y}g\left(\frac{x}{y},Q^2\right)+
\sum_qP_{gq}(y)\frac{x}{y}q\left(\frac{x}{y},Q^2\right)\right]
\label{eq:gluon}
\ee
\be
\frac{\partial \Sigma(x,Q^2)}{\partial
\ln Q^2}=\frac{\alpha_s(Q^2)}{2\pi}\int_x^1dy\left[
P_{qq}(y)\frac{x}{y}\Sigma\left(\frac{x}{y},Q^2\right)+
P_{qg}(y,Q^2)\frac{x}{y}g\left(\frac{x}{y},Q^2\right)\right]
\ee
\be
\frac{q_{NS}(x,Q^2)}{\partial\ln Q^2}=
\frac{\alpha_s(Q^2)}{2\pi}\int_x^1dyP_{qq}(y)\frac{x}{y}q_{NS}\left(\frac{x}{y},Q^2\right)
\ee
where $\Sigma(x,Q^2)\!\!\equiv\!\!\sum_i[q_i(x,Q^2)+\overline q_i(x,Q^2)]$ is
the singlet quark distribution and $q_{NS}=q(x,Q^2)-\overline q(x,Q^2)$ is
the nonsinglet quark distribution.  
The splitting functions read:
\be
P_{qq}(x)=C_F\left[\frac{1+x^2}{(1-x^2)_+}\right],
\ee
\be
P_{qg}(x)=T_R[x^2+(1-x)^2],\,\,T_R=\frac{1}{2},
\ee
\be
P_{gq}(x)=C_F\left[\frac{1+(1-x)^2}{x}\right],
\ee
\bea
P_{gg}(x)=2C_A\bigg[\frac{x}{(1-x)_+}
+\frac{1-x}{x}+x(1-x)\\
+\delta(1-x)\frac{(11C_A-4n_fT_R)}{6}\bigg]
\eea
where
\be
\int_0^1dx\frac{f(x)}{(1-x)_+}=\int_0^1dx\frac{f(x)-f(1)}{1-x}
\ee
and $1/(1-x)_+=1/(1-x)$ for $x<1$.\\
At very small $x$, which is our main interest, the dominant role is played by the gluons. This can be observed
analyzing the low $x$ limit of splitting functions. The splitting function $P_{gg}$
gives the most dominant contribution and may be approximated by
$P_{gg}\!\!\approx\!\!\frac{2N_c}{x}$. Neglecting the quark contribution which
is subleading at low $x$  we obtain from (\ref{eq:gluon}):
\be
\frac{\partial xg(x,Q^2)}{\partial \ln Q^2}=\frac{\alpha_s(Q^2)}{2\pi}\int_x^1dy
P_{gg}(y)\frac{x}{y} xg\left(\frac{x}{y},Q^2\right)
\label{eq:pdfDGLAP}
\ee
This equation can be solved analytically in DLLA (Double Leading Log Approximation). 
Taking the low $x$ limit of $P_{gg}$ and fixed coupling constant
(calculation with running coupling constant does not change the result very
much) and using a saddle point  approximation we obtain: 
\be
xg(x,Q^2)=xg(x,Q_0^2)\exp\sqrt{\frac{N_c\alpha_s}{\pi}\frac{1}{x}\ln\frac{Q^2}{Q_0^2}}
\label{eq:lowxDGLAP}.
\ee
Here we reintroduced $Q_0^2$ for a clearer presentation.
The gluon density (\ref{eq:lowxDGLAP}) grows fast with energy and when used for calculation of cross sections may
violate the unitarity constraints at small $x$ (more detailed discussion about unitarity will be presented later).
\subsection{BFKL approach}
An alternative approach to the evolution in the momentum variable (more suitable for high energy 
processes)
is to consider fixed momentum variable and to perform evolution in the energy variable. 
 Such an evolution
is achieved by the BFKL equation which can be formulated as a Bethe-Salpeter equation. 
The perturbative terms that are summed up are  logarithms of energy accompanied by the strong 
coupling constant $(\alpha_s \ln s/s_0)^n$, $s_0\!\!\approx\!\!M_{hadron}$ ($\ln s/s_0=\ln1/x$). 
In order to justify perturbative approach the coupling constant has to be small.  
The dedicated process where one could expect the BFKL resummation to appear is the scattering of two virtual 
photons at high center of mass energy $\sqrt s$. The BFKL integral equation reads:
\bea
(\omega -\omega(\k)-\omega(\k+\q)){\cal G}_{\omega}^{(2)\{a_i,a_i'\}}(\k,\k',\q)={\cal
G}^{(2)0\{a_i,a_i'\}}(\k,\k',\q)\nonumber\\-
\int\!\frac{d^2\l}{(2\pi)^3}\frac{1}{\k^2(\k-\q)^2}K_{2\rightarrow
2}^{\{a\}\rightarrow\{b\}}(\l,\q-\l;\k,\q-\k){\cal G}^{(2)\{b_i,a_i'\}}_{\omega}(\l,\k',\q)
\label{eq:BFKLgreen}
\eea
where $\k'$,$\k$,$\q$, $\l$ are two dimensional vectors. They represent the transverse components of 
momenta of the gluons that live in a plane transverse to the collision axis. The transverse momenta are not
ordered (in contrast to DGLAP), but are of the same order around $s_0$. The longitudinal components are
However, the longitudinal components are ordered what leads to ordering in rapidity and gives rise to 
BFKL summation.\\ 
In (\ref{eq:BFKLgreen})
\be
{\cal G}_\omega^{(2)\{a_i,a_i'\}}(\k,\k',\q)\!\!\equiv\!\!{\cal G}_\omega^{(2)}(\k,\k',\q)
P^{a_1a_2,a_1'a_2'}
\label{eq:BFKLgreen2}
\ee
is the BFKL Green's function and $P^{a_1a_2,a_1'a_2'}$ is the color projector. 
We are here in particular interested in a color singlet channel such that the
color projector is  given by:
\be
P^{a_1a_2,a_1'a_2'}=\frac{\delta^{a_1a_2}\delta^{a_1'a_2'}}{N_c^2-1}
\ee
In (\ref{eq:BFKLgreen}) the index $i$ runs from $1$ to $2$, and $\omega$ is a variable 
conjugated to the rapidity $Y$ via the Mellin transform:
\be
{\cal G}^{(2)}(Y;\k,\k',\q)=\int\frac{d\omega}{2\pi i}e^{Y\omega}{\cal G}_{\omega}^{(2)}(\k,\k',\q)
\ee
The inhomogeneous term is given by:
\be
{\cal G}^{(2)0\{a_i,a_i'\}}(\k,\k'_1,\r)=
(2\pi)^3 \frac{\delta^{(2)}(\k_1-\k'_1)}{ \k_1^2 (\k_1-\r)^2}P^{\{a_i,a'_i\}}
\ee
and
\be
K_{2\rightarrow
2}^{\{a\}\rightarrow\{b\}}=g^2f_{b_1a_1c}f_{ca_2b_2}\bigg[\q^2-\frac{\k^2(\l-\q)^2}{(\k-\l)^2}-\frac{l^2(\k-\q)}{(\k-\l)^2}\bigg]
\ee
is the real emission part of the BFKL kernel which mediates transition from two-to-two 
reggeized gluons.
The term:
\be
\omega(\k)=-N_c g^2\int\frac{d^2\l}{(2\pi)^3}\frac{\k^2}{\k^2+(\k-\l)^2}\frac{1}{(\k-\l)^2}
\ee
is the virtual part of the BFKL kernel which cancels the singularity of the real part arising
at $\k=\l$.
The cross section for scattering of hadronic objects within the BFKL formalism can be 
written in a factorized form:
\be
\sigma(Y,Q^2)=2\pi\int\frac{d^2\k_1}{(2\pi)^3}\frac{d^2\k_2}{(2\pi)^3}\phi_1^{\{a_i\}}(\k_1,Q)
{\cal G}^{(2)\{a_i,b_i\}}(Y,\k_1,\k_2,\q=0)\phi_2^{\{b_i\}}(\k_2,Q),
\label{eq:reggeprzekroj}
\ee
where the functions $\phi_1^{a_1a_2}\!\!=\!\!\delta^{a_1a_2}\phi_1$,
$\phi_2^{a_1a_2}\!\!=\!\!\delta^{a_1a_2}\phi_2$
are the impact factors and represent coupling of external particles to the Green's function.
Indices $a_i$, $b_i$ refer to the color degrees of freedom.
If we are in particular interested in DIS processes we define the unintegrated gluon density
considering the scattering of a virtual photon $\phi_{T,L}(\k_1,Q) $(transversally
or longitudinally polarized) on a proton and setting in (\ref{eq:reggeprzekroj})
as the $\phi_2$ proton impact factor and rewriting it as
\be
\sigma_{T,L}(Y,Q)=\int\frac{d^2\k}{\k^4}\phi_{T,L}(\k_1,Q)f(Y,\k^2),
\label{eq:photprot}
\ee
where
\be
 f(Y,\k^2)=\int\frac{d^2\k'}{(2\pi)^5}\k^4\phi_p(\k'){\cal G}(Y,\k,\k')
\label{eq:updf}
\ee
is the unintegrated gluon density. The color factor $C_F$, resulting from contracting color 
indices, has been absorbed in the definition of the proton impact factor. From the point of view of QCD
the proton impact factor is of nonperturbative origin and can only be modeled.
The virtual photon impact factor vanishes when momentum variable equals zero, indicating infrared
finites.\\
The usual gluon density is then given by:
\be
xg(x,Q^2)=\int^{Q^2}\!\frac{d\k^2}{\k^2}f(x,\k^2)
\label{eq:updfpdf}
\ee
and the structure functions by:
\be
F_2(x,Q^2)=\frac{Q^2}{4\pi^2\alpha_{em}}\int\frac{d^2\k}{\k^4}\phi(\k,Q)f(x,\k^2),
\label{eq:ktLOx}
\ee
where $\phi(\k,Q)=\phi_{T}(\k,Q)+\phi_{L}(\k,Q)$ and we found
convenient to use $x$ instead of $Y\!\!=\log(1/x)$ in order to use the standard DIS
DGLAP notation.
\begin{figure}[t] \centerline{\epsfig{file=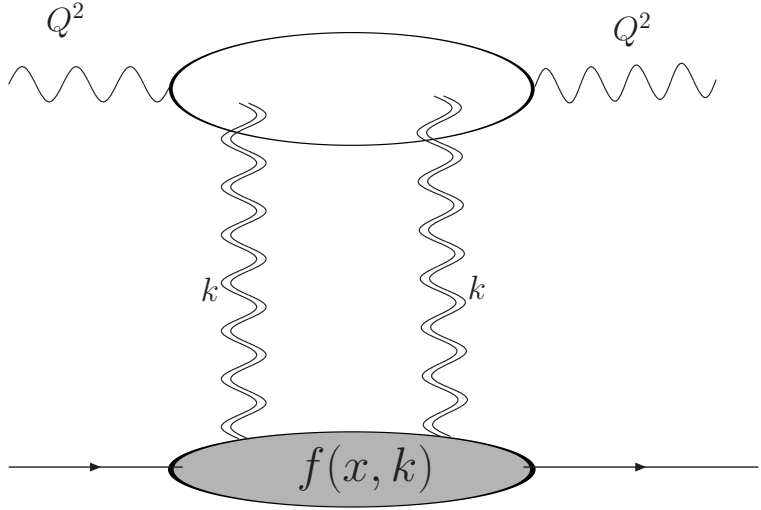,height=7cm}}
\caption{High energy factorization applied to DIS.}
\label{fig:Dipole0}
\end{figure}
Similarly, as in the previous section we may ask what is a prediction for the gluon distribution in 
the unintegrated case.
To answer that question we write the BFKL equation directly as an evolution
equation in $\ln1/x$ and since we are interested in the leading behavior, we
consider scattering in the forward direction: 
\be
\frac{\partial{\cal G}(x,\k,\k')}{\partial\ln1/x}=\frac{N_c\alpha_s}{\pi}
\k^2\int_0^{\infty}\frac{d\kpp^2}{\kpp^2}
\bigg[\frac{{\cal G}(x,\kpp^2,\k'^2)-{\cal
G}(x,\k^2,\k'^2)}{|\k^2-\kpp^2|}+ \frac{{\cal
G}(x,\k^2,\k'^2)}{\sqrt{4\kpp^4+\k^4}}\bigg] \ee
and using relation (\ref{eq:updf}) we write the BFKL equation for 
the unintegrated gluon density as:
\be
\frac{\partial f(x,\k^2)}{\partial\ln
1/x}=\frac{N_c\alpha_s}{\pi}\k^2\int_0^{\infty}\frac{d\kpp^2}{\kpp^2}
\bigg[\frac{f(x,\kpp^2)- f(x,\k^2)}{|\k^2-\kpp^2|}+ \frac{
f(x,\k^2)}{\sqrt{4\kpp^4+\k^4}}\bigg] \label{eq:updfbfkl}
\ee
Such an equation can be solved taking  the Mellin transform with respect to the momentum variable.
The procedure leads to diagonalization of the kernel. The next step is to solve the 
differential equation in rapidity $\ln1/x$ and to invert the Mellin transform. 
To invert the Mellin transform the leading eigenvalue of the
kernel is taken and integration is performed in the saddle point approximation yielding:
\be
f(x,\k^2)= \left(\frac{x}{x_0}\right)^{-\lambda}\left[\frac{\k^2/\k_0^2}{\ln(x_0/x)}\right]^{1/2}
\exp \left[- \frac{\ln^2(\k^2/\k_0^2)}{C 4\ln(x_0/x)}\right]
\label{eq:BFKL2}
\ee
with $\lambda\!=\!\frac{\alpha_sN_c}{\pi}4\ln2\!\approx\!0.5$ for $\alpha_s=0.2$. $x_0$ is 
the value of $x$ at which the evolution starts and $C=3\alpha_s14\zeta(3)/\pi$.
We see that the dominant contribution is given by the first term and the predicted growth of 
the gluon density is very large. When we
apply that result to estimate photon-proton cross section (\ref{eq:photprot}), the $x$ dependence of it
is:
\be
\sigma_{tot}\!\!\approx\!\!\left(\frac{x}{x_0}\right)^{-\lambda}\!\!=\!\!\left(\frac{s}{s_0}\right)^{\lambda}
\ee
That result is in contradiction with Froissart-Martin theorem stating that
hadronic cross section should behave as:
\be
\sigma_{tot}\le const\,\,\ln^2{s}
\label{eq:F-M}
\ee
The authors of that theorem required that the $S$ matrix, containing the information on the scattering 
process, has to be unitary and that the range of interactions is finite.
The other terms in (\ref{eq:BFKL2})
reflect the diffusion properties of the BFKL equation. One sees that $f(x,\k^2)$ is a Gaussian
distribution in $\ln(\k^2/\k_0^2)$, with a width growing as $\sqrt{\ln(x_0/x)}$ when
$x\rightarrow 0$. The diffusion property of the BFKL causes problems due to the possible
diffusion into the nonperturbative regime, where the perturbative approach is not reliable.
To cure those problems and to describe physics at low values of
$x$, a generalization of the BFKL
equation is needed. The goal is to have an evolution in rapidity, so that the essential BFKL
approach is preserved. The generalization can be understood in many ways. In this thesis  we 
are going to consider the so called rescattering corrections to the BFKL equation which
introduces nonlinear effects into the evolution process. This is the simplest modification one can do.
The outline of the approaches which are in line of the one presented in this thesis will be sketched in the
next section.\\ 
It is of particular interest to note here that (\ref{eq:updfbfkl}) 
coincides in the collinear limit (defined as $\k^2\!\!>\!\!\kpp^2$) with the DGLAP equation for 
$xg(x,Q^2)$,
when the dominant  contribution to $P_{gg}$ at low $x$ is taken.
To be explicit we observe that taking collinear limit in BFKL we get:
\be
\frac{\partial f(x,\k^2)}{\partial\ln1/x}=\frac{N_c\alpha_s}{\pi}\int^{\k^2}
\frac{d\kpp^2}{\kpp^2}f(x,\kpp^2)
\ee
Integrating over $\ln k^2$(we set $k_0^2=1GeV^2$) and using
(\ref{eq:updfpdf}) we obtain 
\be
\frac{\partial xg(x,\k^2)}{\partial\ln1/x
\partial\ln\k^2}=\frac{N_c\alpha_s}{\pi}xg(x,\k^2) 
\ee
what coincides with (\ref{eq:pdfDGLAP}) when $P_{gg}$ (the dominant contribution to it) is inserted 
and differentiated with respect to $\ln1/x$.
\section{Rescattering corrections }
The approaches described above which are based on linear evolution equations, 
do not take into account so called rescattering corrections. In case of DGLAP
one usually says that it takes into account so called leading twist
contribution to the structure function.
This means that in the DIS case the
contribution to $F_2$ is assumed to come from two gluons in the $t$-channel. In
case of the BFKL equation one should keep in mind that the gluon is in fact
reggeized what means that it gets contribution from all twists. 
However, it does not include rescattering corrections which manifest themselves in high
energy limit of QCD as a propagation of more than two reggeized gluons in the $t$-channel.  
The physical picture behind the need for including those effects 
follows when one considers parton system at
small $x$. At small $x$ the system  of partons becomes so dense that partons
start to overlap and screen each other. Thus we have to  account for
recombination. Those effects are often called rescattering corrections. The
way to describe them in momentum space in terms of Feynmann diagrams is to
consider more than two gluons in the $t$-channel which goes beyond the leading
order approximation. In particular, within the BFKL framework, the program to
include systematic higher twist effects has been formulated by Bartels and 
the basic elements of hopefully underlying  field theory (conformal), have been
constructed by Bartels \cite{BB} and W\"usthoff \cite{BW},  Bartels and Ewerz
\cite{BE}. This approach is usually called the Extended Generalized Leading
Log Approximation (EGLLA).\\ Another approach within pQCD  to the high energy
evolution and in particular to the unitarisation problem is addressed in
configuration space. It is also possible to formulate in
coordinate space the high energy scattering in terms of dipoles degrees of
freedom. The basic equation, as far as one  will not address the problem of
unitarisation, is the dipole version of the BFKL equation. However, when one 
allows for evolution to the higher energies, one has to include additional
processes which are modeled by the BK equation. Below we give more detailed
outline of those two streams of activity.

\subsection{EGLLA}
The main objects of considerations in this approach are $n$ (reggeized) gluons
amplitudes for which coupled integral equations have been
formulated \cite{BE}. 
The gluons are on the projectile side coupled to the photon impact factor
and propagate in the $t$-channel. The coupling from the target side is not fixed.
We are in particular interested in the Triple Pomeron Vertex which emerges at the level
of four gluons. The relevant equations up two four gluon amplitude are given
by: 
\be
(\omega-\sum_{i=1}^{2}\omega(\k_i))D_2^{a_1a_2}=D_{2;0}^{a_1a_2}+
K_{2\rightarrow 2}^{\{b\}\rightarrow \{a\}}\otimes D_2^{b_1b_2},
\ee
\be
\begin{split}
(\omega-\sum_{i=1}^{3}\omega(\k_i))D_3^{a_1a_2a_3}=D_{3;0}^{a_1a_2a_3}+
K_{2\rightarrow 3}^{\{b\}\rightarrow \{a\}}\otimes D_2^{b_1b_2}+
\sum K_{2\rightarrow 3}^{\{b\}\rightarrow\{a\}}\otimes D_3^{b_1b_2b_3},
\label{eq:dwatrzy}
\end{split}
\ee
\be
\begin{split}
(\omega-\sum_{i=1}^{4}\omega(\k_i))D_4^{a_1a_2a_3a_4}=D_{4;0}^{a_1a_2}+
K_{2\rightarrow 4}^{\{b\}\rightarrow \{a\}}\otimes D_2^{b_1b_2}+
\sum K_{2\rightarrow 3}^{\{b\}\rightarrow\{a\}}\otimes D_3^{b_1b_2b_3}\\
+\sum K^{\{b\}\rightarrow\{a\}}_{2\rightarrow 2}\otimes D_4^{b_1b_2b_3b_4},
\end{split}
\ee
The first of those equations is identical with the BFKL equation.
The inhomogeneous term $D_{2;0}^{a_1a_2}\equiv \phi^{a_1a_2}$ is given by the quark loop.
The superscripts refer to a color degrees of freedom of reggeized gluons.
The convolution symbol $\otimes$ stands for an integral
$\frac{d\l^2}{(2\pi)^3}$ over the loop momentum and propagators
$\frac{1}{\l^2}$ for each of two gluons entering the kernel.

The novel term appearing in (\ref{eq:dwatrzy}) is the kernel $K^{\{b\}\rightarrow \{a\}}_{2\rightarrow 3}$
which mediates transition from two to three gluons. \\
It's structure is similar to the BFKL kernel and reads:
\be
\begin{split}
K_{2\rightarrow 3}^{\{b\}\rightarrow\{a\}}=
g^3f_{b_1a_1c}f_{ca_2d}f_{da_3b_2}
\bigg[(\k_1+\k_2+\k_3)^2-\frac{\q_2^2(\k_1+\k_2)^2}{(\k_3-\q_2)^2}-\frac{\q_1^2(\k_2+\k_3)^2}{(\k_1-\q_1)^2}
\\
+\frac{\q_1^2\q_2^2\k_2^2}{(\k_1-\q_1)^2(\k_3-\q_2)^2}
\bigg]
\end{split}
\ee
The kernel $K^{\{b\}\rightarrow\{a\}}_{2\rightarrow 4}$ which mediates transition from $2$ 
to $4$ reggeized gluons reads:
\be
\begin{split}
K_{2\rightarrow 4}^{\{b\}\rightarrow\{a\}}=g^4f_{b_1a_1c}f_{ca_2d}f_{da_3e}f_{ea_4b_2}
\bigg[(\k_1+\k_2+\k_3+\k_4)^2-\frac{\q_2^2(\k_1+\k_2+\k_3)^2}{(\k_4-\q_2)^2}\\-\frac{\q_1^2(\k_2+\k_3+\k_4)^2}{(\k_1-\q_1)^2}+\frac{\q_1^2\q_2^2(\k_2^2+\k_3^2)}{(\k_1-\q_1)^2(\k_4-\q_2)^2}
\bigg]
\end{split}
\ee
The equation (\ref{eq:dwatrzy}) can be solved with an initial condition provided
by the quark loop with attached three gluons:
\be
D_{3;0}^{a_1a_2a_3}=\frac{1}{2}f_{a_1a_2a_3}[D_{2;0}(\k_1+\k_2,\k_3)-D_{2;0}(\k_1+\k_3,\k_2)+
D_{2;0}(\k_1,\k_2+\k_3)]
\ee
The expression above follows from reggeization property of the gluon at high energies. It states that
coupling of three gluons to the quark loop can be expressed as combination of couplings of two gluons to the
quark loop. 
The solution turns out to be expressible in the same way in terms of two gluon amplitude:
\be
D^{a_1a_2a_3}(\k_1,\k_2,\k_3)=\frac{1}{2}g f_{a_1a_2a_3}[D_2(12,3)-D_2(13,2)+D_2(1,23)]
\ee
where we have applied the shorthand notation $D_2(\k_1+\k_2,\k_3)=D_2(12,3)$ etc.
The structure of the solution expresses the fact that a compound state of three reggeized gluons does not
exist (at least as one considers their coupling to the quark loop).
The equation for four gluons cannot be completely solved.
It can, however, be split into a reggeizing part $D_4^R$ and an irreducible one with
respect to reggeization $D_4^I$
\be
D_4^{a_1a_2a_3a_4}=D_4^{R;a_1a_2a_3a_4}+D_4^{I;a_1a_2a_3a_4},
\ee
where
\be
\begin{split}
D_R^{a_1a_2a_3a_4}(\k_1,\k_2,\k_3,\k_4)=-g^2d^{a_1a_2a_3a_4}[D_{2}(123,4)+D_{2}(1,234)-D_{2}(14,23)]\nonumber\\
-g^2d^{a_2a_1a_3a_4}[D_{2}(134,2)+D_{2}(124,3)-D_{2}(12,34)-D_{2}(13,24)],
\end{split}
\ee
and $d^{a_1a_2a_3a_4}\!=\!\frac{1}{2N_c}\delta_{a_1a_2}\delta_{a_1a_2}+
\frac{1}{4}(d_{a_1a_2k}d_{ka_3a_4}-f_{a_1a_2k}f_{ka_3a_4})$\\
For $D_4^{Ia_1a_2a_3a_4}$ a new integral equation can be derived:
\be
(\omega-\sum_{i=1}^{4}\omega(k_i))D_4^{Ia_1a_2a_3a_4}(\k_1,\k_2,\k_3,\k_4)={\cal V}_{2\rightarrow 4}^
{a'_1,a'_2;a_1a_2a_3a_4}\otimes D_2^{a'_1,a'_2}+\sum K_{2\rightarrow 2}^{\{b\}\rightarrow\{a\}}\otimes D_4^{I b_1b_2b_3b_4}
\ee
In the above expression ${\cal V}_{2\rightarrow 4}^{\{a'_i\};\{a_i\}}$
is an effective vertex which mediates transition from two to four reggeized gluons. 
It represents a compound state of four
gluons and cannot be reduced to a sum of two gluons states.
It, however,  can be expressed in terms of the so called $G$ function which has the momentum
structure as the kernel $K^{\{b\}\rightarrow\{a\}}_{2\rightarrow 3}$ plus trajectory functions.
The explicit formula for that vertex will be presented in the next chapter.
This vertex has the property of
being gauge invariant i.e. it vanishes if any of the momentum arguments is
zero. It is also symmetric with respect to the simultaneous interchange of color
and momentum indices. A further property of the vertex (as in the case of
the BFKL) is its invariance under M\"obius transformations.
The two-to-four transition vertex plays a crucial role in construction of
theoretical framework for saturation physics because it allows to include
rescattering effects. 
\subsection{Dipole approach}
The alternative approach to unitarisation of high energy QCD is persuaded in so called dipole
picture \cite{Mueller94,NikZak}.
In this framework one considers decomposition of the photon wave function into its Fock components. 
In the lowest order, this is the bare quark-antiquark pair $q\overline q$:
\be
|\gamma>=|q\overline q>+...
\ee
The pair of $q\overline q$ forms a color dipole which scatters off the hadronic target.
The dipole formulation of the rescattering problem is often called the $s$-channel approach 
to unitarisation which is alternative to the presented above  $t$-channel approach.
Here emission of a gluon can be viewed as a splitting
of a dipole into two child dipoles in the $s$-channel producing a cascade of
dipoles in $s$-channel. In the previously discussed  EGLLA we introduced new
degrees of freedom in $t$-channel and we required the unitarity to be
fulfilled in the $t$-channel and subchannels. In $s$-channel one starts with
s-matrix $S(\x_{01},\b,Y)$  where $\x_{0\,1}\!\!=\!\!\x_0-\x_1$ is the size of
the dipole and $\x_0$ and $\x_1$ are  the transverse coordinates of the quark
and antiquark and $\b=(\x_0+\x_1)/2$ is the impact parameter. We
consider a frame where the dipole is right moving and the target is left
moving and the all relative rapidity of the dipole and the target is taken by
the target. This assures that there is only a small probability for the
emission of a gluon from the dipole before the interaction. We may ask now how
the $S(\x_{01},b,Y)$ evolve when we increase the rapidity by small amount $dY$.
Here we decide to change (increase) the rapidity of the dipole which consequently
emits a gluon.
\begin{figure}[t] \centerline{\epsfig{file=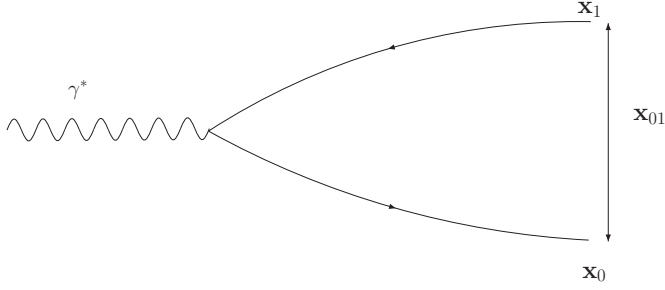,height=4cm}}
\caption{Heavy quark-antiquark dipole {\it onium}.} \label{fig:Dipole0}
\end{figure}
\begin{figure}[t]
\centerline{\epsfig{file=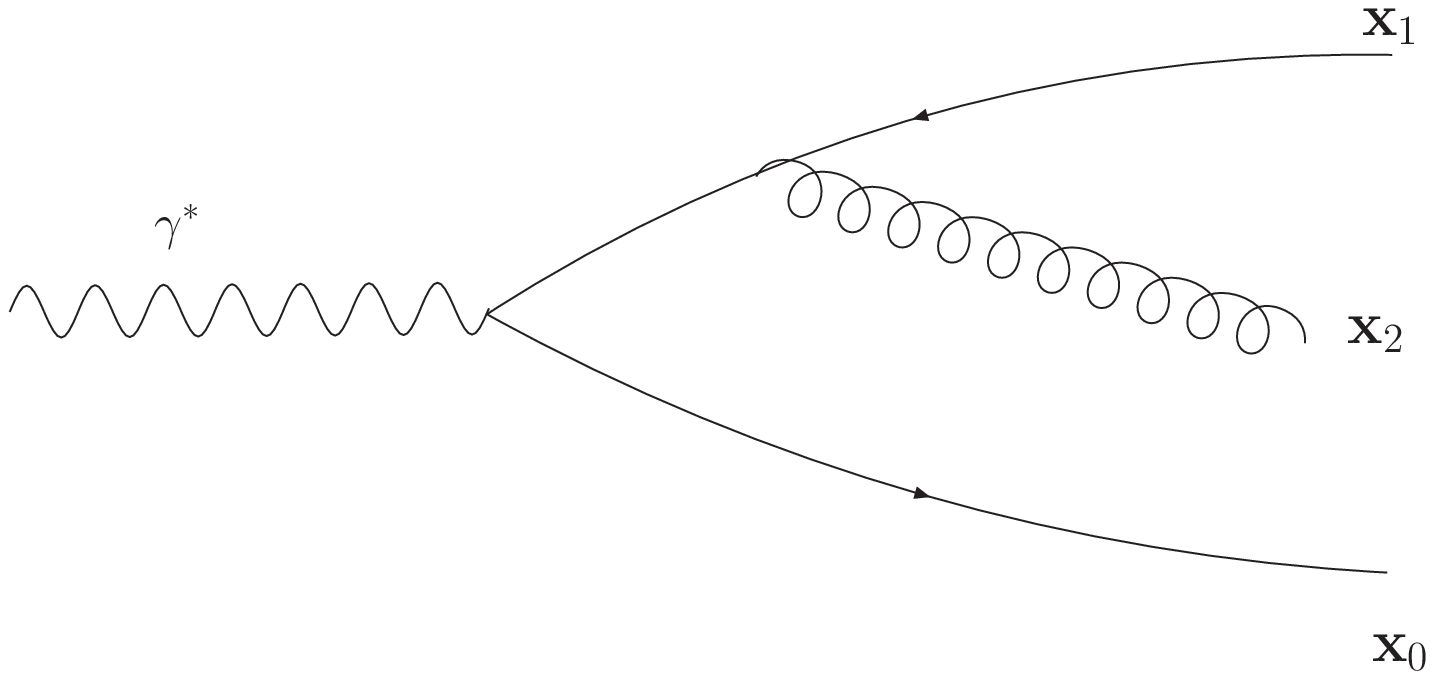,height=3.5cm}\hfill
\epsfig{file=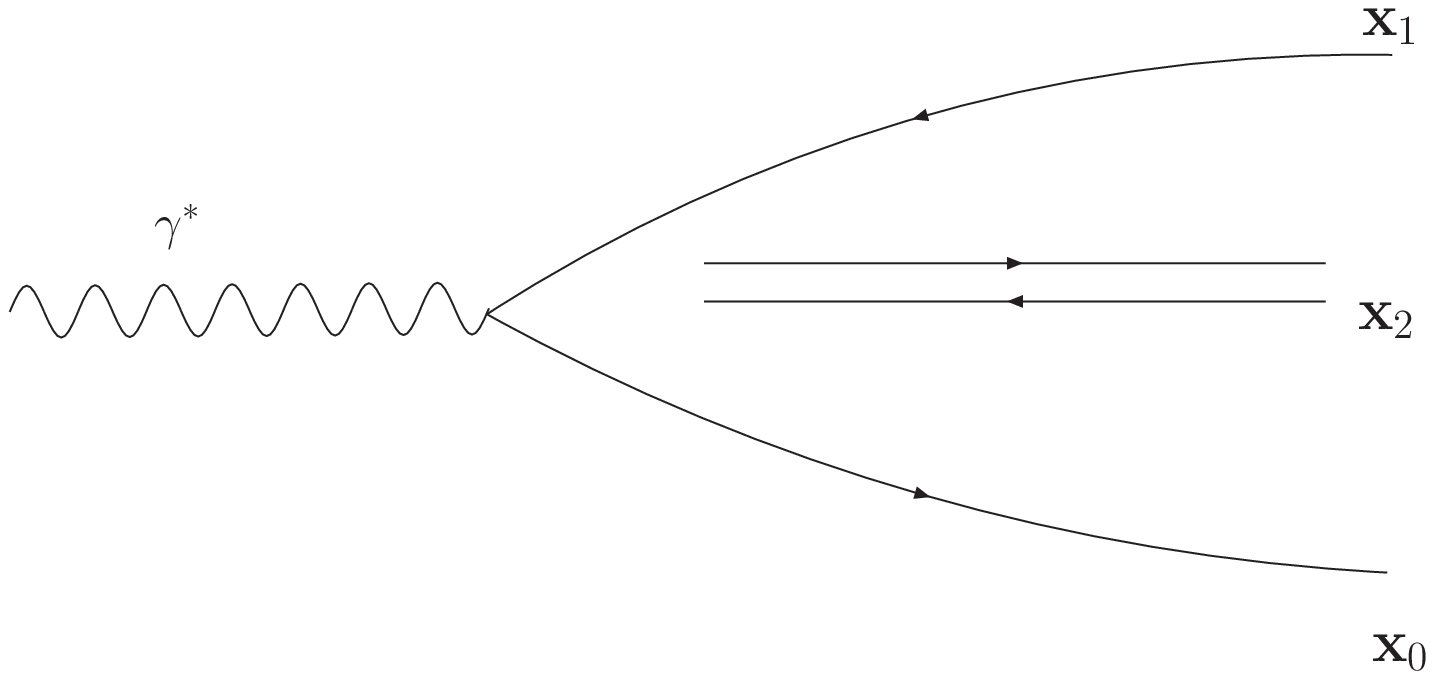,height=3.5cm}}
\caption{Onium with an additional single soft gluon.}
\label{fig:Dipole1}
\end{figure}
Due to the Fierz decomposition, the emission of a gluon in the large $N_c$
limit can be viewed as the splitting of a parent dipole into two daughter
dipoles. The probability for the production of quark-antiquark-dipole state
from the initial quark-antiquark dipole is given by:
\be
dP=\frac{\alpha_sN_c}{2\pi^2}dx_2dY\frac{x_{01}^2}{x_{02}^2x_{12}^2}
\ee
When this probability is multiplied by the $S$ matrix it the gives change in the $S$ matrix so that one can
write evolution equation \cite{MuellerBK}:
\be
\frac{\partial S(\x_{01},\b,Y)}{\partial Y}=\frac{\alpha_s N_c}{2\pi^2}\int d^2x_2
\frac{x_{01}^2}{x_{02}^2x_{12}^2}[S(\x_{02},\b+\frac{1}{2}\x_{12},Y)S(\x_{12},\b-
\frac{1}{2}\x_{20},Y)-
S(\x_{01},\b,Y)]
\ee
where the first term corresponds to the scattering of the two dipole state on the target under
the factorization assumption: 
\be
S^{(2)}(\x_{02},\x_{12},\b,Y)=
S(\x_{02},\b+\frac{1}{2}\x_{12},Y)S(\x_{12},\b-\frac{1}{2}\x_{20},Y).
\ee
This assumption is called  mean field
approximation to the gluonic field in the target. The exact treatment leads to an infinite
set of coupled integro-differential equations where the equation for scattering of the two dipole 
state on the target involves the scattering of three dipoles. The other term corresponds to virtual
contribution. One can rewrite that equation in more useful for our purposes form.
Invoking the relation between the $S$ matrix and the forward scattering amplitude $N$, $S=1-N$ 
we obtain:
\be
\begin{split}
\frac{\partial N(\x_{01},\b,Y)}{\partial Y}=\frac{\alpha_sN_c}{2\pi^2}\int d^2x_2
\frac{\x_{01}^2}{\x_{02}^2\x_{12}^2}[N(\x_{02},\b+\frac{1}{2}\x_{12},Y)+
N(\x_{12},\b+\frac{1}{2}\x_{20},Y)\\
-N(\x_{01},\b,Y)-N(\x_{02},\b+\frac{1}{2}\x_{21},Y)N(\x_{12},\b+\frac{1}{2}\x_{20},Y)]
\label{eq:BKfull}
\end{split}
\ee
That formulation in terms of the forward scattering amplitude makes an easier link to approach in
 the $t$-channel. The direct comparison is, however, impossible because of 
different spaces of formulation. The impact factor in the momentum space
corresponds to the Fourier transform of square of the wave function with the phase
factors. The forward scattering amplitude $N$ can be related via a Fourier
transform to the dipole density in the momentum space which then can be further
transformed to obtain the unintegrated gluon density. We discuss that in detail
in chapter 3. The linear part of that equation is the dipole formulation
of the BFKL equation \cite{Mueller94}. Those terms are leading when the
amplitude is small (i.e. the number of  dipoles is small) and the quadratic
term can be ignored. In the case of scattering off the nuclei one can  interpret
that contribution as single scattering of a single dipole off a nucleon in
a nuclei. The nonlinear term introduces the possibility of simultaneous
scattering of two dipoles off two nuclei and becomes important when the
amplitude grows. That term contains the same TPV that we
introduced before, but the most general  proof (with dependence on momentum
transfer) is not trivial due to complicated structure of the TPV.  In the
proof, Bartels, Lipatov and Vacca wrote most general fan diagram equation in
configuration space. Then they performed the Fourier transform of the momentum
expression of the TPV. The next step required restrictions on impact factors
that they belong to the M\"obius representation. With that assumption their
fan diagram equation became equivalent to the BK equation \cite{BLV}. We are going to discuss similar 
case later under the restriction that there is no momentum transfer.
It has been shown \cite{GBS,GBMS} that BK equation unitarises the cross section
for  fixed impact parameter and also suppress the diffusion into the infrared
region which was draw bag  of the BFKL equation.\\
It, however, violates the unitarity for a large impact parameter \cite{GBS,Kovner} which is due to lack of 
a confining mechanism included in the BK equation.
In this thesis, in part devoted to phenomenological applications we study properties of the BK
equation in situations where the impact parameter dependence is provided by initial conditions.
This allows us to avoid problems coming from diffusion to large dipoles.
Further type of limitations are linked to the mean field approximation. The BK
equation misses the effects
of pomeron loops which are important at low densities.
We show a how single pomeron loop can be constructed within the QCD, but the problem of introducing it
into the evolution equation we leave  as a task for the future.

\chapter{The Triple Pomeron Vertex}
The Triple Pomeron Vertex in the perturbative QCD has attracted significant
attention in the recent years. This particular is due to  nonlinear evolution
equations, e.g. the introduced previously BK equation,
where the nonlinearity is given by the TPV. More recently,
also the question of pomeron loops  has been addressed:
here again the TPV plays a central role. Whereas in many studies and applications it is
convenient to use the coordinate representation, it is important to
understand the structure also in momentum space.

In this chapter we will investigate aspects of the TPV, related to its dependence
on momenta arguments
starting from the momentum space representation of the $2 \to 4$ reggeized gluon
transition vertex.
So far, most of the studies of the nonlinear evolution equations have been
done in the context of deep inelastic scattering where a virtual photon
scatters off a single nucleon or off a nucleus. In both cases the momentum
scale of the photon is much larger than the typical scale of the hadron or
nucleus, i.e. one is dealing with asymmetric in configurations.
As a first step in investigations the TPV, therefore, we will focus on the
limit where the transverse momenta in the vertex are strongly ordered:
this includes both the collinear and the anticollinear limits of the TPV.
Furthermore we will also review and discuss the connection with other
nonlinear evolution equation that have been discussed in the literature (GLR, GLR-MQ)
\section{The $2 \to 4$ gluon transition vertex}
The $2\rightarrow 4$ gluon transition vertex introduced in previous chapter
\begin{figure}[t!]
\centerline{\epsfig{file=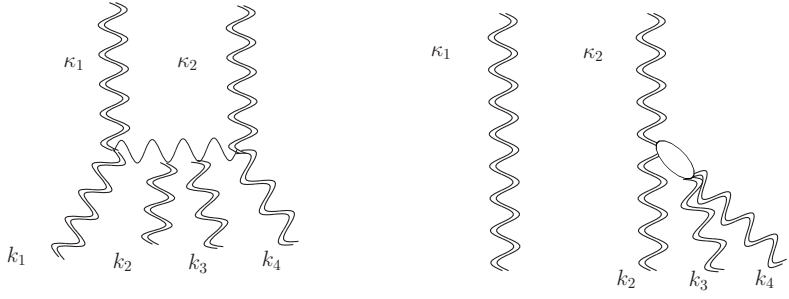,height=4cm}}
\caption{\em Example of diagrams that contribute to the $2 \to 4$ gluon
transition vertex (wavy vertical lines represent reggeized gluons):
real emission (left), a disconnected contribution (right).}
\label{fig:diagramy}
\end{figure}
consists of three pieces:
\be
\begin{split}
\verta(\kpp_1,\kpp_2;\k_1,\k_2,\k_3,\k_4)=\frac{\sqrt{2}\pi\delta^{a'_1a'_2}}{N_c^2-1}
\Bigg[\delta^{a_1a_2}\delta^{a_3a_4}V(\kpp_1,\kpp_2,\k_1,\k_2,\k_3,\k_4)\\+
\delta^{a_1a_3}\delta^{a_2a_4}V(\kpp_1,\kpp_2,\k_1,\k_3,\k_2,\k_4)
+\delta^{a_1a_4}\delta^{a_2a_3}V(\kpp_1,\kpp_2,\k_1,\k_4,\k_2,\k_3)\Bigg],
\label{eq:fvertex}
\end{split}
\ee
where $\kpp_1+\kpp_2 = \k_1+\k_2+\k_3+\k_4=\q$
and the subscripts $a_i'$, $a_i$ refer to the color degrees of freedom of the
reggeized gluons. It is convenient to express the 'basic vertex function'
$V(\kpp_1,\kpp_2,\k_1,\k_2,\k_3,\k_4)$ with the help of the function
$G(\kpp_1,\kpp_2,\k_1,\k_2,\k_3)$ 
\be
\begin{split}
V(\kpp_1,\kpp_2,\k_1,\k_2,\k_3,\k_4)=
\frac{1}{2}g^4\bigg[
G(\kpp_1,\kpp_2,\k_1,\k_2+\k_3,\k_4)+G(\kpp_1,\kpp_2,\k_2,\k_1+\k_3,\k_4)\\
+G(\kpp_1,\kpp_2,\k_1,\k_2+\k_4,\k_3)
+G(\kpp_1,\kpp_2,\k_2,\k_1+\k_4,\k_3)-G(\kpp_1,\kpp_2,\k_1+\k_2,\k_3,\k_4)\\
-G(\kpp_1,\kpp_2,\k_1+\k_2,\k_4,\k_3)
-G(\kpp_1,\kpp_2,\k_1,\k_2,\k_3+\k_4)-G(\kpp_1,\kpp_2,\k_2,\k_1,\k_3+\k_4)\\
+G(\kpp_1,\kpp_2,\k_1+\k_2,-,\k_3+\k_4)\bigg]
\label{eq:vertex}
\end{split}
\ee
The function $G(\kpp_1,\kpp_2,\k_1,\k_2,\k_3)$ \cite{BV,JPV} generalizes
the $G$ function introduced  in \cite{BW} to the non-forward direction. This function can again
be split up into two pieces:
\be
G(\kpp_1,\kpp_2,\k_1,\k_2,\k_3)=G_1(\kpp_1,\kpp_2,\k_1,\k_2,\k_3)
+G_2(\kpp_1,\kpp_2,\k_1,\k_2,\k_3),
\label{eq:funkg1}
\ee
where the first part describes the emission of a real gluon:
\bea
G_1(\kpp_1,\kpp_2,\k_1,\k_2,\k_3)=
\frac{(\k_2+\k_3)^2\kpp_1^2}{(\kpp_1-\k_1)^2}+\frac{(\k_1+\k_2)^2\kpp_2^2}{(\kpp_2-\k_3)^2}-
\frac{\k_2^2\kpp_1^2\kpp_2^2}{(\kpp_1-\k_1)^2(\kpp_2-\k_3)^2}\nonumber\\-(\k_1+\k_2+\k_3)^2
\label{eq:funkg2}
\eea
and the second one describes the virtual contribution:
\bea
G_2(\kpp_1,\kpp_2,\k_1,\k_2,\k_3)&=&-\frac{\kpp_1^2\kpp_2^2}{N_c}\Big([\omega(\k_2)-
\omega(\k_2+\k_3)]\delta^{(2)}(\kpp_1-\k_1)
\nonumber\\&&+[\omega(\k_2)-\omega(\k_1+\k_2)]\delta^{(2)}(\kpp_1-\k_1-\k_2)\Big)
\label{eq:funkg3}
\eea
As it has been already stated the vertex in
(\ref{eq:fvertex}) is completely symmetric under the permutation of the four
gluons. Moreover, it has been shown to be invariant under M\"obius
transformations ~\cite{BLW}, and it vanishes when $\kpp_i$ or $\k_i$
goes to zero.

\begin{figure}[t!]
\centerline{\epsfig{file=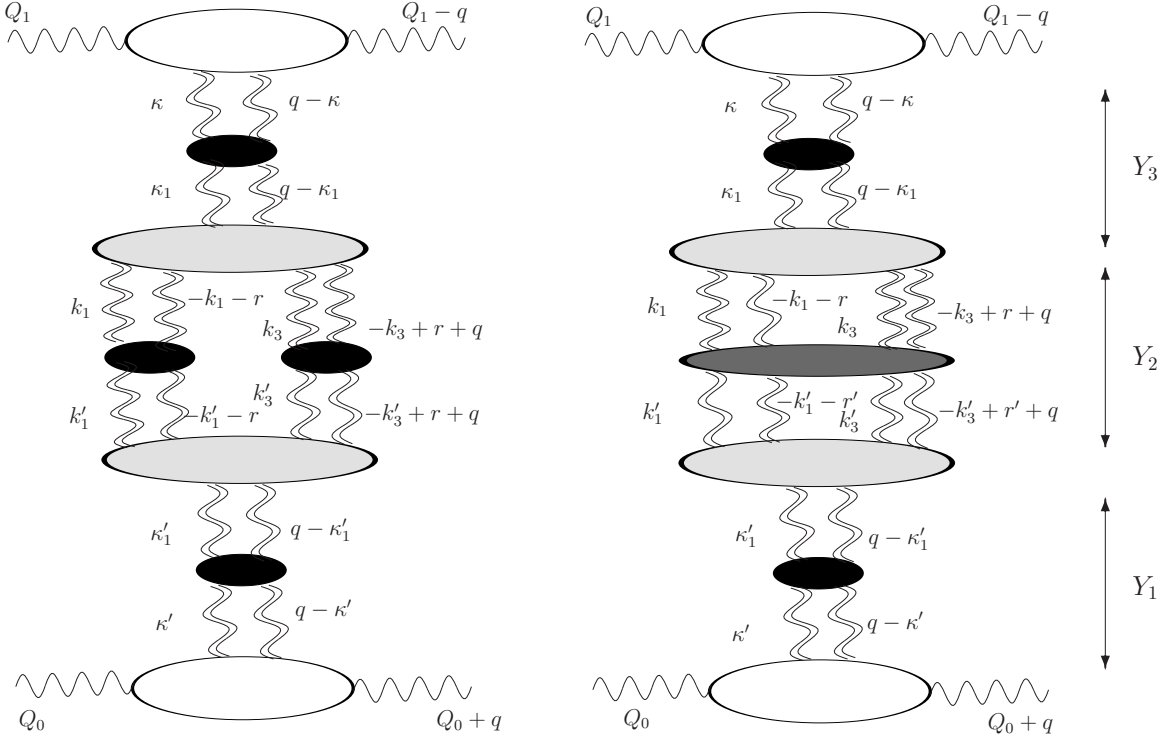,height=10cm}}
\caption{\em Contributions to the elastic scattering of virtual photons
which contain the $2 \to 4$ gluon vertex: a pomeron loop (left),
a four gluon BKP state (right). Dark blobs represent Green's functions of
reggeized gluons.}
\label{fig:pomloopBKP3}
\end{figure}

This vertex can be used, for example, to construct a pomeron loop. A simple
example for a pomeron loop in case of an elastic scattering of two virtual photons, is shown in
Fig. \ref{fig:pomloopBKP3}(left).
With the coupling of the pomeron loop to the external photons given by the impact factors,
$\phi^{a_1,a_2}$, the expression for it reads:
\bea
A(s,t)=-i2s\pi\frac{1}{2!}
\int_0^{Y} dY_3 \int_0^Y dY_2 \int_0^Y dY_1\;
\delta(Y-Y_1-Y_2-Y_3)\nonumber\\
\int\!\!\!\frac{d^2\kpp}{(2\pi)^3}\frac{d^2\kpp_1}{(2\pi)^3}
\phi^{a'_1a'_2}(\kpp,\q-\kpp)
{\cal G}^{(2)a_1'a_2',a_1^{''}a_2^{''}}(Y_3;\kpp,\kpp_1,\q)
\nonumber\\
\int \frac{d^2\r}{(2 \pi)^3} \int \frac{d^2\k_1}{(2\pi)^3}
\frac{d^2\k_3}{(2\pi)^3}{\cal V}^{a_1''a_2'';a_1a_2a_3a_4}(\kpp_1,\q-\kpp_1;\k_1,-\k_1-\r,\k_3,-\k_3+\r+\q)\nonumber\\
\int\frac{d^2\k_1'}{(2\pi)^3}\frac{d^2\k_3'}{(2\pi)^3}
{\cal G}^{(2)a_1a_2b_1b_2}(Y_2;\k_1,\k_1',\r)
{\cal G}^{(2)a_3a_4b_3b_4}(Y_2;\k_3,\k_3',\r+\q) \nonumber\\
\int\frac{d^2\kpp'_1}{(2\pi)^3}\frac{d^2\kpp'}{(2\pi)^3}
{\cal V}^{b_1b_2b_3b_4;b_1''b_2''}(\k_1',-\k_1'-\r,\k_3',-\k_3'+\r+\q;\kpp_1',q-\kpp_1')\nonumber\\
{\cal G}^{(2)b_1''b_2'',b'_1b'_2}(Y_1;\kpp_1',\kpp',\q)
\phi^{b'_1b'_2}(\kpp',\q-\kpp').
\label{eq:loop1}
\eea
Here $s$ is the squared center of mass energy, $Y=\ln(s/s_0)$ is the total
rapidity, $Y_1$, $Y_2$, $Y_3$ are the rapidity intervals as depicted in \ref{fig:pomloopBKP3} 
which sum up to the total rapidity $Y$.
${\cal G}^{(2)\{a_i,b_i\}}$ (\ref{eq:BFKLgreen2}) is the (non-amputated) BFKL Green function which
satisfies the BFKL equation.
The combinatorial factor $\frac{1}{2!}$ in front of the integral
is due to the symmetry under the interchange of the two pomerons which has
its origin in the symmetry property of the $2 \to 4$ vertex
with respect to the interchange of outgoing gluons.
The procedure that leads to (\ref{eq:loop1}) is outlined in the appendix.

As a second example, we write the expression for the propagation of the BKP
\cite{BKP1,BKP2} state between the the $2\rightarrow 4$ gluon vertices
Fig. \ref{fig:pomloopBKP3}(right).
Such a state contains the pairwise interaction of all four reggeized $t$-channel
gluons. This is in contrast with the pomeron loop where gluons are forced to form pomerons
therefore not all possible interactions are taken into account.
Its Green's function satisfies the following evolution equation:
\bea
(\omega -\omega(\k_1)-\omega(\k_2)-\omega(\k_3)-\omega(\k_4))
{\cal G}^{(4)\;\{a_i\},\{a'_i\}}_{\omega} (\{\k_i\},\{\k'_i\})=&\nonumber\\
{\cal G}^{(4)0\;\{a_i\},\{a'_i\}}(\{\k_i\},\{\k'_i\})\;
-\; \sum_{(ij)} \frac{1}{\k_i^2 \k_j^2}
K_{2\rightarrow 2}^{\{a\}\rightarrow\{b\}}\,\otimes\,
{\cal G}^{(4)\;\{b_i\}\{a'_i\}}_{\omega} (\{\k_i\}\{\k'_i\})
&
\eea
where we have used the shorthand notation $\{\k_i\}=(\k_1,\k_2,\k_3,\k_4)$ etc.
The sum extends over all pairs $(ij)$ of gluons, and the kernel
$ K_{2\rightarrow 2}^{\{a\}\rightarrow\{b\}}$ includes the color tensor
$f_{a_i b_i c} f_{a_j b_j c}$ and the convolution symbol $\otimes$ stands for
$\int\frac{d\k^2}{(2\pi)^3}$.
The inhomogeneous term has the form:
\bea
\delta^{(2)}(\sum \k_i-\sum \k_i') {\cal G}^{(4)0\;\{a_i\}\{a'_i\}}
(\{\k_i\},\{\k'_i\})= (2 \pi)^9 \prod_1^4
\frac{\delta_{a_i a'_i} \delta^{(2)}(\k_i-\k'_i)}{\k_i^2} .
\eea
The expression for the propagation of the BKP state between the $2 \to 4 $
vertices reads:
\bea
A(s,t)&=&-i2s\pi\frac{4}{4!}
\int_0^{Y} dY_3\int_0^Y dY_2\int_0^Y dY_1 \delta(Y-Y_1-Y_2-Y_3) \nonumber \\
&&\int \frac{d^2\kpp}{(2\pi)^3} \frac{d^2\kpp_1}{(2\pi)^3}
\phi^{a'_1a'_2}(\kpp,\q-\kpp){\cal G}^{(2)a_1'a_2';a_1''a_2''}(Y_3;\kpp,\kpp_1,\q)
\nonumber\\
&&\int \prod_{i=1}^4 \left( \frac{d^2k_i}{(2\pi)^3} \right) (2\pi)^3
\delta^{(2)}(\sum \k_i - \q)  \int \prod_{i=1}^4 \left(
\frac{d^2\k_i'}{(2\pi)^3}\right) (2\pi)^3  \delta^{(2)}(\sum \k_i' - \q)
\nonumber \\ &&{\cal V}^{a_1''a_2'';a_1a_2a_3a_4}(\kpp_1,\q-\kpp_1;\k_1,\k_2,\k_3,\k_4)
{\cal G}^{(4);\{a_i\}\{b_i\}}(Y_2;\{\k_i\} \{\k_i'\})
\nonumber \\
&&\int \frac{d^2\kpp'}{(2\pi)^3} \frac{d^2\kpp_1'}{(2\pi)^3}
{\cal V}^{b_1b_2b_3b_4;b_1''b_2''}(\k_1',\k_2',\k_3',\k_4';\kpp'_1,\q-\kpp'_1)\nonumber\\
&&{\cal G}^{(2)b_1''b_2'',b_1'b_2'}(Y_1;\kpp'_1,\kpp',q)
\phi^{b'_1b'_2}(\kpp',q-\kpp')
\label{eq:loop2}
\eea
The statistical factor $\frac{1}{4!}$ appears due to symmetry of
the expression under interchange of the four gluons. It is easy to see that,
when going from the BKP state to the state of two noninteracting BFKL Pomerons, the statistic
factor $\frac{1}{4!}$ is replaced by $\frac{1}{2!}$.
Again, a more detailed discussion of this diagram is presented in the appendix.

In the following we focus on the pomeron loop and
investigate, for zero momentum transfer 
$\q=0$, in the kinematic limit where, say, the momentum scale of the upper
photon is much larger than the lower one. This implies that, at the upper
TPV, the momentum from above, $\kpp_1$, is larger than the momenta from
below, $\k_1$, $\k_3$, or $\r$ ('collinear limit'). Conversely, for the
lower TPV we have the opposite situation: the momenta $\k_1'$, $\k_3'$,
and $\r$ are larger than $\kpp_1'$ ('anticollinear limit').
Let us become a bit more formal. We expand the amplitude of Fig. \ref{fig:pomloopBKP3} 
in powers of $Q_0^2/Q_1^2$ ('twist expansion'). The object of our interest is the
self-energy of the Pomeron Green's function, $\Sigma(\kpp_1,\kpp'_1)$.
In eq.(\ref{eq:loop1}), $\Sigma(\kpp_1,\kpp'_1)$ is defined to be
line 3 - 5, i.e. the convolution of the two TPV's with the two BFKL Green
functions between them. It has the dimension $\k^2$, and it is convenient to
define the dimensionless object $\tilde\Sigma(\frac{\kpp_1}{\kpp'_1})=
\frac{\Sigma(\kpp_1,\kpp'_1)}{\sqrt{\kpp_1^2\kpp_1^{'2}}}$
with the Mellin transform:
\be
\tilde\Sigma(\gamma)=\int_0^{\infty}dk^2\tilde\Sigma(k^2)(k^2)^{-\gamma-1}.
\label{eq:mellin1}
\ee
The inverse Mellin transform reads:
\be
\tilde\Sigma(k^2)=\int_C \frac{d\gamma}{2\pi i}(k^2)^{\gamma}
\tilde \Sigma(\gamma),
\label{fig:mellinfig}
\ee
where $k^2=\frac{\kpp_1^2}{\kpp_1^{'2}}$ and the contour crosses the real axis
between $0$ and $1$ see Fig. \ref{fig:mellinfig}.
Our analysis will then reduce to the study of the singularities
of the function $\tilde\Sigma(\k^2)$. The twist expansion corresponds to
the analysis of the poles located to the left of the contour
in the $\gamma$ plane: the pole at $\gamma=0$ is the leading twist pole,
the pole at $\gamma=-1$ belongs to twist 4 and so on.
For the upper TPV in Fig. \ref{fig:pomloopBKP3}, the analysis of the twist
expansion requires the 'collinear limit', for the lower TVP the
'anticollinear' one.
\begin{figure}[t!] \centerline{\epsfig{file=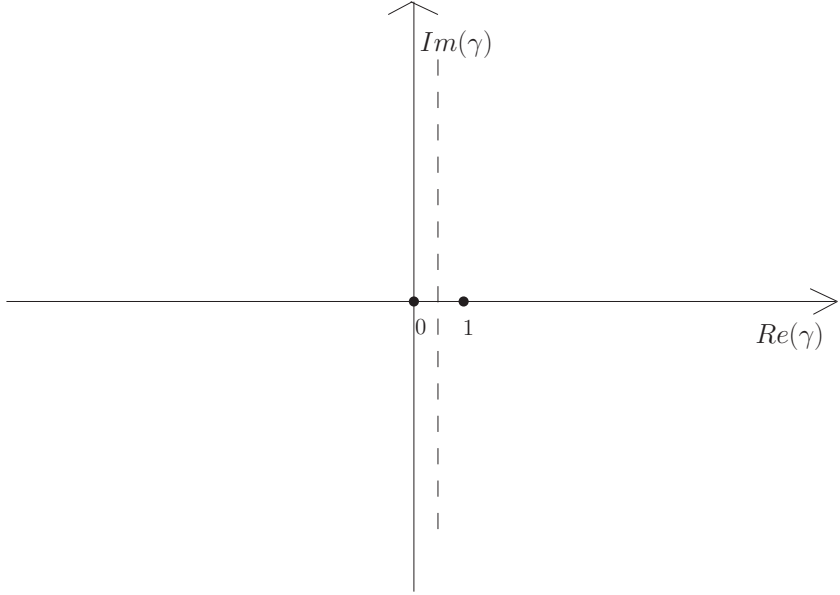,height=8cm}}
\caption{\em Singularities in the $\gamma$ plane.}
\label{fig:mellinfig}
\end{figure}
\section{The collinear limit}
In this section we are going to study the collinear limit of the TPV.
The ordering of the transverse momenta is the following:
$|\k|\!\!\gg\!\!|\k_1|,|\k_2|,|\k_3|,|\k_4|$
(and expansion parameters are $|\k_1|/|\k|$, $|\k_2|/|\k|$, $|\k_3|/|\k|$,
$|\k_4|/|\k|$).
In our investigations we will be interested in attaching color singlet objects to the vertex 
so we project (\ref{eq:fvertex}) onto the color singlets using
$P^{a_1a_2a_3a_4}\!=\!\delta^{a_1a_2}\delta^{a_3a_4}/(N_c^2-1)$.
In the multicolor limit $N_c\rightarrow \infty$ we obtain:
\be
\begin{split}
P^{\{a\}}{\cal V}^{\{a';a\}}(\kpp_1,\kpp_2;\k_1,\k_2,\k_3,\k_4)=
\delta^{a_1',a_2'}\frac{\sqrt{2}\pi}{N_c^2-1}\bigg[(N_c^2-1)V(1,2,3,4)
+V(1,3,2,4)\\+V(1,4,2,3)\bigg]
\simeq\dell V(1,2,3,4)
\equiv{\cal V}_{L0N_c}^{\{a'\}}(1,2,3,4)
\label{eq:kovl}
\end{split}
\ee
where $\tilde\delta^{a_1',a_2'}=\sqrt{2}\pi\delta^{a_1',a_2'}$
, $V(1,2,3,4)\equiv V(\kpp_1,\kpp_2;\k_1,\k_2,\k_3,\k_4)$.

\subsection{The real part}
Let us begin the analysis by expanding the real part of the $G$ function (\ref{eq:funkg1}) in the
collinear limit. In the analysis we are going to limit ourselves to
the forward case.  In the forward configuration we are going to use the
simplified notation $G(\kpp,-\kpp;\k_1,-\k_1-\k_3,\k_3)\equiv G(\k_1,\k_3)$
and similarly for the vertex  and basic vertex function
$V(\kpp,-\kpp;\k_1,\k_2,\k_3,\k_4)\!\equiv\!V(\k_1,\k_2,\k_3,\k_4)$.
According to these definitions the $G_1$ (\ref{eq:funkg2}) function reads:
\be
\begin{split}
G_1(\k_1,\k_3)=
\frac{\k_1^2\k^2}{(\k-\k_1)^2}+
\frac{\k_3^2\k^2}{(\k+\k_3)^2}-\frac{(\k_1+\k_3)^2
\k^4}{(\k-\k_1)^2(\k+\k_3)^2} 
\label{eq:greal}
\end{split}
\ee
where we used $\kpp_1\!=\!\k$, $\kpp_2\!=\!-\k$.
In the collinear limit the momenta of outgoing gluons in (\ref{eq:greal}) satisfy conditions 
$|\k_1|\!\!<\!<\!\!|\k|$, $|\k_3|\!\!<\!<\!\!|\k|$. Performing the expansion up to second order terms
we obtain from (\ref{eq:greal}):
\be
\begin{split}
G_1(\k_1,\k_3)\simeq 2\k^2\Bigg[-\frac{\k_1\cdott
\k_3}{\k^2}-\frac{\k_1\cdott \k\;\k_3^2}{\k^4}+\frac{2\k_1\cdott \k_3\;
\k_3\cdot \k}{\k^4} -\frac{2\k_1\cdott \k_3\;\k_1\cdott \k}{\k^4}
+\frac{\k_3\cdott \k\;\k_1^2}{\k^4}\\+\frac{2 (\k_1\cdott
\k_3)^2-\k_1^2\;\k_3^2}{\k^4}\Bigg]
\end{split}
\ee
The first interesting term, which can be identified as the twist two contribution,
is:
\be
G_{1}(\k_1,\k_3)\simeq-2\,\k^2\,\frac{\k_1\cdott \k_3}{\k^2}
\ee
where, since we are interested in convoluting the TPV
with forward and angular averaged BFKL kernel we averaged over the azimuthal angle of $\k$. 
Such contribution has the potential to give us the logarithmic
integral over the transverse momentum. The presence of the momentum transfer would cause loss of a
logarithmic contribution.
We kept the same notation for the averaged function.
After inserting this expression into (\ref{eq:vertex}) the whole expression vanishes.
Passing to the twist four
\begin{equation}
G_{1}(\k_1,\k_3)\simeq 2\k^2\left[\frac{2 (\k_1\cdott \k_3)^2-\k_1^2
\k_3^2}{\k^4}\right]
\end{equation}
we obtain for the TPV:
\begin{equation}
\cVr(\k_1,\k_2,\k_3,\k_4)=\tilde\delta^{a_1'a_2'} 8\frac{g^4}{2} \k^2\frac{
(-\k_1\cdott \k_2\,\k_3\cdott \k_4+
\k_1\cdott \k_4\,\k_2\cdott \k_3+\k_1\cdott \k_3\,\k_2\cdott \k_4)}{\k^4}
\end{equation}
The expression above is the master formula for twist four contribution. In the above
we introduced a shorthand notation. The superscript $r$ stands for the real emission
and similarly we will also use $v$ for the disconnected (virtual) contributions in subsequent sections.
In a spirit of our strategy we are going to consider the forward configuration of the TPV in the collinear
limit i.e. when momenta are:
\be
\k_1=-\k_2,\,\,\,\,\k_3=-\k_4
\ee
Putting $\k_1=\l$, $\k_2=-\l$, $\k_3=\m$, $\k_4=-\m$ we obtain:
\begin{equation}
\cVr(\l,-\l,\m,-\m)=\tilde\delta^{a_1'a_2'} 8
\frac{g^4}{2}\k^2\frac{2(\l\cdott \m)^2 - \l^2 \m^2}{\k^4}
\end{equation}
Having the expression for the vertex we are going to attach the BFKL ladders to its external legs.
To achieve that, we multiply the vertex by propagators and act on it with the
BFKL kernels (one for each leg). As a result we get the vertex and two interacting
gluons attached to it.
\begin{figure}[t!]
\centerline{\epsfig{file=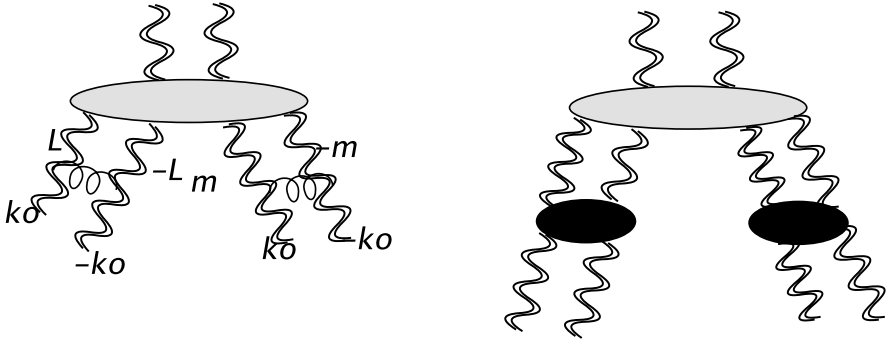,height=4.5cm}}
\caption{\em TPV with two interacting gluons attached to it(left).
TPV with ladders of gluons(right).}
\label{fig:drabiny}
\end{figure}
This procedure can be iterated and as a result we will obtain a vertex with attached
BFKL ladders Fig. \ref{fig:drabiny}(right). We wish to obtain this result
within leading logarithm accuracy, and our goal is to get
maximal power of the logarithm from the convolutions of the vertex with propagators and kernels.
To do that we should act on the twist four contribution to the vertex with the twist four operator,
which in our case is the product of two twist two contributions to the BFKL kernel
(twist two for each leg).
\begin{figure}[t!]
\centerline{\epsfig{file=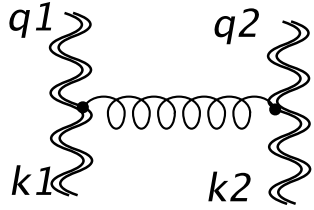,height=3cm}}
\caption{\em Kinematics used in \ref{eq:eqBFKL}.}
\label{fig:BFKLreal}
\end{figure}
Let us compute  the collinear approximation to the BFKL kernels which,
 convoluted with the TPV, will give the maximal power of the logarithm.
The expression for the emission part of the BFKL kernel Fig. \ref{fig:BFKLreal}
has already been presented, but we will recall it here (we will not write
below $K_{BFKL}$ to simplify the notation): 
\be
K(\q_1,\q_2;\k_1,\k_2)=(\k_1+\k_2)^2 -\frac{\q_2^2
\k_1^2}{(\k_2-\q_2)^2}-\frac{\q_1^2 \k_2^2}{(\k_1-\q_1)^2}
\label{eq:eqBFKL}
\ee
Assuming zero momentum transfer and $\q^2_1\!>\!\!\!>\!\k^2_2$ we get:
\be
K_{coll}\!\simeq\!-2\k_1^2
\ee
where we focused on the emission part since the virtual part does not contribute in the
strong ordering approximation. The subscript $coll$ stands for collinear.
Using this formula in the case of kinematics represented in Fig.
\ref{fig:drabiny}(left) we get the following expression for convolution of
the vertex with the kernels (one kernel for each leg of the vertex):
\be
(K_{coll1}K_{coll2})\otimes
\cVr  \simeq  \tilde\delta^{a_1'a_2'}16\pi \k^2\frac{g^4}{2}
\int_{\k_0^2}^{\k^2}\frac{d^2\l}{(2\pi)^3}\int_{\k_0^2}^{\k^2}\frac{d^2\m}{(2\pi)^3}
\frac{2\k_0^2}{\l^4}\frac{2\k_0^2}{\m^4}
\frac{(2(\l\cdott \m)^2 - \l^2 \m^2)}{\k^4}
\label{eq:lmvet},
\ee
where $\k_0$ is the lower scale which we do not specify at present.
We notice that integrals over $\l$ and $\m$ give logarithms, but what is more
striking is that angular integral over angle between $\l$ and $\m$ reduces
the expression above and in particular the Triple Pomeron Vertex to zero. 
This result can also be obtained without expanding. 
To see that, it is enough to angular average over the azimuthal angles of momenta in the vertex (see appendix):
\bea
V(\k,-\k;\l,-\l,\m,-\m)=4\frac{g^4}{2}\bigg[2\k^2\theta(\l^2-\k^2)\theta(\m^2-\k^2)\nonumber\\
+\ln\left(\frac{\l^2}{\m^2}\right)\delta(\l^2-\k^2) \theta(\m^2-\l^2)
+\ln\left(\frac{\m^2}{\l^2}\right)\delta(\m^2-\k^2)\theta(\l^2-\m^2)\bigg]
\label{eq:melG2}
\eea
The presence of $\theta$ functions forbids
collinear configuration.  The physical meaning of that result is
clear. If the two pomerons entering the vertex from below have smaller momenta than the
pomeron from above, they cannot resolve it and cannot merge because they do
not feel 'color' and the vertex vanishes\footnote {This has first been noticed in ~\cite{GLR}, see our
discussion in chapter devoted to nonlinear evolution equations.}. 
\subsection{The virtual part} So far we have investigated contributions that
come from the real part of the vertex. What is remaining is the disconnected
part. Virtual pieces will be investigated convoluted with an impact
factor. To deal with infrared finite quantities we choose to work with the
impact factor of the photon. The function $G_2(\k_1,\k_3)$ (\ref{eq:funkg3}) in the forward
direction reads: 
\be
G_2(\k_1,\k_3)=-2\k^4\ln\frac{|\k_1|}{|\k_1+\k_3|}\delta(\k_1^2-\k^2)
-2\k^4\ln\frac{|\k_3|}{|\k_1+\k_3|}\delta(\k_3^2-\k^2)
\label{eq:gvirt}
\ee
and the photon impact factor (transverse which is the leading one) is:
\be
\phi_{a_1'a_2'}(\k,Q)=\delta^{a_1'a_2'}\alpha_s\alpha_{em}\sum_q e_q^2\int_0^1 d\tau
d\rho\frac{[\rho^2-(1-\rho)^2]
[\tau^2-(1-\tau)^2]\k^2}{\rho(1-\rho)Q^2+\tau(1-\tau)\k^2}
\ee
In this expression $\rho$ denotes the longitudinal component of $k$ (in Sudakov decomposition) while
second integration variable $\tau$ is the Feynmann parameter.
However, for
our investigations it is enough to
limit ourselves to twist expansion.
To perform twist expansion of the impact factor one has to perform the Mellin
transform of it with respect to $\k^2/Q^2$.
With the Mellin transform (\ref{eq:mellin1}) one has:
\be
\phi_{a_1'a_2'}(\k,Q)=\int\frac{d\gamma}{2\pi
i}\left(\frac{\k^2}{Q^2}\right)^{\gamma}\phi_{a_1'a_2'}(\gamma)
\ee
and we obtain:
\be
\phi_{a_1'a_2'}(\gamma)=\delta^{a_1'a_2'}{\cal C}\frac{\Gamma(-\gamma+3)}
{(-\gamma+\frac{5}{2})}\frac{\Gamma(-\gamma+1)}{-\gamma+1}
\frac{\Gamma(\gamma)}{\gamma}\frac{\Gamma(\gamma+2)}{\Gamma(\gamma+\frac{3}{2})} \ee
Expanding that function and closing the contour of inverse Mellin transform to
the left we obtain the following collinear expansion of the impact factor:
\bea
\phi_{a_1'a_2'}(\l,Q_1)=\delta^{a_1'a_2'}\phi(\l,Q_1)\simeq
\delta^{a_1'a_2'}\B\Bigg\{\left[\frac{14}{9}-\frac{4}{3}\ln\left(\frac{\l^2}{Q_1^2}\right)\right]
\frac{\l^2}{Q_1^2} +
\frac{2}{5}\left(\frac{\l^2}{Q_1^2}\right)^2+\Bigg\}...
\label{eq:imp}
\eea
while in the anticollinear case one has:
\be
\phi_{b_1'b_2'}(\p,Q_0)=\delta^{b_1'b_2'}\phi(\p,Q_0)\simeq
\delta^{b_1'b_2'} \B\Bigg\{-\frac{2}{5}\left(\frac{\p^2}{Q_0^2}\right)^{-1}
+\left[-\frac{14}{9}-\frac{4}{3}\ln\left(\frac{\p^2}{Q_0^2}\right)\right]+...\Bigg\}
\ee
where $\B=\sum_f e_f^2\alpha_s\alpha_{em}$. The momentum scales are ordered as follows
$|Q_1|\!\gg\!|\l|\!\gg\!|\p|\!\gg\!|Q_0|$.
$Q_1^2$ stands for the virtuality of the photon which is larger then the momentum $\l^2$: 
while $Q_0^2$ stands for a virtuality which is smaller then $\p^2$. From now on we
will use the notation: $\phi(\k,Q)\equiv\phi(\k)$.\\
The virtual terms in the leading part of the vertex can be combined to give:
\be
\begin{split}
\phi_{\{a'\}}(\k)\otimes\cVv(\l,-\l,\m,-\m)\simeq-\frac{1}{(2\pi)^2}
\tilde\delta^{a_1'a_1'}
\frac{g^4}{2}\Bigg[\left(\ln\frac{\l^2}{(\l+\m)^2}+
\ln\frac{\l^2}{(\l-\m)^2}\right)\phi(\l,Q_1)\\
+\left(\ln\frac{\m^2}{(\l+\m)^2}+\ln\frac{\m^2}{(\l-\m)^2}\right)\phi(\m,Q_1)\Bigg]
\label{eq:lvirt}
\end{split}
\ee
Assuming that $|\l|\!>\!|\m|$ and inserting the twist four contribution to $\imp(\l)$
($\B\frac{2}{5}(\frac{\l^2}{Q_1^2})^2$)  we can expand
(\ref{eq:lvirt}) in terms of $|\m|/|\l|$: 
\be
\begin{split}
\phi_{\{a'\}}(\k)\otimes\cVv(\l,-\l,\m,-\m)\simeq
-2\B\frac{1}{(2\pi)^2}\delta^{a_1'a_1'}\frac{g^4}{2}\frac{2}{5}\Bigg[-2\frac{\m^2}{\l^2}
\frac{\l^4}{Q_1^4}+
4\frac{(\l\cdott\m)^2}{\l^4}\frac{\l^4}{Q_1^4}\\
+2\ln\left(\frac{\m^2}{\l^2}\right)\frac{\m^4}{Q_1^4}
-2\frac{\m^2}{\l^2}\frac{\m^4}{Q_1^4}
+4\left(\frac{\l\cdott\m}{\l^2}\right)^2 \frac{\m^4}{Q_1^4}\Bigg]
\end{split}
\ee
After convolution with BFKL kernels we obtain:
\be
\phi_{\{a'\}}(\k)\otimes\cVv(\l,-\l,\m,-\m)\otimes(K_{acoll1}K_{acoll2})\simeq
-2\B\frac{1}{(2\pi)^6}\tilde\delta^{a_1'a_1'}
\frac{g^4}{2}\frac{2}{5}\frac{\k_0^6}{Q_1^6}\ln\frac{\k_0^2}{Q_1^2}
\ee
We see that this expression does not give us the expected power (third) of logarithm.
Thus we arrive at the conclusion that  we do not get a contribution from the TPV in the collinear
limit in the leading logarithmic approximation.

\section{The anticollinear limit}
\subsection{Real parts}
\begin{figure}[t!] \centerline{\epsfig{file=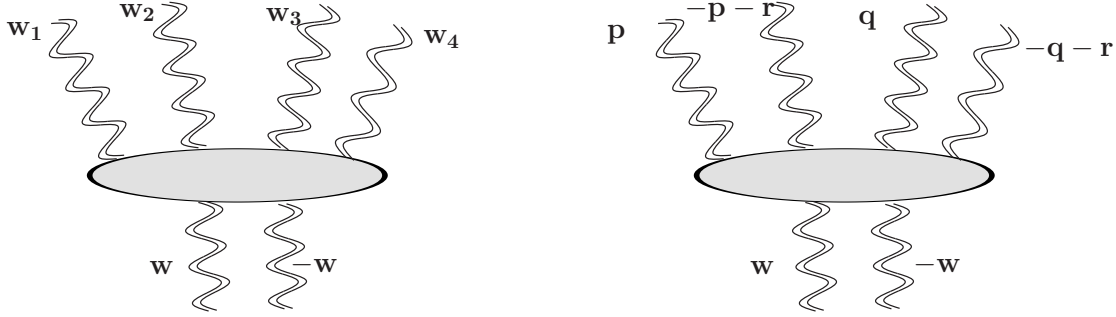,height=4.5cm}}
\caption{\em Configuration of the momenta that is considered in the anticollinear limit.}
\label{fig:TPVrealanticol}
\end{figure}

Let us now
investigate the anticollinear limit. In the anticollinear configuration we allow for momentum
to be transfered between the legs of the vertex. The momentum transfer here, as we will see, does not lead
to loss of a logarithm. In this configuration
$\w\!\!<\!<\!\!|\w_1|\!,|\w_2|\!,|\w_3|\!,|\w_4|$  see Fig.
\ref{fig:TPVrealanticol}. To study the emission part of the TPV it is convenient to rewrite the
$G_1$ function in the form: 
\be
G_1(\w_1,\w_3)=2\w^2\left[\frac{1}{(1-\frac{\w\cdot\w_1}{\w_1^2})^2}+
\frac{1}{(1+\frac{\w\cdot\w_3}{\w_3^2})^2}-\frac{\left(\frac{\w\cdot\w_1}{\w_1^2}+\frac{\w\cdot
\w_3}{\w_3^2}\right)^2}
{(1-\frac{\w\cdot\w_1}{\w_1^2})^2(1+\frac{\w\cdot\w_3}{\w_3^2})^2}\right]
\label{eq:Ganticol}
\ee
the expansion parameters are $|\w|/|\w_1|$ and $|\w|/|\w_3|$. Performing
the expansion we obtain:
\be
G_1(\w_1,\w_3)\simeq 2\w^2\left[1-\frac{\w\cdott\w_3}{\w_3^2}+\frac{\w\cdott
\w_1}{\w_1^2}-\frac{\w_1\cdott\w_3}{\w_1^2\w_3^2}\w^2\right]
\label{eq:rozwG}
\ee
Using (\ref{eq:Ganticol}) and averaging it over the azimuthal angle of $\w$ 
we obtain for the TPV:
\be
\cVrr(\p,-\p-\r,\q,-\q+\r)\simeq\tilde\delta^{b_1'b_2'}4\frac{g^4}{2} 2\w^2
\ee
This term also appears when forward configuration is considered. Upon angular
averaging over angles (without any expansion) this term survives, showing that
the anticollinear configuration is preferred by the TPV (see appendix). 
However, in order to get the required logarithmic contribution after
convolution with the BFKL kernels we need to consider higher
order terms in \ref{eq:rozwG}. The resulting contribution is the following: 
\be
\cVrr(\w_1,\w_2,\w_3,\w_4)\simeq \tilde\delta^{b_1'b_2'}\frac{g^4}{2}2\w^4    
               \Bigg[-\frac{\w_1\cdott \w_3}{\w_1^2
\w_3^2}-\frac{\w_2\cdott\w_3}{\w_2^2 \w_3^2} -\frac{\w_1\cdott \w_4}{\w_1^2
\w_4^2}-\frac{\w_2\cdott \w_4}{\w_2^2 \w_4^2} \nonumber\ee \be
\begin{split}
+\frac{\w_1\cdott (\w_3+\w_4)}{\w_1^2 (\w_3+\w_4)^2}+
                     \frac{\w_2\cdott (\w_3+\w_4)}{\w_2^2 (\w_3+\w_4)^2}+
                     \frac{\w_3\cdott (\w_1+\w_2)}{\w_3^2 (\w_1+\w_2)^2}+
                     \frac{\w_4\cdott (\w_1+\w_2)}{\w_4^2 (\w_1+\w_2)^2}\\
                   -\frac{(\w_1+\w_2)\cdott(\w_3+\w_4)}
                      {(\w_1+\w_2)^2(\w_3+\w_4)^2}\Bigg]
\end{split}
\label{eq:mastac}
\ee
To proceed further we need the
anticollinear limit of the BFKL kernel. Using (\ref{eq:eqBFKL}), setting
$\k_1=\w_1$, $\k_2=-\w_1-\r$, $\q_1=\p$, $\q_2=-\q_1-\r$ and
requiring that $|\q_1|\!\gg\!|\w_1|\!,|\r|$ we obtain:
\be
K_{acoll}=-2\w_1\cdott\w_2,
\label{eq:approxkern}
\ee
where $\w_2=-\w_1-\r$.
As already mentioned before we are interested in a particular configuration, where
there is a momentum flow
between the legs of the vertex, but where there is no momentum
flow out of the vertex. Such a configuration is realized by setting:
$\w_1\!\!=\!\!\p$, $\w_2\!\!=\!\!-\p\!\!-\!\!\r$,$\w_3\!\!=\!\!\q$,
$\w_4\!\!=\!\!-\q\!\!+\!\!\r$.
In order to obtain the logarithmic contribution after the acting of the BFKL kernels on the TPV,
 we are going to consider the following cases:
\begin{itemize}
\item The configuration where $|\r|\!\!>|\p|\!>\!|\q|$. Using (\ref{eq:mastac}) we obtain for the vertex:
\be
\cVrr(\p,-\p-\r,\q,-\q+\r)\simeq-\delta^{b_1'b_2'}4\frac{g^4}{2}2\w^4\frac{\p\cdott
\q}{\p^2 \q^2}
\label{eq:lncrerpq}
\ee
while the propagators read:
\be
\frac{1}{\p^2(\p+\r)^2}\simeq \frac{1}{\p^2\r^2}
\ee
\be
\frac{1}{\q^2(\q+\r)^2}\simeq\frac{1}{\q^2\r^2}
\ee
using (\ref{eq:approxkern}) we obtain for the kernels:
\be
K_{1}=2\p\cdott(-\p-\r)\simeq-2\p\cdott \r\equiv K_{acoll1}
\ee
\be
K_{2}=2\q\cdott(\q+\r)\simeq 2\q\cdott \r\equiv K_{acoll2}
\ee
Convoluting those expressions we obtain:
\be
{\cal V}^{r\{b'\}}_{LON_c}
\otimes(K_{acoll1}K_{acoll2})\simeq\!\tilde\delta^{b'_1b'_2}\frac{g^4}{2}
\!\!\int_{\w^2}^{\w_0^2}\!\!\!\!
\frac{d^2\r}{(2\pi)^3}\!\!\!\int_{\w^2}^{\r^2}\!\!\!\frac{d^2\p}{(2\pi)^3}\!\!\!\int_{\w^2}^{\p^2}\!\!\!\frac{d^2\q}{(2\pi)^3}\w^4
\!\frac{2\p\cdott \r}{\p^2\r^2}\!\frac{2\q\cdott \r}{\q^2 \r^2}\!\frac{2\p\cdott \q}{\p^2
\q^2}\!=\nonumber
\ee
\be
\delta^{b_1'b_2'}\!\frac{2\pi^3}{(2\pi)^9}\!\frac{g^4}{2}
\!\frac{\w^4}{3!}\!\!\!\left(\ln\frac{\w_0^2}{\w^2}\right)^3,
\label{eq:tot11}
\ee
where $|\!\w_0\!|$ is specified only by the condition
that it should be smaller than the momentum scales $|\l|$ and $|\m|$ which were considered in the 
collinear limit. Due to the additional integration over the momentum transfer this expression
has different dimension than considered in the collinear limit.
Convolution with the impact factor
$\imp_{\{b'\}}(\w)=-\delta^{b_1'b_2'}\frac{2}{5}\B\ln Q_0^2/\w^2$  yields: 
\be
\begin{split}
\imp_{\{b'\}}\otimes \cVrr\otimes(K_{acoll1}K_{acoll2})\simeq-\tilde\delta^{b_1'b_1'}\frac{2}{5}\B
\frac{\pi^4}{(2\pi)^8}\frac{g^4}{2}
\int_{Q_0^2}^{\w_0^2}\frac{d^2\w}{(2\pi)^3}\frac{1}{\w^4}\frac{Q_0^2}{\w^2}
\frac{\w^4}{3!}\left(\ln\frac{\w_0^2}{\w^2}\right)^3\\
=-\delta^{b_1'b_1'}\frac{2}{5}\B\frac{2\pi^4}{(2\pi)^{12}}
\frac{g^4}{2}\frac{Q_1^2}{4!}\left(\ln\frac{\w_0^2}{Q_0^2}\right)^4
\label{eq:kas1}
\end{split}
\ee
where in order to get the logarithmic contribution we took the lowest order term in the expansion of
$\imp$.
In the configuration $|\q|\!\!\!>\!\!\!|\p|$ the same result is obtained.
\item  Repeating the similar analysis in the case when $|\q|\!\!>\!\!|\p|\!\!>\!\!|\r|$, we obtain:
\be
\imp_{\{b'\}}\otimes \cVrr\otimes(K_{acoll1}K_{acoll2})
=-\delta^{b_1'b_1'}\frac{2}{5}\frac{\B}{2}\frac{(2\pi)^4}{(2\pi)^{12}}
\frac{g^4}{2}\frac{Q_1^2}{4!}\left(\ln\frac{\w_0^2}{Q_1^2}\right)^4
\label{eq:cancel}
\ee
\item And when $|\q|\!>\!|\r|\!>\!|\p|$, we obtain:
\be
\imp_{\{b'\}}\otimes \cVrr\otimes(K_{acoll1}K_{acoll2})
=-\delta^{b_1'b_1'}\frac{2}{5}\frac{\B}{4}\frac{(2\pi)^4}{(2\pi)^{12}}
\frac{g^4}{2}\frac{Q_1^2}{4!}\left(\ln\frac{\w_0^2}{Q_1^2}\right)^4
\label{eq:kas2}
\ee
The configuration  $|\p|\!>\!|\r|\!>\!|\q|$ gives the same contribution.
\end{itemize}

\subsection{Virtual parts}
Let us now analyse the contribution coming from the virtual parts of the vertex
in the anticollinear limit.
Using (\ref{eq:vertex}) and (\ref{eq:gvirt}) we list below a set of functions which contribute in the case when $|\p|,|\q|\!\!>|\r|$:
The other configurations will not contribute.
\be
\impf G_2(\p,\r)=-\delll\frac{1}{(2\pi)^2}\ln\frac{|\p|}{|\p+\r|}\imp(\p)-
\delll\frac{1}{(2\pi)^2}\ln\frac{|\r|}{|\p+\r|}\imp(\r)\\
\label{eq:seti5}
\ee

\be
\impf
G_2(-\p-\r,\r)=-\delll\frac{1}{(2\pi)^2}\ln\frac{|\p+\r|}{|\p|}\imp((\p+\r)^2)-
\delll\frac{1}{(2\pi)^2}\ln\frac{|\r|}{|\p|}\imp(\r)\\
\label{eq:seti6}
\ee

\be
\impf G_2(\q,-\r)=-\delll\frac{1}{(2\pi)^2}\ln\frac{|\q|}{|\q-\r|}\imp(\q)-
\delll\frac{1}{(2\pi)^2}\ln\frac{|\r|}{|\q-\r|}\imp(\r)\\
\label{eq:seti7}
\ee

\be
\impf
G_2(-\q+\r,-\r)=-\delll\frac{1}{(2\pi)^2}\ln\frac{|\q-\r|}{|\q|}\imp((\q-\r)^2)-
\delll\frac{1}{(2\pi)^2}\ln\frac{|\r|}{|\q|}\imp(\r)\\
\label{eq:seti8}
\ee
Analyzing the case when $|\q|\!>\!\!\!|\p|>\!\!\!|\r|$, we obtain contributions
from (\ref{eq:seti5}), (\ref{eq:seti6}). As an example,
let us look for contribution arising from (\ref{eq:seti5}).
Expanding $1/2\ln\frac{\r^2}{(\p+\r)^2}\imp(\r)$ in terms of $\r^2/\p^2$ we
obtain:
\be
\frac{1}{2}\ln\frac{\r^2}{(\p+\r)^2}\imp(\r)\simeq
\B\frac{2}{5}\frac{1}{4}\frac{Q_0^2}{\r^2}\left[\ln\frac{\r^2}{\p^2}-\frac{\r\cdott\p}{\p^2}
+\frac{\r^2}{\p^2}-2\left(\frac{\r\cdott\p}{\p^2}\right)^2\right]
\ee
Proceeding similarly with the other $G_2$ function we obtain
\be
\impf{\cal V}^{r\{b'\}}_{LON_c}=-\delta^{b_1'b_1'}\frac{\B}{2}\frac{2}{5}\frac{1}{(2\pi)^2}\frac{Q_0^2}{\r^2}
\ln\frac{\r^2}{\p^2}
\ee
Convoluting it with kernels and setting the lowest scale equal to $Q_0^2$ we find:
\be
\impf
\cVvb\otimes(K_{acoll1}K_{acoll2})\simeq\delta^{b_1'b_1'}\frac{2}{5}\frac{\B}{2}
\frac{(2\pi)^4}{(2\pi)^{12}}\frac{Q_0^2}{\w_0^2}\left(\ln\frac{\w_0^2}{Q_0^2}\right)^4
\ee
which is exactly the same expression as (\ref{eq:cancel}), but with the opposite sign.
The remaining $G$ function gives contribution in the case where
$|\p|\!\!>\!\!\!|\q|\!\!>\!\!|\r|$ which cancels againist with analogous coming
from the real part.

Let us shortly summarize our results before we continue the finite $N_c$ analysis.
The contribution coming from the anticollinear part of the vertex is given by (\ref{eq:kas1})
when
$|\r|\!\!>\!\!|\p|\!>\!\!|\q|$ and  $|\r|\!\!>\!\!|\q|\!\!>\!\!|\p|$. In the case
when $|\p|\!>\!|\q|\!>\!|\r|$ and $|\q|\!>\!|\p|\!>\!|\r|$ we do not get
contribution, because the appropriate terms cancel each other.
If $|\q|\!\!>\!\!|\r|\!\!>\!\!|\p|$ and $|\p|>\!\!|\r|\!\!>\!\!|\q|$ we finally obtain
(\ref{eq:kas2}).
Our goal was to find the terms of the twist expansion of the TPV which, after convolution with the BFKL kernels would generate the maximal possible  power of the logarithm. In the
collinear case we expected  to find two logarithms and after
convolution with an impact factor, a third one. Disconnected pieces did not give
contributions either.
In the anticollinear case, however, we find the logarithmic
contributions in the two of the specified regions.
In the forthcoming section  we are going to investigate  contributions to the  vertex
that are suppressed in the large $N_c$ limit.
\section{Finite $N_c$}
\subsection{The collinear limit}
Analysis of previous sections repeated for subleading parts in $N_c$ gives:
\be
\csVr(1,3,2,4)\simeq
\frac{\delta^{a_1'a_2'}}{N_c^2-1}\frac{g^4}{2}\k^2\frac{8 (\k_1\cdott \k_2
\k_3\cdott\k_4- \k_1\cdott\k_3\k_2\cdott\k_4+\k_1\cdott\k_4\k_2\cdott
\k_3)}{\k^4}
\ee
Substituting $\k_1=\l$, $\k_2=-\l$, $\k_3=\m$, $\k_4=-\m$ we obtain:
\be
\csVr(\l,\m,-\l,-\m)\simeq
\frac{\delta^{a_1'a_2'}}{N_c^2-1}\frac{g^4}{2}\k^2
\frac{8\l^2\m^2}{\k^4}
\ee
And analogously for the second subleading part. Analysis of the second
and third non leading term give the same results so we limit ourselves to the second one.
The convolution with the two BFKL kernels  gives:
\be
(K_{coll1}K_{coll2})\otimes
\csVr\simeq\frac{\delta^{a_1'a_2'}}{N_c^2-1}\k^2\frac{g^4}{2}\int_{\k_0^2}^{\k^2}\frac{d^2\l}
{(2\pi)^3}
\int_{\k_0^2}^{\l^2}\frac{d^2\m}{(2\pi)^3}\frac{2\k_0^2}{\l^4}\frac{2\k_0^2}{\m^4}
\frac{(8 \l^2\,\m^2)}{\k^4}
\ee
$$
=\frac{\delta^{a_1'a_2'}}{N_c^2-1}\frac{(2\pi)^2}{(2\pi)^6}
8\frac{g^4}{2}\frac{\k_0^4}{\k^2}\frac{1}{2!}\left(\ln\frac{\k^2}{\k_0^2}\right)^2 $$
Convolution with the impact factor gives:
\be
\impfa
\csVr(K_{acoll1}K_{acoll2})\simeq\frac{\delta^{a_1'a_1'}}{N_c^2-1}\frac{2}{5}4\B\frac{(2\pi)^3}{(2\pi)^9}
\frac{g^4}{2}\frac{\k_0^4}{Q_1^4}\frac{1}{3!}\left(\ln\frac{Q_1^2}{\k_0^2}\right)^3
\label{eq:totsubreal}
\ee

As we have already observed in order to obtain the logarithmic contribution from disconnected
piece, we are forced to expand $G$ function in terms of ratio $|\l|/|\m|$ or reverse.
It turns out that the virtual pieces of subleading in $N_c$ part of the vertex give
contribution. The contribution comes from the following functions:
\be
\impfa
G_2(\l,-\l-\m)=-\delta^{a'_1a'_2}\frac{1}{(2\pi)^2}\ln\frac{|\l|}{|\m|}\imp(\l)-
\delta^{a'_1a'_2}\frac{1}{(2\pi)^2}\ln\frac{|\l+\m|}{|\m|}\imp(\l+\m)
\ee
\be
\impfa
G_2(-\l,\l+\m)=-\delta^{a'_1a_2}\frac{1}{(2\pi)^2}\ln\frac{|\l|}{|\m|}\imp(\l)-
\delta^{a'_1a'_2}\frac{1}{(2\pi)^2}\ln\frac{|\l+\m|}{|\m|}\imp(\l+\m)
\ee
\be
\impfa
G_2(\m,-\l-\m)=-\delta^{a'_1a'_2}\frac{1}{(2\pi)^2}\ln\frac{|\m|}{|\l|}\imp(\m)-
\delta^{a'_1a'_2}\frac{1}{(2\pi)^2}\ln\frac{|\l+\m|}{|\l|}\imp(\l+\m)
\ee
\be
\impfa
G_2(\m,\l+\m)=-\delta^{a'_1a'_2}\frac{1}{(2\pi)^2}\ln\frac{|\m|}{|\l|}\imp(\m)-
\delta^{a'_1a'_2}\frac{1}{(2\pi)^2}\ln\frac{|\l+\m|}{|\l|}\imp(\l+\m)
\ee
Expanding the functions above in terms of $|\m|/|\l|$, collecting all terms and
integrating over angles  yields for convolution of $\csVv (\l,\m,-\l,-\m)$ with
the impact factor and kernels, the following result:
\be
\impfa\csVv\otimes(K_{coll1}K_{coll2})\simeq-\frac{\tilde\delta^{b_1'b_1'}}{N_c^2-1}
6\B\frac{2}{5}\frac{g^4}{2}\frac{(2\pi)^3}{(2\pi)^9}
\frac{\k_0^4}{Q_1^4}\frac{1}{3!}\left(\ln\frac{Q_1^2}{\k_0^2}\right)^3
\label{eq:totsubvirt}
\ee
The power of the logarithm is the same as in the real part, when we compare (\ref{eq:totsubvirt}) with
(\ref{eq:totsubreal}). The total contribution reads:
\be
\impfa\csVv\otimes(K_{coll1}K_{coll2})\simeq-\frac{\tilde\delta^{a_1'a_1'}}{N_c^2-1}
2\B\frac{2}{5}\frac{g^4}{2}\frac{(2\pi)^3}{(2\pi)^9}
\frac{\k_0^4}{Q_0^4}\frac{1}{3!}\left(\ln\frac{Q_1^2}{\k_0^2}\right)^3
\label{eq:totsubvirt}
\ee
\subsection{The anticollinear limit}
Here the results differs only in the kinematical region
where $|\q|\!\!\!>\!\!\!|\p|\!\!\!>\!\!\!|\r|$ and $|\p|\!\!\!>\!\!\!|\q|\!\!\!>\!\!\!|\r|$ i.e. 
we do not get a piece from the vertex that convoluted with kernels give logarithms.
The other regions give the similar results as the leading $N_c$.
To be explicit:
\begin{itemize}
\item $|\r|\!>\!|\p|\!>\!|\q|$ and $|\r|\!>\!|\q|\!>\!|\p|$
\be
\impf\cVrrr\otimes(K_{acoll1}K_{acoll2})\simeq-
\frac{\delta^{b_1'b_1'}}{N_c^2-1}\frac{2}{5}\B\frac{2\pi^4}{(2\pi)^{12}}
\frac{g^4}{2}\frac{Q_0^2}{4!}\left(\ln\frac{\w_0^2}{Q_0^2}\right)^4
\label{eq:anticolfiniNc1}
\ee
which is exactly the same as (\ref{eq:tot11}), but suppressed by $N_c^2-1$.
\item $|\p|>\!|\r|\!>\!|\q|$ and $|\q|\!>\!|\r|\!>|\p|$
\be
\impf\cVrrr\otimes(K_{acoll1}K_{acoll2})
=-\frac{\delta^{b_1'b_1'}}{N_c^2-1}\frac{2}{5}\frac{\B}{4}\frac{(2\pi)^4}{(2\pi)^{12}}
\frac{g^4}{2}\frac{Q_0^2}{4!}\left(\ln\frac{\w_0^2}{Q_0^2}\right)^4
\label{eq:anticolfiniNc2}
\ee
which is exactly the same as (\ref{eq:kas2}), but suppressed by $N_c^2-1$.
\end{itemize}
Repeating the similar analysis for the virtual part we observe that none of the specified regions gives the
leading  logarithmic contribution and the real part gives the total contribution.
The considered limits do not cover all possible cases. We could for example consider situations when
the momentum transfer is the largest scale in the problem.
\chapter{Nonlinear evolution equations}
\section{The Balitsky-Kovchegov equation for the unintegrated gluon density}
In this section we are going to derive the BK equation.
Let us consider the state of $k$ reggeized gluons in the Heinsenberg picture which is 
labeled by color and momentum degrees of freedom:
\be
|k\rangle=|\k_1..\k_k;a_1,..a_k\rangle
\ee
The normalization is:
\be
\langle k|k'\rangle=\delta_{k'k}\frac{1}{k'!}
\sum_{\sigma(k)}\prod_{i=1}^{k'}\left((2\pi)^{3}\delta(\k_i-\k_i')\k_i^{2}
\delta^{a_ia_i^{'}}\right)
\ee
where the sum is over the permutations of outgoing gluons.
Using this states as a base we assume now that the proton
state can be written as:
\be
|p\rangle=\sum_{k'=1}^\infty\!c_{k'}|\!k'\rangle
\ee
Evolution of this state is given by:
\be
e^{-yH}|p\rangle=|p(y)\rangle,
\ee
where the Hamiltonian is given by:
\be
H=H_{2\rightarrow 2}+H_{2\rightarrow 4}+H_{4\rightarrow 2}
\ee
Its matrix element are defined via:
\be
\begin{split}
\langle k|H_{2\rightarrow
2}|k'\rangle=\delta_{kk'}\bigg[\frac{1}{(k'-2)!}\sum_{\sigma(k'-2)}\!\sum_{i>j=1}^{k'}f_{a_ib_i'c}f_{cb_j'a_j}
K(\k_i,\k_j;\k_i',\k_j')
\prod_{l\neq ij}^{k'}\delta(\k_l-\k_l')\delta^{a_lb_l'}\\
+\sum_i^{k'}\prod_l^{k'}\delta_{a_lb_l'}\k_i^2\omega(\k_i')
\delta(\k_i-\k_i')\bigg]
\label{eq:ham1}
\end{split}
\ee
This part corresponds to the BKP interaction. 
The last two terms account for gluons which propagate freely without interaction. 
\be
\begin{split}
\langle k|H_{4\rightarrow 2}|k'\rangle=
\delta_{k\,k'-2}\sum_{s\!>\!t}^{k}\!
\sum_{i\!>\!j\!>\!l\!>\!r\!=1}^{k'}\!\!\!\!\!{\cal V}^{a_sa_t;a_i'a_j'a_l'a_r'}
(\k_s,\k_t;\k_i',\k_j',\k_l',\k_r')\\
\delta(\k_i'+\k_j'+\k_l'+\k_r'-\k_s-\k_t)\prod_{p\neq i,j,l,r}^{k'}\delta(\k_p-\k'_p)\delta^{a_pa'_p}
\label{eq:ham2}
\end{split}
\ee
this part of the Hamiltonian corresponds to the changing number of gluons interaction.
It allows for
four gluon to fuse into two gluons.\\
Finally the part which corresponds to the transition of two gluons into four gluons is given by:  
\be
\begin{split}
\langle k|H_{2\rightarrow 4}|k'\rangle=\delta_{k\,k'+2}\sum_{s\!>\!t=1}^{k'}
\sum_{i\!>\!j\!>\!l\!>\!r\!=1}^{k}
{\cal V}^{a_ia_ja_la_r;a_s',a_t'}(\k_i,\k_j,\k_l,\k_r;\k_s',\k_t')\\
\delta(\k_i+\k_j+\k_l+\k_r-\k_i-\k_j)\prod_{p\neq i,j,l,r}^{k'}\delta(\k_p-\k'_p)\delta^{a_pa'_p}
\end{split}
\label{eq:ham3}
\ee
Let us define the wave function of $k$ gluons in the proton at rapidity $y$ 
in the following way: 
\be
{\cal F}_{k}^{\{a_i\}}=\langle k|e^{-yH}|p\rangle\equiv(2\pi)^3{\cal
F}_{k}^{\{a_i\}}(y,\k_1,\k_2...\k_n) 
\ee

Upon differentiating it with respect to $y$ we obtain:
\bea
\frac{\partial {\cal F}^{\{a_i\}}_{k}}{\partial y}=-\langle
k|He^{-yH}|p\rangle =-\sum_{k'}\langle k|H|k'\rangle\langle
k'\!|e^{-yH}|\!p\rangle\nonumber\\= -\sum_{k'}\langle k| H|\!k'\rangle{\cal
F}^{\{a_i'\}}_{k'} \label{eq:coupledeq}
\eea
The unity operator is given by:
\be
\sum_{k'}|\!k'\!\rangle\langle\!k'\!|=\sum_{k'}^{\infty}\prod_{i=1}^{k'}\int\frac{d^2\k_i'}{(2\pi^3)}
\frac{1}{\k_i^{'2}}|\!\k_i';a_i'\rangle\langle\!\k_i';a_i'\!|,
\ee
where the summation on the left hand-side means also integration over the continuous degrees of 
freedom. The expression (\ref{eq:coupledeq}) is a set of the infinite, many coupled equations. 
It cannot be closed because, for instance equation for two gluon wave function involves 
contribution coming from four gluon wave function.
\be
\frac{\partial{\cal F}^{a_1a_2}_2}{\partial y}=-(\langle 2|H_{2\rightarrow 2}|2\rangle\otimes{\cal
F})_2^{a_1a_2}-
(\langle2|H_{4\rightarrow 2}|4\rangle\otimes {\cal F}_4)^{a_1a_2}
\ee
The term proportional to (\ref{eq:ham3}) vanishes since 
it requires zero gluons in the initial state. The minus sign appearing in front of the linear term is 
due to our definition of the BFKL kernel. In forward case one gets the usual relative sign difference
between linear and nonlinear parts. 
One can, however, write closed equation using factorisation ansatz: 
\be
{\cal F}_4(y,\k_1,\k_2,\k_3,\k_4)^{a_1a_2a_3a_4}=c{\cal F}_2(y,\k_1,\k_2)^{a_1a_2}
{\cal F}_2(y,\k_3,\k_4)^{a_3a_4}
\label{eq:factorisation}
\ee
This assumption is strictly speaking justified for a large nuclei where it is natural to
consider coupling of one pomeron to one nucleon. At present we, however, assume that we can use this
approximation for the proton therefore we obtain: 
\be
\frac{\partial{\cal F}^{a_1a_2}_2}{\partial y}=
-(\langle 2|H_{2\rightarrow 2}|2\rangle\otimes{\cal
F}_2)^{a_1a_2}-c(\langle 2|H_{4\rightarrow 2}|4\rangle\otimes
({\cal F}_2 {\cal F}_2))^{a_1a_2}
,
\label{eq:balbk}
\ee
The factorisation (\ref{eq:factorisation}) gives some freedom of choosing the
parameter $c$ \cite{Janik}.
We will choose the constant $c$ equal to $1/\sqrt{2}$ 
to get the usual version of BK equation. 
To obtain the BK equation for the unintegrated gluon density let us define the unintegrated gluon 
density via:
\be
{\cal F}_2^{a_1a_2}(y,\k_1,\k_2)=
(2\pi)^3{\cal F}_2(y,\k_1,\k_2)\frac{\delta^{a_1a_2}}{2N_c}=\langle k|e^{-yH}|p\rangle
\label{eq:defgluonu}
\ee 
In the next step we act on (\ref{eq:balbk}) with $\delta^{a_1a_2}$ (such color tensor
is present for instance in photon impact factor). Using (\ref{eq:ham1}), (\ref{eq:ham2}),
(\ref{eq:ham3}), (\ref{eq:balbk}), (\ref{eq:defgluonu}) we obtain:  
\be 
\begin{split}
\frac{\partial{\cal F}_2(x,\k,\q)}{\partial \ln1/x}=
- N_c\!\!\int\!\!\frac{d^2\l}{(2\pi)^3}K_{BFKL}(\l,\q-\l;\k,\q-\k)\frac{{\cal
F}_2(x,\l,\q)}{\l^2(\q-\l)^2}\nonumber\\
\end{split}
\ee
\be
\begin{split}
-\frac{1}{2!}\pi\int\frac{d^2\r}{(2\pi)^3}\frac{d^2\l}{(2\pi)^3}\frac{d^2\m}{(2\pi)^3}
V(\k,-\k+\q;\l,-\l-\frac{\q}{2}+\r,\m,-\m-\frac{\q}{2}-\r)\\
\times
\frac{{\cal F}_2(x,\l,\frac{\q}{2}+\r)}{\l^2(-\l+\frac{\q}{2}+\r)^2}
\frac{{\cal F}_2(x,\m,\frac{\q}{2}-\r)}{\m^2(-\m+\frac{\q}{2}-\r)^2}
\label{eq:BKK}
\end{split}
\ee
Factors $C_F$ arising from contractions of color tensors were absorbed in
definition of function ${\cal F}_2$. The factorial indicates the symmetry of
the TPV.  
We put  $\k_1\equiv \k$ and for the second momentum argument 
$\k_2\!\!\equiv\!\!-\k_1+\q$ we wrote just $\q$ to indicate that
there is a dependence on the second  momentum variable. The momenta incoming into
the vertex from below is labeled  $\k_1'\!\!=\!\!\l$, $\k_2'\!\!=\!\!-\l\!-\!\q/2$,
$\k_3'\!\!=\!\!\m$, $\k_4'\!\!=\!\!-\m-\q/2-\r$. The variable $\r$ stands for the loop
momentum. 
If we restrict ourselves to the zero momentum transfer at the top of the
fan diagrams (which corresponds to neglect the impact parameter dependence)
and, inside the fan diagrams, carry out the integrations over
momentum transfer  under the factorization assumption:
\be
{\cal F}_2(x,\k,\r)={\cal F}(x,\k)F(\r,R),
\ee
where in the present case
\be
F(\r,R)=\frac{e^{\frac{-\r^2 R^2}{4}}}{2\pi}
\label{eq:formfactor}
\ee
$F(\r,R)$ is the form factor of the proton. The parameter $R$ has meaning of
a radius of the proton.
After assumptions done above we obtain  somewhat simpler equation:
\be
\begin{split}
\frac{\partial {\cal F}(x,\k)}{\partial \ln1/x}=
-N_c\int\frac{d^2\l}{(2\pi)^3}K_{BFKL}(\l,-\l;\k,-\k)\frac{{\cal F}(x,\l)}{\l^4}\\
-\frac{1}{2!}\pi\frac{1}{2\pi
R^2}\int\frac{d^2\l}{(2\pi)^3}\frac{d^2\m}{(2\pi)^3} V(\k,-\k;\l,-\l,\m,-\m)
\frac{{\cal F}(x,\l)}{\l^4}\frac{{\cal F}(x,\m)}{\m^4}
\label{eq:ffaneq}
\end{split}
\ee
The assumed factorization  limits the momentum transfer across the ladders to small values.
Note that this procedure has been developed for deep inelastic scattering
on a large nucleus and that nonzero momentum transfer allows to keep track of the
impact parameter dependence in configuration space.
In the next step we perform integrations over azimuthal angle of $\l$, $\m$ and $\k$. 
We use $f(x,\k^2)$ for angular averaged
function ${\cal F}(x,\k)$.
\begin{figure}[t!]
\centerline{\epsfig{file=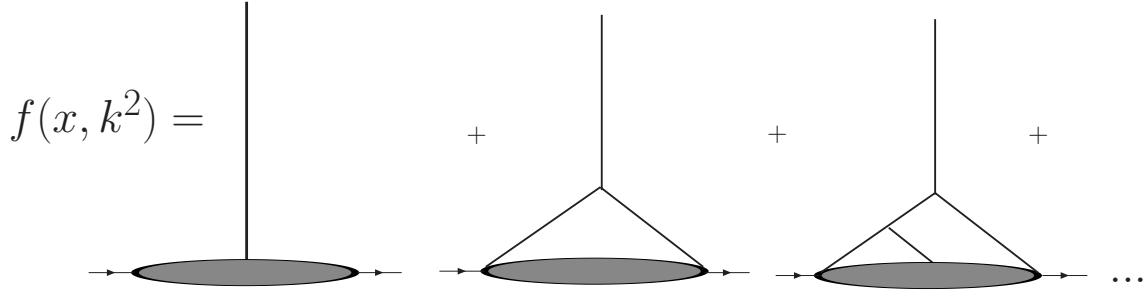,height=4cm}}
\caption{\em Diagrammatic representation of the integral version of the equation (\ref{eq:faneq}).}
\label{fig:TPVreal}
\end{figure}
After this procedure the linear part of the equation (\ref{eq:ffaneq}) is just the BFKL equation for
the unintegrated gluon density while in the nonlinear part the basic vertex
function takes the form (see appendix for details of this calculation): 
\bea
V(\k,-\k;\l,-\l,\m,-\m)=4\frac{g^4}{2}\bigg[2\k^2\theta(\l^2-\k^2)\theta(\m^2-\k^2)\nonumber\\
+\ln\left(\frac{\l^2}{\m^2}\right)\delta(\l^2-\k^2) \theta(\m^2-\l^2)
+\ln\left(\frac{\m^2}{\l^2}\right)\delta(\m^2-\k^2)\theta(\l^2-\m^2)\bigg]
\label{eq:melG2}
\eea
The fan diagram equation reads:
\be
\begin{split}
\frac{\partial f(x,\k^2)}{\partial \ln 1/x}= \frac{N_c\alpha_s}{\pi}\k^2\int_0^{\infty}\frac{d\kpp^2}{\kpp^2}
\bigg[\frac{f(x,\kpp^2)- f(x,\k^2)}{|\k^2-\kpp^2|}+ \frac{
f(x,\k^2)}{\sqrt{(4\kpp^4+\k^4)}}\bigg]\\
-\frac{\alpha_s^2}{2R^2} \Bigg\{2\k^2
\int_{\k^2}^{\infty}\frac{d\l^2}{\l^4}f(x,\l^2)\int_{\k^2}^{\infty}\frac{d\m^2}{\m^4}f(x,\m^2)\\
+f(x,\k^2)\int_{\k^2}^{\infty}\frac{d\l^2}{\l^4}\ln\left(\frac{\l^2}{\k^2}\right)f(x,\l^2)
+f(x,\k^2)\int_{\k^2}^{\infty}\frac{d\m^2}{\m^4}\ln\left(\frac{\m^2}{\k^2}\right)f(x,\m^2)\Bigg\}
\label{eq:faneq}
\end{split}
\ee

When applying this equation to the scattering of a virtual photon on a
nucleus one may ask what is the most dominant contribution. In the DGLAP approach 
(see chapter 2) one has strong ordering in momentum, i.e  virtualities of gluons closer to
photon quark vertex are larger. Here (on a level of the evolution equation) we can also consider such configuration where the transverse momenta
are strongly ordered: the upper momenta should be smaller than the lower ones.
However, making use of our results for the collinear limit of the TVP and of
the structure of the angular averaged vertex, where the $\theta$ functions
does not allow the collinear configuration, we arrive at the conclusion that,
there is no such contribution. In contrast, the
momenta are ordered in the opposite direction, i.e. at the nonlinear vertex of
the equation, the momentum above have to be smaller than those below. In more
physical terms, the recombination of two smaller gluons ends up in a larger
gluon. Note that this only holds, after the large-$N_c$ limit has been taken
and the angular averaging has been performed.\\ 
\section{Comparison with other equations} 
One can obtain (\ref{eq:faneq}) from the BK equation under the assumption
that the target is  large in comparison to the size
of the dipole which scatters on it. The crucial point is the space of
formulation and identification of the correct degrees of freedom. The BK equation
was formulated in configuration space so the first step to compare it with
(\ref{eq:faneq}) is to perform its Fourier transform.  The next point is to
notice that the equation obtained via Fourier transform has to be further
transformed in order to take form of (\ref{eq:faneq}). This is achieved by
invoking relations linking dipole density, dipole cross section and
the unintegrated gluon density.   To derive  this one has to apply relations
between a scattering amplitude of a color dipole  $N(\x_{01},\b,x)$, the
dipole cross section  $\sigma(\x_{01},x)$ (where $\x_{01}$ is a size of the
initial dipole) and the unintegrated gluon distribution $f(x,\k^2)$
\cite{KKM01,KK03,KS}. Let us start with the Balitsky-Kovchegov equation for the
dipole amplitude. \be \begin{split} \frac{\partial N(\x_{01},\b,x)}{\partial\ln
1/x}=\frac{\alpha_sN_c}{2\pi^2}\int d^2x_2
\frac{\x_{01}^2}{\x_{02}^2\x_{12}^2}[N(\x_{02},\b+\frac{1}{2},x)+
N(\x_{12},\b+\frac{1}{2}\x_{20},x)\\
-N(\x_{01},\b,x)-N(\x_{02},\b+\frac{1}{2}\x_{21},x)N(\x_{12},\b+\frac{1}{2}\x_{20},x)] \label{eq:BKfull} \end{split} \ee
The assumption that target is large in comparison to the size of dipoles 
allows us to  simplify BK equation and rewrite it in slightly different form:
\be
\begin{split}
\frac{N(\x_{01},\b,x)}{\partial \ln 1/x}=\frac{\alpha_sN_c}{\pi^2}\int_{\rho}d^2\x_2\left[
\frac{x_{01}^2}{x_{02}^2x_{12}^2}-2\pi\delta^2(\x_{01}-\x_{02})\ln\frac{x_{01}}{\rho}N(\x_{02},b,x)
\right]\\
-\frac{\alpha_s N_c}{2\pi^2}\int d^2\x_2\frac{x_{01}^2}{x_{02}^2x_{12}}N(\x_{02},b,x)N(\x_{12},b,x)
\end{split}
\label{eq:kovev}
\ee
what allows to write the dipole amplitude in a factorised form:
\be
N({\bf x},{\bf b},x)= N({\bf x},x)S({\bold b})
\ee
where $S(\b)$ is the profile function of the proton related to form factor
(\ref{eq:formfactor}) via two dimensional Fourier transform.
The factorization ansatz will be used later.
Performing Fourier transform of (\ref{eq:kovev}) \cite{Kov}
we obtain \cite{Kov,BR}:
\be
\frac{\partial\Phi(x,\k,\b)}{\partial\ln1/x}=
\overline\alpha\int_0^{\infty}\frac{d\k'^2}{\k'^2}\left[\frac{\k'^2\Phi(x,\k,\b)-\k^2\Phi(x,\k,\b)}
{|\k'^2-\k^2|}+\frac{\Phi(x,\k,\b)}{\sqrt{4\k'^4+\k^2}}\right]
-\overline\alpha\Phi^2(x,\k,\b)
\label{eq:bkphi},
\ee
where 
\be
\Phi(x,\k,\b)=\int\frac{d^2\x_{01}}{2\pi}e^{-i\k\cdot\x_{01}}\frac{N(\x_{01},\b,x)}{\x_{01}^2}
\label{eq:fourphi}
\ee
and inverse transform is given by:
\be
N(\x_{01},\b,Y)=\x_{01}^2\int\frac{d^2\k}{2\pi}e^{i\k\cdot\x_{01}}\Phi(x,\k,\b)
\ee
Multiplying both sides of (\ref{eq:bkphi}) by $\k^2$ and using Mellin
representation:
\be
\Phi(x,\gamma,\b)=\int_0^{\infty} d\k^2 (\k^2)^{-\gamma-1}\k^2\Phi(x,\k^2,\b)
\ee
we obtain:
\be
\frac{\partial\Phi(x,\gamma,\b)}{\partial \ln 1/x}=K(\gamma)\Phi(x
\gamma,b)-\overline\alpha\int_0^{\infty}d\k^2\Phi^2(\k^2,
\b,x)\k^2(\k^2)^{-\gamma -1}
\label{eq:basiceq}
\ee
where:
\be
K(\gamma)=\frac{N_c\alpha_s}{\pi}[2\psi(1)-\psi(\gamma)-\gamma(1-\psi)]
\ee
is the BFKL eigenvalue.
To  find the equation for the unintegrated gluon density we apply relation between the dependent on
the impact parameter gluon
density and the dipole scattering amplitude:
\be
N(\x_{01},\b,x)=\frac{4}{N_c\pi^2}\int\frac{d\k}{\k^3}[1-J_0(|\k||\x_{01}|)]\alpha_s f(x,\k^2,\b)
\label{eq:dipxsec}
\ee
where
\be
\int d^2bf(x,\k^2,\b)=\int d^2bf(x,\k^2)S(b)=f(x,\k^2)
\ee
using (\ref{eq:fourphi}) we obtain:
\be
\Phi(x,\l,\b)=\frac{1}{2}\int \frac{d^2\x_{01}}{2\pi \x_{01}^2}\,
e^{i\l\cdot\b}\,\frac{8\pi^2}{N_c}\int\frac{dk}{k^3}[1-J_{0}(k\x_{01})]\alpha_{s}f(x,\k^2,\b)  \;
. \ee
Integration over angles yields:
\be
\Phi(x,\l^2,\b)=\frac{2\pi^2}{N_c}\int_{l^2}^{\infty}\frac{d\k^2}{\k^4}\int_{0}^{\infty}\frac{dx_{01}}{x_{01}}J_{0}(|\l|)[1-J_{0}(|\k||\x_{01}|)]
\alpha_{s}f(x,\k^2,\b) \; ,
\label{eq:dipglu}
\ee
and the integral over $\x_{01}$ gives
\be
\Phi(x,\l^2,\b)={\pi^2 \over N_c}\int_{\l^2}^{\infty}{d\k^2\over
k^4}\ln\left(\frac{\k^2}{\l^2}\right) \alpha_s f(x,\k^2,\b)  \; .
\ee
Now we need  to invert the  operator
\be
\hat{O} = \frac{\pi^2 \alpha_s} {N_c}\int_{\l^2}^{\infty}{d\k^2\over
k^4}\ln\left(\frac{\k^2}{\l^2}\right) g(\k^2) \; ,
\ee
where $g(\k^2)$ is a test function.
Multiplying both sides of  (\ref{eq:dipglu}) by $\l^2$ and performing the Mellin transform
with respect to $\l^2$ we obtain \cite{KS}:
\be
\begin{split}
\Phi(x,\gamma,\b)\equiv\int d\l^2 \l^2\Phi(x,\l^2,\b)(\l^2)^{-\gamma-1}=\\\int d\l^2
\l^2{\pi^2 \over N_c}\int_{\l^2}^{\infty}{d\k^2\over
\k^4}\ln\left(\frac{\k^2}{\l^2}\right) \alpha_s f(x,\k^2,\b)(\l^2)^{-\gamma-1}
=\frac{\alpha_s\pi^2}{N_c}f(\gamma,\b)\frac{1}{(1-\gamma)^2},
\label{eq:dipglu}
\end{split}
\ee
and equivalently
\be
f(x,\gamma,\b)=\frac{N_c}{\alpha_s\pi^2}(1-\gamma)^2\Phi(x,\gamma,\b) \; .
\ee
The inverse Mellin transform gives the momentum space version of this expression:
\be
f(x,\l^2,\b)=\frac{N_c}{\alpha_s\pi^2}\int\frac{d\gamma}{2\pi
i}(\l^2)^{\gamma}(1-\gamma)^2\Phi(\gamma,x)=
\frac{N_c}{\alpha_s\pi^2}(1-\l^2\frac{d}{d\l^2})^2\l^2\Phi(x,\l,\b)\; .
\label{eq:relacja}
\ee
Inserting (\ref{eq:dipglu}) in (\ref{eq:basiceq}), and inverting the Mellin transform we obtain:
\be
\frac{\partial f(x,\k^2,\b)}{\partial\ln1/x}=\frac{N_c \alpha_s}{\pi}K\otimes
f(x,\k^2,\b)
\label{eq:BK222}
\ee
$$
-\pi\alpha_s^2\left[2\k^2\left(\int_{\k^2}^{\infty}\frac{d\k'^2}{\k'^4}f(x,\k'^2,\b)\right)^2
+2f(x,\k^2)\int_{\k^2}^{\infty}\frac{d\k'^2}{\k'^4}\ln\left(\frac{\k'^2}{\k^2}\right)f(x,\k'^2,\b^2)\right]
$$
This equation can be further transformed, applying the factorization ansatz $f(x,\k^2,\b)=f(x,\k^2)S(\b)$
and taking for example a Gaussian profile function:
\be
S(\b)=\frac{e^{\frac{-\b^2}{R^2}}}{\pi R^2}
\label{eq:profil1}
\ee
After using this and integrating over the impact parameter, we obtain:
\be
\frac{\partial f(x,\k^2)}{\partial\ln1/x}=\frac{N_c \alpha_s}{\pi}K\otimes
f(x,\k^2)
\label{eq:BK222}
\ee
$$
-\frac{\alpha_s^2}{2R^2}\left[2\k^2\left(\int_{\k^2}^{\infty}\frac{d\k'^2}{\k'^4}f(x,\k'^2)\right)^2
+2f(x,\k^2)\int_{\k^2}^{\infty}\frac{d\k'^2}{\k'^4}\ln\left(\frac{\k'^2}{\k^2}\right)f(x,\k'^2)\right]
$$

At this stage we would like to comment on the relation between (\ref{eq:BK222}) and the other known 
nonlinear evolution equations appearing in considerations of high energy limit of QCD. 
The first nonlinear evolution equation which was a milestone in physics of
saturation 
is the GLR-MQ \cite{GLR}, \cite{MQ} (Gribov, Levin, Ryskin, Mueller, Qiu) equation,
which is given by (equation number (2.41) in \cite{GLR} and (30) in \cite{MQ}): 
\be
\frac{\partial^2xg(x,\k^2)}{\partial \ln(1/x)\partial \ln\k^2}=\frac{\alpha_s N_c}{\pi}xg(x,\k^2)-
C\frac{\alpha_s^2}{\k^2R^2}[xg(x,\k^2)]^2
\ee
in the equation above the constant C is not the same in both of cited papers, however, for our discussion its value is not
crucial.
This equation can be rewritten for the unintegrated gluon density with the collinear approximated
BFKL kernel:
\be
\frac{\partial f(x,\k^2)}{\partial \ln1/x}=\frac{N_c\alpha_s}{\pi}\int_{k_0^2}^{\k^2}\frac{d\k'^2}{\k'^2} f(x,\k'^2)-
C\frac{\alpha_s^2}{\k^2R^2}\bigg[\int_{\k_0}^{\k^2}\frac{d\k^{'2}}{\k^{'2}}f(x,\k^{'2})\bigg]^2
\label{eq:GLRMQ}
\ee
The nonlinearity is 
claimed to be given by the TPV at the collinear limit. 
From our analysis one, however, comes to the conclusion that at collinear limit
(in angular averaged case) TPV vanishes and only anticollinear pole contributes.
This can also be read off from comparing nonlinear terms in (\ref{eq:BK222}) 
and (\ref{eq:GLRMQ}).  The structure of integrals is totally different and 
one cannot transform one into another. 
The equation that has similar properties to (\ref{eq:BK222}) is the GLR
equation (equation number (2.108) in \cite{GLR}) derived from summing up at low $x$ 
(single logs are summed) fan
diagrams.  It is written directly for the unintegrated gluon density
(the gluon density in the GLR
notation includes propagator). That equation is an attempt to generalize the BFKL equation for physics
of dense systems. GLR obtained TPV
(they did not consider disconnected pieces, the pre factors are different and they work in large
$N_c$ from the beginning) and approximated it by a constant.
This equation reads:
\be
\frac{\partial\Phi'(x,\k^2)}{\partial\ln1/x}=\frac{N_c\alpha_s}{\pi}\int^{\infty}_0\frac{d\l^2}{\l^2}\left[\frac{\Phi(x,\l^2)-
\Phi'(x,\k^2)}{|\l^2-\k^2|}-\frac{\Phi'(x,\k^2)}{\sqrt{4\l^4+\k^4}}\right]
-g\frac{1}{4\pi R^2}\left(\frac{\alpha_s}{4\pi}\right)^2\Phi^{'2}(x,\k^2)
\ee
where $g$ is the TPV vertex in the local approximation.
The relation to integrated gluon distribution is given by:
\be
xg(x,Q^2)=\int_{k_0^2}^{Q^2}d\k^2\Phi'(x,\k^2)
\ee
Gribov, Levin and Ryskin before applying local approximation to their nonlinear 
part of their equation wrote expression for the TPV convoluted with gluon densities from below:
\be
V\otimes\left(\Phi'(x,\l^2) \Phi'(x,\m^2)\right)=\int\frac{d\m^2}{\l^2}\frac{d\l^2}{\l^2}
\alpha_s(\m^2)\alpha_s(\l^2)\Phi'(x,\m^2)
\Phi'(x,\l^2)\theta(\l^2-\k^2)
\theta(\m^2-\k^2)
\label{eq:vertglr}
\ee
The structure of this expression (where to have direct comparison one should
replace  $\Phi'(x,\k^2)$ with $f(x,\k^2)/\k^2$, we did not do that since we want to refer to
the original formulation) 
is similar to the 
structure of the real emission contributions in the nonlinear part of (\ref{eq:BK222}). 
The lack of the disconnected contribution does not lead to singularities since  
(\ref{eq:vertglr}) 
is infrared safe.    
From this structure, GLR  arrive at similar conclusion as we that in order for gluons to fuse they 
have to resolve individual gluons in the ladder from above.\\
It will be interesting to compare precisely all of the listed above equations and to check validity of
approximations which were done. At this moment one can guess that although very different structure of
nonlinearity in (\ref{eq:BK222}) and (\ref{eq:GLRMQ}) the behavior of solutions will be similar
in the saturated phase because the nonlinear terms have similar scaling.
More precise studies are, however, necessary.

\chapter{Phenomenological applications of the fan diagram equation}
\section{Model of subleading corrections to BK}
\begin{figure}[h!]
\centerline{\epsfig{file=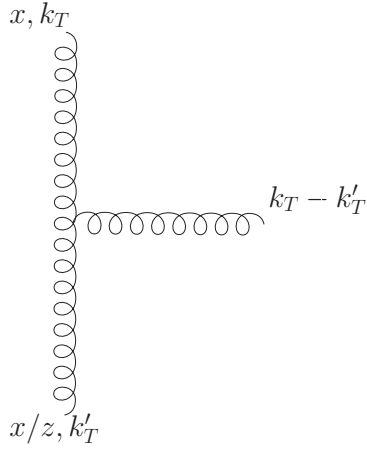,height=6cm}}
\caption{\em Kinematic variables appearing in the kinematic constraint}
\label{fig:kincon}
\end{figure}

For phenomenological applications of BK it is desirable to have a
formalism which  embodies the resummation of the subleading corrections in
$\ln 1/x$.
Attempts in this direction already exist, see for example
\cite{KKM01,KK03,LevLub,GBMS,BraunRC,Trian,MartinSat}.
Equation (\ref{eq:BK222}) contains the BFKL kernel at leading logarithmic (LOln1/x)
accuracy.    This is a coarse approximation as far as a description of the
HERA data is concerned. It is well known  \cite{NLLBFKL} that the  NLOln1/x
corrections  to the BFKL equation are quite large. To make the equation  more
realistic,   one  can \cite{KKM01,KK03,KS}
implement  in the linear term  of (\ref{eq:BK222}) a unified BFKL-DGLAP framework developed in  \cite{KMS}.
This framework unifies the BFKL and the DGLAP equations in alternative to a CCFM way \cite{CCFM}. This unification is
however, on a different level of rigor then the CCFM (direct summation of diagrams).
Kwieci\'nski, Martin and Sta\'sto observed that the BFKL equation may be refined if one includes
relevant parts of the DGLAP splitting functions and other contributions of
subleading in $\ln1/x$ order. This contribution leads to coupled integral
equations which at low $x$ limit become equivalent to the BFKL and at collinear
limit become equivalent to DGLAP evolution for the gluon density. Let us be more
precise. In the KMS scheme \cite{KMS}, the BFKL kernel becomes modified by 
the consistency constraint Fig. \ref{fig:kincon} \cite{AGKS,KMScc} 
\be
\k'^2  <  \k^2 / z \; ,
\label{eq:kincon}
\ee
imposed  onto the real-emission part of the kernel in Eq.~(\ref{eq:BK222})
\be
\int_0^{\infty} \frac{d\k'^2}{\k'^2} \,
\bigg\{ \, \frac{f(\frac{x}{z},\k^{\prime 2},\b) \, \theta(\frac{\k^2}{z}-\k'^2)\,
-\, f(\frac{x}{z},\k^2,\b)}{|\k'^2-\k^2|}   +\,
\frac{f(\frac{x}{z},\k^2,\b)}{|4\k^{\prime  4}+\k^4|^{\frac{1}{2}}} \, \bigg\} \,
.\ee
The consistency constraint on the real gluon emission(\ref{eq:kincon})
resumes a large part of
the subleading corrections in $\ln1/x$ \cite{Salam98,CCS99}.
The physical meaning of it is that the virtuality of the gluon is
dominated by its transverse momentum.

Additionally, the non-singular (in $x$) part of the leading order (LO) DGLAP
splitting function is included into the evolution:
\be
\int_x^1 \frac{dz}{z} K\otimes f \rightarrow \int_x^1 \frac{dz}{z} K \otimes f +
\int^{\k^2}\frac{d\k'^2}{\k'^2} \int_x^1 dz\bar{P}_{gg}(z)
f(\frac{x}{z},\k^{\prime 2},\b) \; ,   
\ee
where
\be
\bar{P}_{gg}(z) = P_{gg}(z) - \frac{2N_c}{z} \; .
\ee
Additionally, we assume that in our evolution equation $\alpha_s$  runs with
scale $\k^2$ which is another source of important NLOln1/x corrections.
The improved nonlinear equation for the unintegrated gluon density is as follows \cite{KK03}:
\begin{multline}
f(x,\k^2) \; = \; \tilde f^{(0)}(x,\k^2,) \\
+\,\frac{\alpha_s(\k^2) N_c}{\pi} k^2 \int_x^1 \frac{dz}{z}
\int_{\k_0^2}^{\infty} \frac{d\k'^2}{\k'^2} \,   \bigg\{ \,
\frac{f(\frac{x}{z},\k^{\prime 2}) \, \theta(\frac{\k^2}{z}-\k'^2)\,   -\,
f(\frac{x}{z},\k^2)}{|\k'^2-\k^2|}   +\,
\frac{f(\frac{x}{z},\k^2)}{|4\k^{\prime   4}+\k^4|^{\frac{1}{2}}} \, \bigg\}
\\  + \, \frac{\alpha_s (\k^2)}{2\pi} \int_x^1 dz \,  \bigg[
\left(P_{gg}(z)-\frac{2N_c}{z}\right) \int^{\k^2}_{\k_0^2} \frac{d \k^{\prime
2}}{\k^{\prime 2}} \, f(\frac{x}{z},\k'^2)] \\ 
-\frac{\alpha_s^2(\k^2)}{R^2}\left[2\k^2\left(\int_{\k^2}^{\infty}\frac{d\k'^2}{\k'^4}f(x,\k'^2)\right)^2 +
2f(x,\k^2)\int_{\k^2}^{\infty}\frac{d\k'^2}{\k'^4}\ln\left(\frac{\k'^2}{\k^2}\right)f(x,\k'^2)\right];
\label{eq:fkovres}
\end{multline}
above we assumed the target to have cylindrical profile with radius $R=4GeV^{-1}$
\be
S(\b)=\frac{\theta(R^2-\b^2)}{\pi R^2}
\ee
and we integrated over the impact parameter.
In the KMS approach the inhomogeneous term was defined in terms of
the integrated gluon distribution
\be
\tilde f^{(0)}(x,\k^2)=\frac{\alpha_S(\k^2)}{2\pi}\int_x^1
dzP_{gg}(z)\frac{x}{z}g\left(\frac{x}{z},\k_0^2\right)
\label{eq:input}
\ee
taken at scale $\k_0^2=1 \gev^2$. This scale was also used as a cutoff in the
linear version of the evolution equation (\ref{eq:fkovres}). In the linear case this provided a
very good description of $F_2$ data with a minimal number
of physically motivated parameters, see  \cite{KMS}.
The initial integrated density at scale $\k_0^2$ was parameterized as
\be
xg(x,\k_0^2)=N(1-x)^{\rho} \; ,
\label{eq:gluony1}
\ee
where $N=1.57$ and $\rho=2.5$.

Let us finally note that we include most of subleading corrections only in the linear 
part of the BK equation. The nonlinear term gets contribution only from running of the coupling constant.
Effects of energy conservation
in nonlinear term are important as has been shown in analysis of BK in the configuration space
(but subleading)\cite{chachamis}  

One could try to implement the kinematical constraint effects in the TPV
in a similar manner as in the linear part i.e:
\be
\begin{split}
V\otimes f(x,\k^2)^2=
\frac{2\k^2\alpha_s^2}{R^2}
\left[\int_{\k^2}^{\infty}\frac{d\l^2}{\l^4}f(x,\l^2)\theta\left(\frac{\k^2}{z}-\l^2\right)\right]^2\nonumber\\
+f(x,\k^2)\int_{\k^2}^{\infty}\frac{d\l^2}{\l^4}\ln\left(\frac{\l^2}{\k^2}\right)f(x,\l^2)
\theta\left(\frac{\k^2}{z}-\l^2\right)
\end{split}
\ee
This contribution is not considered here and we postpone it to 
further studies. 
\section{Numerical analysis}

\label{sec:num}
\subsection{The unintegrated and integrated gluon densities}

In this section we introduce the method of solving Eq.~(\ref{eq:fkovres}) and we
present the  numerical results for the unintegrated gluon distribution function
$f(x,\k^2)$ and the integrated  gluon density $xg(x,Q^2)$. The method of solving
(\ref{eq:fkovres}), developed in \cite{KK03},  relies on reducing it to an
effective evolution equation in $\ln 1/x$
 with the boundary condition
at some moderately small value  of $x$ (i.e. $x=x_0 \sim 0.01$).

To be specific,  we make the following approximations:
\begin{enumerate}
\item The  consistency constraint $\theta(\k^2/z-\k^{\prime 2})$ in the  BFKL kernel  is replaced by the following
effective ($z$ independent)  term
\begin{equation}
\theta(\k^2/z-\k^{\prime 2}) \rightarrow \theta(\k^2-\k^{\prime 2}) +
\left({\k^2\over \k^{\prime 2}}\right)^{\omega_{eff}}\theta(\k^{\prime 2}-\k^2)\,.
\label{bfkleff}
\end{equation}
This  is motivated by the structure of the consistency constraint
in the moment space, i.e.
\begin{equation}
\omega \int_0^1{dz \over z}z^{\omega}\theta(\k^2/z-\k^{\prime 2})=\theta(\k^2-\k^{\prime 2}) +
\left({\k^2\over \k^{\prime 2}}\right)^{\omega}\theta(\k^{\prime 2}-\k^2) \; ,
\label{ccmom}
\end{equation}
\item The splitting function is approximated in the following way
\begin{equation}
\int_x^1{dz\over z}[zP_{gg}(z)-2N_c]
f\left({x\over z},\k^{\prime 2}\right) \rightarrow \bar P_{gg}(\omega=0) f(x,\k^{\prime 2}) \; ,
\label{dglapeff}
\end{equation}
where $\bar P_{gg}(\omega)$ is a moment function
\begin{equation}
\bar P_{gg}(\omega)=\int_0^1{dz\over z}z^{\omega}[zP_{gg}(z)-2N_c] \; ,
\label{pggmom}
\end{equation}
and
\begin{equation}
\bar{P}_{gg}(\omega=0)=-\frac{11}{12} \; .
\label{eleventwelve}
\end{equation}
This approximation corresponds to retaining only  the leading term in the expansion
of $\bar P_{gg}(\omega)$ around $\omega=0$, see \cite{LOWX}.
\end{enumerate}

\begin{figure}[th!]
  \begin{picture}(490,170)
    \put(40, -126){
      \includegraphics{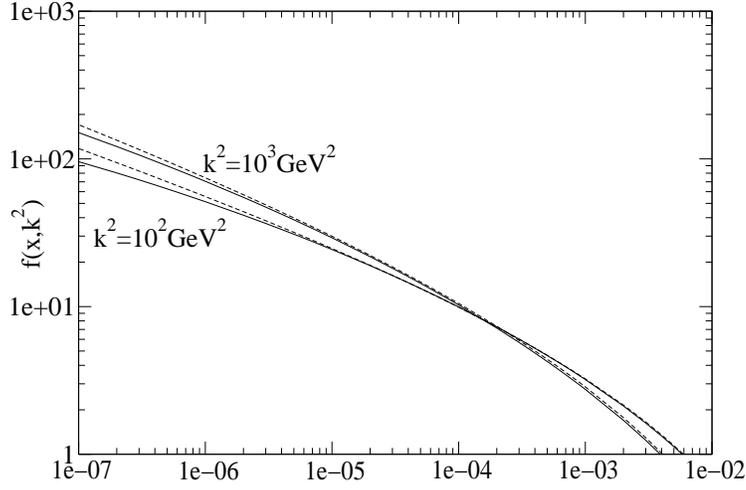}
    }
    \put(40, -416){
      \includegraphics{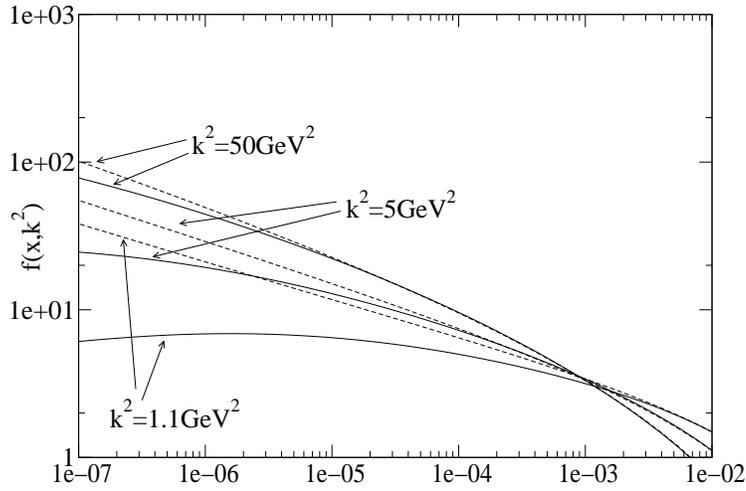}
    }
        \end{picture}
\vspace{13cm}
\caption{\footnotesize{The unintegrated
gluon distribution $f(x,k^2)$ obtained from Eq.(\ref{eq:eqdorozw}) as a  function
of $x$ for  different values $k^2 = 10^2 \ \gev^2$ and $k^2 = 10^3  \ \gev^2$
(up) and for $k^2 = 1.1 \ \gev^2$, $k^2 = 5 \ \gev^2$, $k^2 = 50  \  \gev^2$ (down).
The solid lines correspond to the solution of the nonlinear equation
(\ref{eq:eqdorozw}) whereas the dashed lines correspond to the linear  BFKL/DGLAP term
in (\ref{eq:eqdorozw})}}
\label{fig:fx}
\end{figure}

\begin{figure}[th!]
  \begin{picture}(490,170)
    \put(40, -126){
      \includegraphics{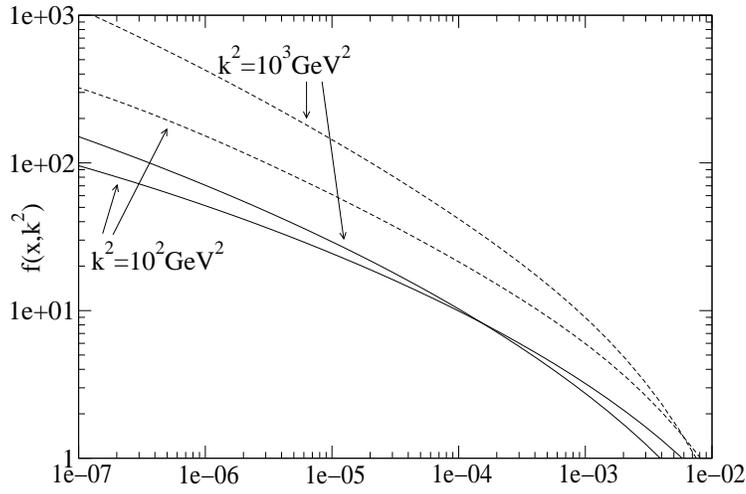}
    }
    \put(40, -416){
      \includegraphics{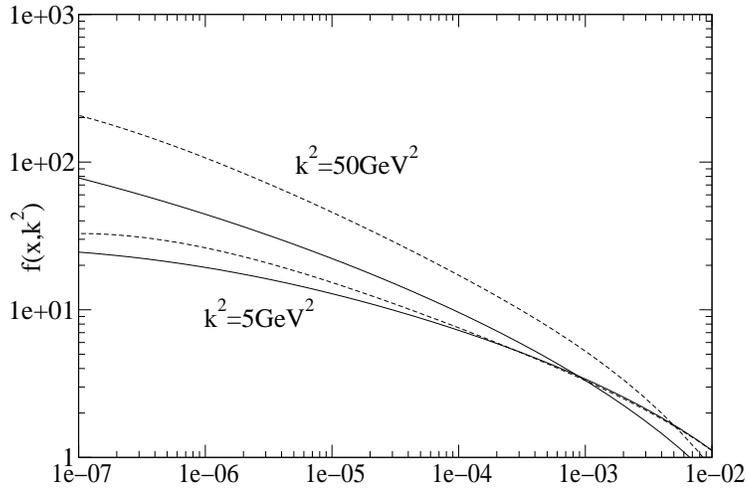}
    }
        \end{picture}
\vspace{13cm}
\caption{\footnotesize{The same as Fig.~\ref{fig:fx}(for clearer presentation we skip the plot with 
$k^2=1.1GeV^2$) but now the modified BK equation
(\ref{eq:eqdorozw}) (solid lines) is compared with the
original BK equation (\ref{eq:eqdorozw}) without subleading corrections
(dashed lines).}}
\label{fig:fx2}
\end{figure}

Using these approximations in (\ref{eq:fkovres}) we obtain \cite{KS}:
$$
\frac{\partial f(x,\k^2)}{\partial \ln(1/x)}=\\
$$
\be
+ {N_c\alpha_s(k^2)\over \pi} \k^2\int_{\k_0^2}^{\infty}{d\k^{\prime 2}\over
\k^{\prime 2}}\left \{{f\left({x},\k^{\prime 2}\right)(\theta\left(\k^2-\k^{\prime 2}
\right) +\left(\frac{\k^2}{\k^{\prime 2}}\right)^{\omega_{eff}}\theta\left(\k^{\prime 2}-\k^2\right)) - f\left(x,\k^2\right)\over |\k^{\prime 2}-\k^2|} +
{f\left({x},\k^2\right)\over [4\k^{\prime 4}+\k^4]^{{1\over 2}}}\right\}
\nonumber\ee
$$
+{\alpha_s(k^2)\over 2 \pi}P_{gg}(0)
\int_{k_0^2}^{\k^2}{d\k^{\prime 2}\over \k^{\prime 2}}f\left({x},\k^{\prime 2}\right)\\
$$
\be
 -\frac{\alpha_s^2(\k^2)}{R^2}\left[2\k^2\left(\int_{\k^2}^{\infty}\frac{d\k'^2}{\k'^4}
 f(x,\k'^2)\right)^2
+2f(x,\k^2)\int_{\k^2}^{\infty}\frac{d\k'^2}{\k'^4}\ln\left(\frac{\k'^2}{\k^2}\right)f(x,\k'^2)\right]
\label{eq:eqdorozw}
\ee
First, the equation (\ref{eq:eqdorozw}) was solved   with the non-linear term
neglected starting from the initial conditions   at $x=10^{-2}$ given by
(\ref{eq:gluony1}).   The parameter $\omega_{\rm eff}$ was adjusted in such a way
that the solution of the linear part of (\ref{eq:eqdorozw})  matched the
solution of   the original equation in the BFKL/DGLAP framework \cite{KMS}.
This procedure gives $\omega_{\rm eff}=0.2$ and the solution of the
linear part of (\ref{eq:eqdorozw}) reproduces the  original results of  \cite{KMS}
within 3\%  accuracy in the region $10^{-2}>x>10^{-8}$ and $2
\gev<\k^2<10^6\gev$. This matching procedure has also the advantage that
the quark contribution  present in the original BFKL/DGLAP
framework  is effectively  included by fitting the  value of $\omega_{\rm eff}$.  
The full non-linear equation (\ref{eq:eqdorozw}) was then solved using
the same initial conditions and setting   $R=4\gev^{-1}$.
In Fig.~\ref{fig:fx}, we plot the unintegrated gluon distribution function as a
function of $x$ for different values of $\k^2$.   This figure compares  the results of two
calculations, based on the linear  and   nonlinear equations.
 The   differences are not significant, however  there is
some suppression due to the nonlinearity at the smallest values of $x \le 10^{-5}$.
The subleading   corrections strongly decrease the value of the
intercept with respect to the   LOln1/x  value and   the
nonlinear term becomes important only at   very low values of $x$.
As is evident from Fig.~\ref{fig:fx2}, the subleading corrections cause a large suppression
in the normalization, also at moderate values of $x$.
\newpage
\newpage
\newpage
\begin{figure}[th!]
  \begin{picture}(490,170)
    \put(40, -126){
      \includegraphics{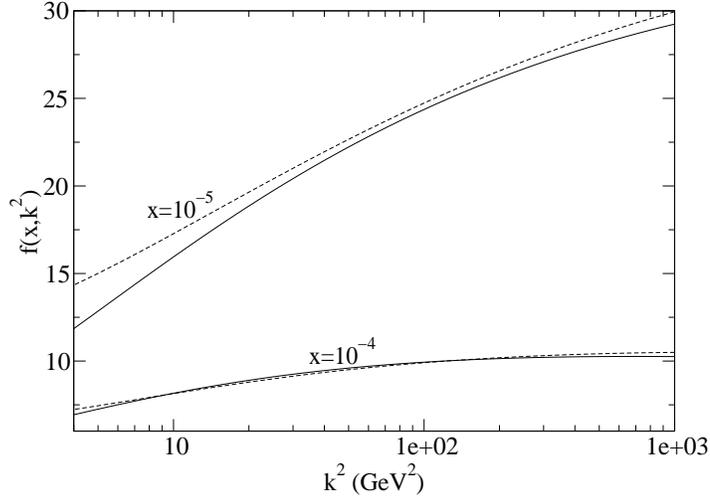}
    }
    \put(40, -411){
      \includegraphics{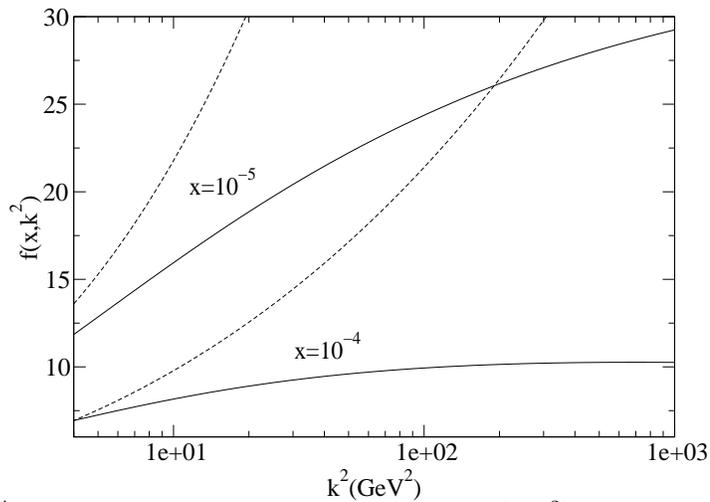}
    }
        \end{picture}
\vspace{13cm}
\caption{\footnotesize{The unintegrated gluon distribution $f(x,k^2)$
 as a function of $k^2$ for two values of $x=10^{-5}$
and $10^{-4}$  .
Up: solid lines correspond to the solution of the nonlinear
equation (\ref{eq:eqdorozw}) whereas dashed lines correspond to linear
BFKL/DGLAP term in (\ref{eq:eqdorozw}).
Down: solid lines correspond to the solution of the nonlinear equation (\ref{eq:eqdorozw})
whereas dashed lines correspond to the solution of the original BK equation without
the NLL1/x modifications in the linear part (\ref{eq:eqdorozw}).
}}
\label{fig:fk}
\end{figure}
\newpage
\begin{figure}[th!]
  \begin{picture}(490,170)
    \put(40, -126){
      \includegraphics{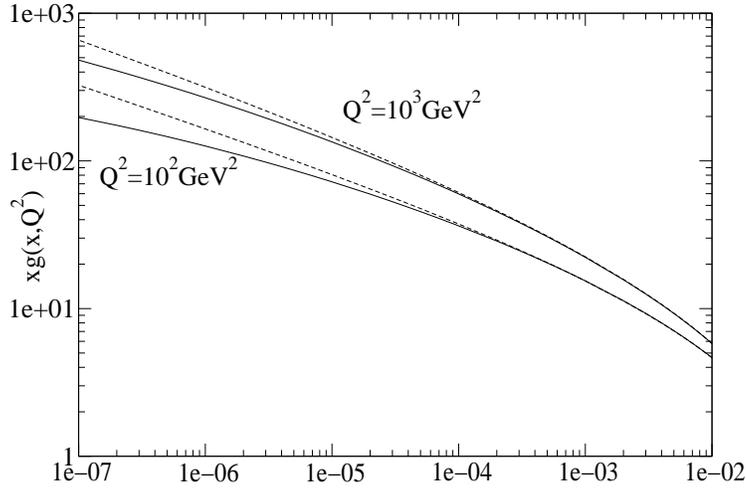}
    }
    \put(40, -416){
      \includegraphics{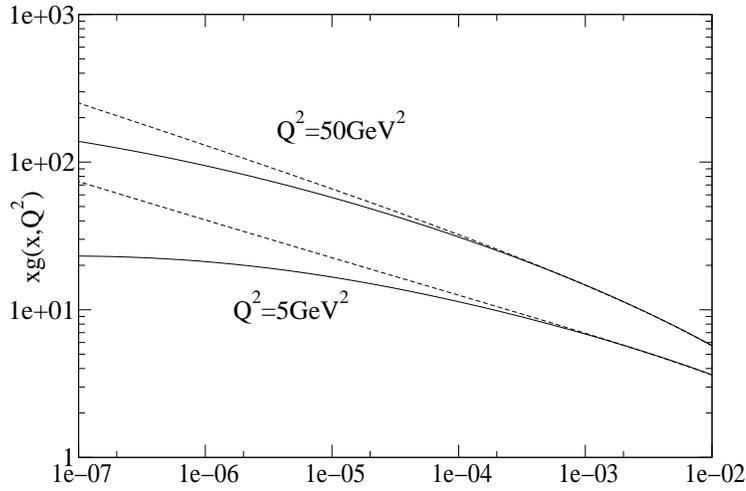}
    }
        \end{picture}
\vspace{13cm}
\caption{\footnotesize{The integrated gluon distribution $xg(x,k^2)$ as a
function of $x$ for  values of $Q^2 = 10^2 \gev^2$ and $Q^2 =
10^3$ (up) and for $Q^2 = 5 \  \gev^2$ and $Q^2 =
50$ (down) obtained from integrating $f(x,k^2)$Eq.(\ref{eq:eqdorozw}). Dashed
lines   correspond to solution of linear BFKL/DGLAP evolution equation.}}
\label{fig:xg}
\end{figure}
\newpage
This is due to the fact that
the non-singular in $x$ part of the $P_{gg}$ splitting function was included
into the evolution. This term is negative and is important at large and moderate values
of $x$.
The same conclusions can be reached by  investigating the plots in Fig.~\ref{fig:fk}
where the unintegrated density is shown as a function of the transverse momentum $\k^2$
for fixed values of $x$.
The nonlinear effects seem to have a moderate impact in that region.
On the other hand the subleading corrections are substantial. For example, at $x=10^{-5}$ and $\k^2=10 \, {\rm GeV^2}$
the reduction in magnitude of the unintegrated  gluon density is about  $25$.
Fig.~\ref{fig:xg} shows the integrated gluon density $xg(x,Q^2)=\int^{Q^2}d\k^2f(x,\k^2)/\k^2$.
The change from the power behavior   at small $x$   is clearly
visible in the nonlinear case. Additionally the differences   between    the
distributions in the linear and nonlinear case seem to be more   pronounced
for the quantity $xg(x,Q^2)$.   This is due to the fact that in   order to
obtain the gluon density $xg(x,Q^2)$   one needs to integrate over
scales up to $Q^2$ including small values of $\k^2$, where the     suppression
due to the nonlinear term is bigger.


\subsection{The saturation scale $Q_s(x)$}

\begin{figure}[th!]
  \begin{picture}(490,170)
    \put(40, -126){
      \includegraphics{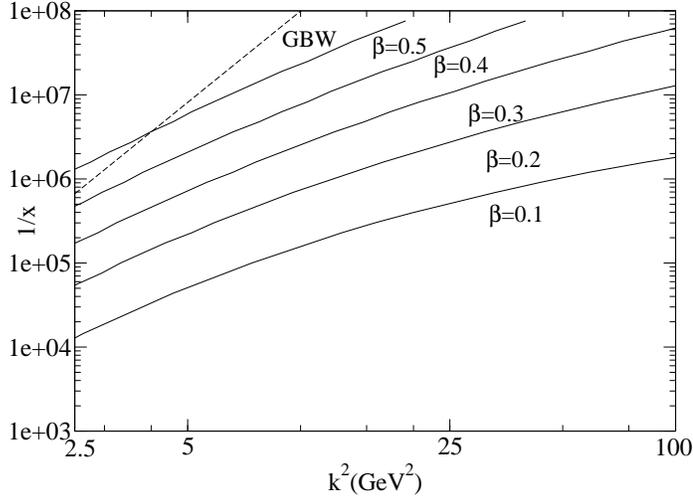}
    }
    \put(40, -411){
      \includegraphics{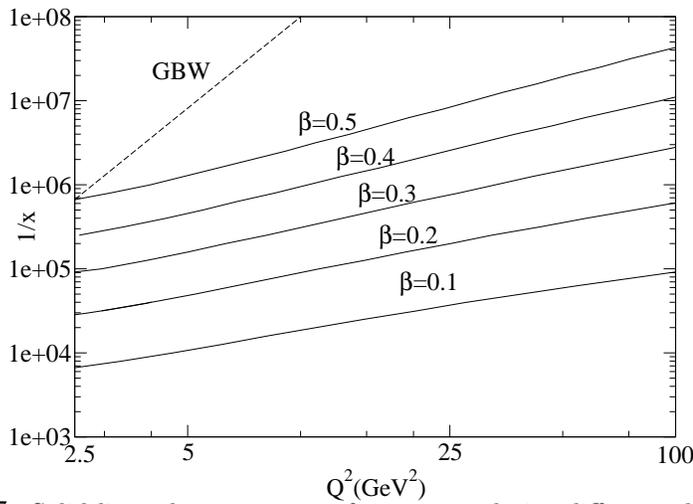}
    }
        \end{picture}
\vspace{13cm}
\caption{\footnotesize{Solid lines show contours of constant relative difference between solutions to linear and nonlinear equations, Eq.~(\ref{eq:qsat}).
Lines from bottom to top correspond to 10\%, 20\%, 30\%, 40\%, 50\% difference.
Left: Contours in the case of the unintegrated gluon distribution $f(x,k^2)$;
right: contours in the case of the integrated gluon distribution $xg(x,Q^2)$.
Dashed line in both case corresponds to the saturation scale from
Golec-Biernat and W\"usthoff model \cite{GBW}.}}
\label{fig:qsatbeta}
\end{figure}

In order  to quantify the strength of the nonlinear term, one  introduces
the  saturation scale $Q_s(x)$. It divides the space in $(x,\k^2)$
into regions of the dilute and dense partonic system. In the case when $\k^2\!\!<\!\!
Q_s^2(x)$, the solution of the nonlinear BK equation exhibits the geometric scaling.
This  means that  it is dependent only on  one variable $N(r,x)=N(r Q_s(x))$ or in momentum space 
$f(k^2/Q_s(x))$.
To calculate the saturation scale in our model with resumed NLln1/x effects we follow
\cite{MartinSat}. There, the saturation scale was calculated
from the formula
\be   
-\frac{d \omega(\gamma_c)}{d\gamma_c} \; = \;
\frac{\omega_s(\gamma_c)}{1-\gamma_c} \; ,  
\label{eq:qsats}
\ee
This formula has been obtained by interpreting the quantity $\Phi(x,\k^2)$ or equivalently $f(x,\k^2)$
as field describing particle with coordinates ($\ln1/x$, $\ln\k^2/\k_0^2$). With such a field one can accociate
group velocity and phase velocity. Studying the solution of linearized BK one 
realizes that
in the particular case, when phase velocity and wave velocity become equal, the equation
admits so called wave front solution. This happens for some anomalous dimension $\gamma_c$
($\gamma$ is an argument of Mellin transform of the kernel). It was realized that the slightly modified
traveling wave solution of the linear equation can be used as ansatz for nonlinear equation.
The crucial point is that the exponent entering this ansatz is $\gamma_c$. It has interpretation of
critical exponent relevant for saturation. Using this exponent it is possible to calculate effective slope of
pomeron and saturation scale. For LOln1/x BFKL one obtains $\gamma_c=0.23$
\begin{figure}[th!] \centerline{\epsfig{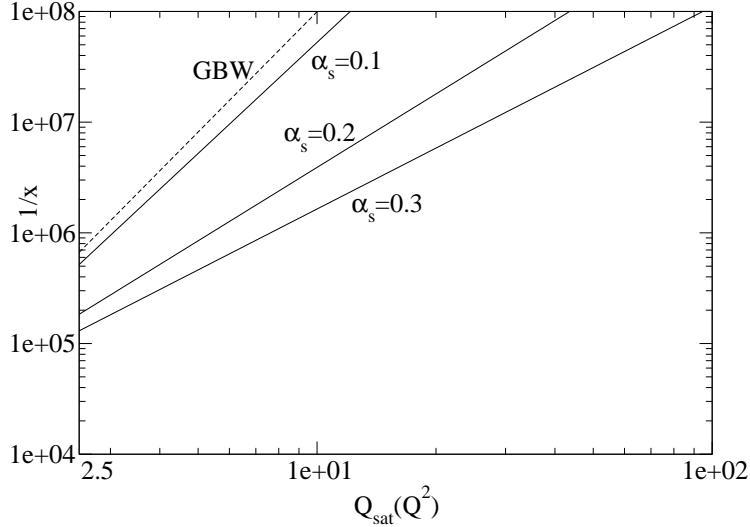}}
\caption{\em Saturation line obtained from (\ref{eq:omegas}, \ref{eq:chieigen}).}
\label{fig:linia}
\end{figure}
The effective Pomeron intercept $\omega_s$   is a solution to the equation
\be
\omega_s(\gamma) \; = \; \bar{\alpha}_s \, \chi(\gamma,\omega_s) \; ,
\label{eq:omegas}
\ee
where $\chi(\gamma,\omega)$ is  the kernel eigenvalue of the resumed model.
In our case the eigenvalue has the following form
\be
\chi(\gamma,\omega) \; = \; 2\Psi(1)-\Psi(\gamma)-\Psi(1-\gamma+\omega) +
\frac{\omega}{\gamma} \bar{P}_{gg}(\omega) \; .  
\label{eq:chieigen}
\ee
The solution for the saturation scale  obtained from solving (\ref{eq:qsats},\ref{eq:omegas}) 
using eigenvalue (\ref{eq:chieigen})
is shown in Fig.~\ref{fig:linia}
and gives $\lambda=\frac{\omega_s(\gamma_c)}{1-\gamma_c}=0.30,0.45,0.54$ for
three values of $\alpha_s=0.1,0.2,0.3$, respectively \cite{KS}. These results are similar to
results obtained in \cite{MartinSat}. We compare our results with the saturation
scale   from the Golec-Biernat and W\"usthoff model. Normalization of the
saturation scale is set to match GBW saturation scale at $x_0 = 0.41 \times
10^{-4}$.
The saturation scale $Q_s(x)$ can be also obtained directly from the numerical solution 
to the nonlinear equation by locating, for example, the maximum of the
unintegrated gluon density as a function of rapidity.
At large $x$ for fixed gluon momentum BFKL/DGLAP effects lead to a
strong growth of the gluon density. At certain value of $x$ the nonlinear
term becomes equal to linear and cancels out with it. That effect leads to
occurrence of a maximum which leads to a natural
definition of a saturation scale. We are going to discuss this issue further in context of the gluon density
with explicit dependence on impact parameter.
For the purpose of phenomenology we  attempt here to  estimate the effect
of the nonlinearity in a different, probably more quantitative way \cite{KS}.
We  study the relative difference between the solutions to the linear and nonlinear equations
\begin{equation}
\frac{|f^{\rm lin}(x,\tilde{Q}_s(x,\beta)^2)-f^{\rm nonlin}(x,\tilde{Q}_s(x,\beta)^2)|}
{f^{\rm lin}(x,\tilde{Q}_s(x,\beta)^2)} \; = \; \beta
\label{eq:qsat}
\end{equation}
where $\beta$ is a constant of order $0.1-0.5$.
Since this definition of the saturation scale is different from the one
used in the literature and is likely to posses different $x$ dependence, we denote  it 
as $\tilde{Q}_s$.
In Fig.~\ref{fig:qsatbeta}(up) we show a set $\tilde{Q}_s$ which are solutions to
Eq.~(\ref{eq:qsat}) for different choices of $\beta$ together with the
saturation scale calculated from the original saturation model by Golec-Biernat and 
W\"usthoff \cite{GBW}. Solid lines given by Eq.(\ref{eq:qsat}) show where the nonlinear solution
for the unintegrated gluon starts to deviate from the linear one by
 10\%, 20\%,\dots, 50\%. It is interesting  that
contours $\tilde{Q}_s(x)$ defined in  (\ref{eq:qsat}) have much stronger
$x$ dependence   than saturation scale  $Q_s(x)$ defined by
Eq.(\ref{eq:qsats}) and the one from  GBW model.   In particular
$\tilde{Q}_s(x,\beta) \, > \, Q_s(x)$ for given $x$ (at very small values of $x$).
This might be a hint that saturation corrections can  become  important much
earlier (i.e. for lower energies) than it would be expected from the usual definition of
the saturation scale $Q_s(x)$.   In Fig.~\ref{fig:qsatbeta}(down)
we also show contours in the case of the integrated gluon distribution function, that is the
solution to (\ref{eq:qsat}) with $f(x,\k^2)$ replaced by $xg(x,Q^2)$.
As already seen from the previous plot, Fig.~\ref{fig:xg}, the differences in 
the integrated gluon are more pronounced. For example in the case of $Q^2=25 \gev^2$
and $x\simeq 10^{-5}-10^{-6}$  we expect about 15\% to 30\% difference in the normalization.
 Again, by looking solely at the position of the critical line, one would expect the nonlinear 
 effects
 to be completely negligible in this region since at $x=10^{-6}$  the  corresponding 
 $Q_s^2(x) \simeq 2.8 \gev^2$ (taking  $Q_s^2(x)=Q_{s,0}^2 (x/x_0)^{-\lambda}$ 
 with  normalization  $Q_{s,0}^{2}=1 \gev^2$ at 
 $x_0\simeq 4 \times 10^{-5}$ and $\lambda \simeq 0.28$, \cite{GBW}). 
 This rough analysis shows that one cannot think of saturation scale
as a definite  and sharp border between very dilute and dense system.
 The transition
between these two regimes appears to be rather smooth and the nonlinear term
of the equation seems to have quite a large impact on the normalization
even in the 'linear' regime defined as $Q^2 \gg Q_s^2(x)$.
In practice, the estimate of the saturation effects is even more complicated since the
unintegrated gluon density has to be convoluted with some impact factor, and the integration
over the range of scales must be performed.

\newpage
\subsection{Dipole cross section $\sigma(r,x)$}

\begin{figure}[t!]
\centering{
\includegraphics[width=0.7\textwidth]{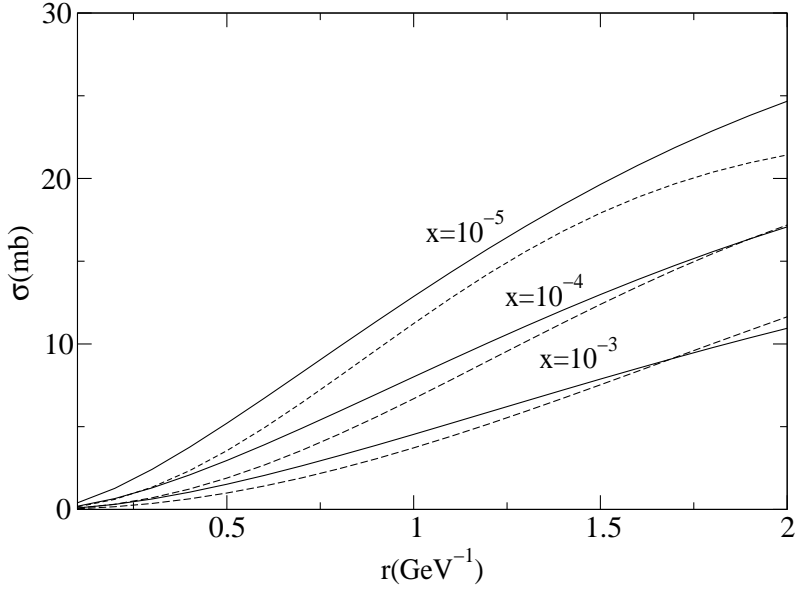}\\}
\caption{\em The dipole cross section obtained from
modified BK (solid line) compared to GBW dipole model (dashed line).}
\label{fig:dipol}
\end{figure}

It is interesting to see what is the behavior of the dipole cross section $\sigma(r,x)$ \cite{KS}obtained 
from the unintegrated gluon density via:
\be
\sigma(r,x)=\frac{8\pi^2}{N_c}\int\frac{d\k}{\k^3}(1-J_0(|\k||\r|))\alpha f(x,\k^2),
\ee
this equation follows from (\ref{eq:dipglu}) via integration over impact parameter. 
In this calculation we assume that $\alpha_s$ is running with the scale $\k^2$ and we use $r$
size of the initial dipole, i.e. $r\!\equiv\!\x_{01}$.
Calculation of the dipole cross section requires the knowledge of the unintegrated gluon density for
all scales $0\!<\!\k^2\!<\!\infty$.
Since in our formulation the unintegrated gluon density is known for
$k^2>k_0^2$ we need to parametrise $f(x,k^2)$ for lowest values of $\k^2\!\!<\!\!\k_0^2$. We use the matching
condition
\be
xg(x,\k_0^2)=\int_{0}^{\k{_0}^2}\frac{d\k^2}{\k^2}f(x,\k^2)  \; ,
\ee
and following \cite{GBBK} we assume that $f(x,k^2)\sim \k^4/\k_0^4$ for low $\k^2$.
This gives  (compare Eq.(\ref{eq:gluony1}) )
\be
f(x,\k^2)=4N(1-x)^{\rho}(\k/\k_0)^4 \; .
\ee
In Fig.~\ref{fig:dipol} we present the dipole cross section as a function of the
dipole size r for three values of $x=10^{-3}\,10^{-4}$ and $10^{-5}$. For comparison
we also present  the dipole cross section obtained from GBW parametrisation. To
be self-consistent, we cut the plot at $r=2 \ \gev^{-1}$ because we assumed in the
derivation  of formula (\ref{eq:BK222}) that the dipoles are small in comparison
to the target size (we assume proton radius to be $4\, {\rm GeV}^{-1}$).
This cut allows us to obtain a model independent result since we observe that
different parametrisation of $f(x,\k^2)$ for $\k^2\!\!<\!\!\k_0^2$ give essentially
the same contribution for $r<2\, {\rm GeV}^{-1}$.
We observe that our extraction of the dipole cross section gives similar result to the GBW parametrisation.
The small difference in the normalization is probably due to the different values of $x_g$ which probe the gluon
distribution (or alternatively the dipole cross section).
In the GBW model the dipole cross section is taken at the value $x_g=x$ which is the standard 
Bjorken $x = Q^2/2 p\cdot q$.
On the other hand, in the formalism presented in Ref.~\cite{KMS}
one takes into account the exact kinematics (energy conservation) in the photon impact factor.
It is a part of the subleading effect in the impact factor and it increases the value of $x_g \sim 5 x$.
Therefore, in our formalism the normalization of the unintegrated gluon  is increased so that
the convolution with the impact factor and the resulting structure function remains the same.

\section{Description of the $F_2$ structure function}
In this section we are going to use the BK equation with subleading 
corrections to describe the $F_2$ structure function \cite{KuM}.
To do so, first of all we are going to use more realistic profile function of
the nucleon. Instead of the cylindrical profile of the nucleon we are going to
use Gaussian one.
Our motivation is to control the impact parameter
dependence in a more detail in order to have more general formalism applicable 
for prediction for more exclusive
processes as for example diffractive Higgs production. The description of $F_2$ is performed to fix
parameters of the input.
The Gaussian profile is more realistic than the cylindrical since it introduces dependence of
screening corrections on the distance from the center of the target and does not
lead to over-saturation of the gluon density at very large rapidity.
With that setup the BK equation reads:
$$
\frac{\partial f(x,\k^2,b)}{\partial \ln(1/x)}=\\
$$
\be
\begin{split}
+ {N_c\alpha_s(k^2)\over \pi} \k^2\int_{k_0^2}^{\infty}{d\k^{\prime 2}\over
\k^{\prime 2}}\bigg \{{f\left({x},\k^{\prime 2},b\right)(\theta\left(\k^2-\k^{\prime 2}
\right) +\left(\frac{\k^2}{\k^{\prime 2}}\right)^{\omega_{eff}}\theta\left(\k^{\prime 2}-\k^2
\right)) - 
f\left(x,\k^2,b\right)\over |\k^{\prime 2}-\k^2|}\\
+{f\left({x},\k^2,b\right)\over [4\k^{\prime 4}+\k^4]^{{1\over 2}}}\bigg\}
+{\alpha_s(\k^2)\over 2 \pi}P_{gg}(0)
\int_{k_0^2}^{k^2}{d\k^{\prime 2}\over \k^{\prime 2}}f\left({x},\k^{\prime 2},b\right)
\nonumber\\
\end{split}
\ee
\be
\begin{split}
 -2\pi\alpha_s(\k^2)\left[2\k^2\left(\int_{\k^2}^{\infty}\frac{d\k'^2}{\k'^4}f(x,\k'^2,b)\right)^2
+2f(x,\k^2,b)\int_{\k^2}^{\infty}\frac{d\k'^2}{\k'^4}\ln\left(\frac{\k'^2}{\k^2}\right)f(x,\k'^2,b)\right]
\label{eq:eqdorozw2}
\end{split}
\ee
the inhomogeneous term stands for the input gluon distribution and is given by:
\be
\tilde f^{(0)}(x,\k^2,b)=\frac{\alpha_s(\k^2)}{2\pi}\, S(b)\,
\int_x^1 dz P_{gg}(z)\frac{x}{z}g\left(\frac{x}{z},\k_0^2\right) \,.
\label{eq:input}
\ee
As usual the unintegrated gluon density is obtained via the integration over distance
from the center of the target:
\be
f(x,\k^2)=\int\!d^2b f(x,\k^2,b)
\ee
the input profile is 
\be
S(b)=\frac{\exp (\frac{-b^2}{R^2})}{\pi\,R^2}
\label{eq:zalodb}
\ee
We take $R=2.8GeV^{-1}$ which follows from the measurement of
diffractive $J/\psi$ photo-production off proton.
\begin{figure}[t!]
\centerline{\epsfig{file=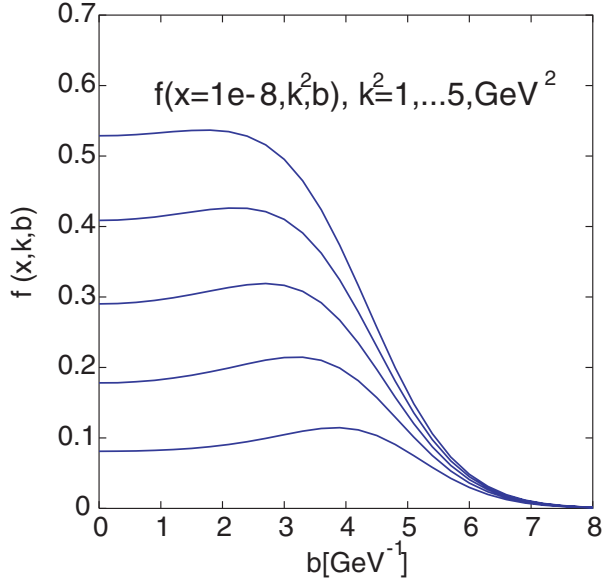,height=8cm}}
\caption{\em The unintegrated gluon density as a function of b for fixed $k^2$ and $x=10^{-8}$}
\label{fig:bdepglue}
\end{figure}
\begin{figure}[th!]
  \begin{picture}(490,0)
    \put(100, -170){
      \includegraphics{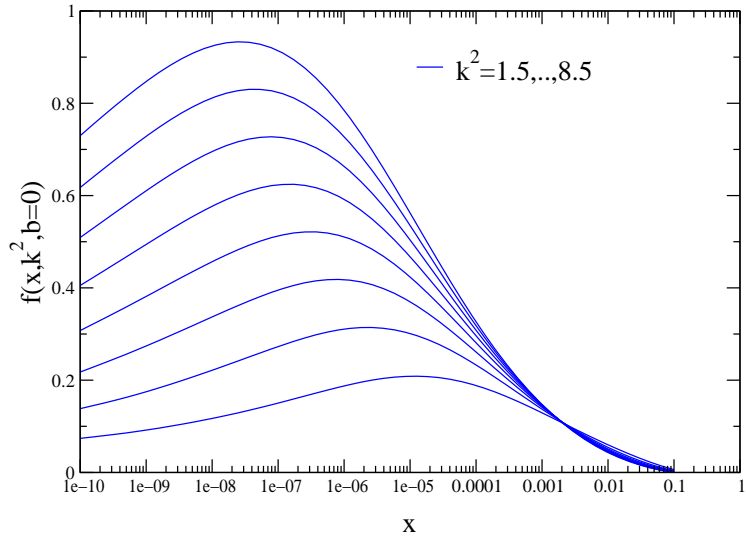}
    }
    \put(100, -480){
      \includegraphics{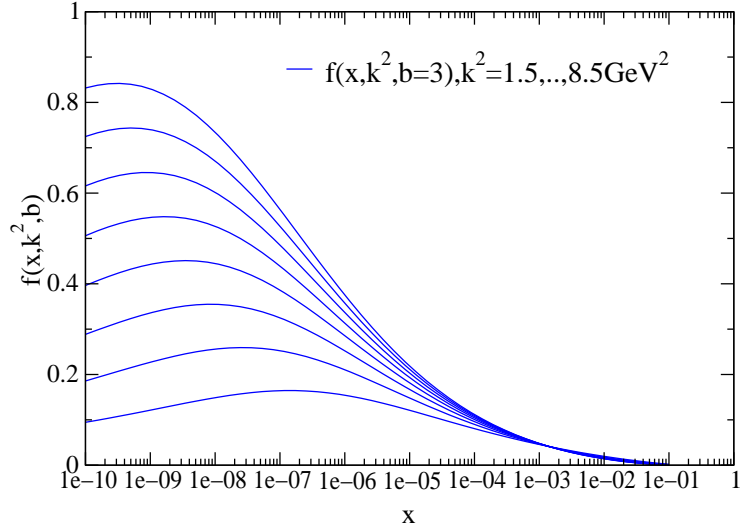}
    }
        \end{picture}
\vspace{16cm}
\caption{\footnotesize{The unintegrated gluon density as a function of $x$ for b=0 and fixed $\k^2$(up),
The unintegrated gluon density as a function of $x$ for $b=3$ and fixed $\k^2$ (down).}}
\label{fig:funkx}
\end{figure}
\begin{figure}[th!]
  \begin{picture}(490,0)
    \put(100, -170){
      \includegraphics{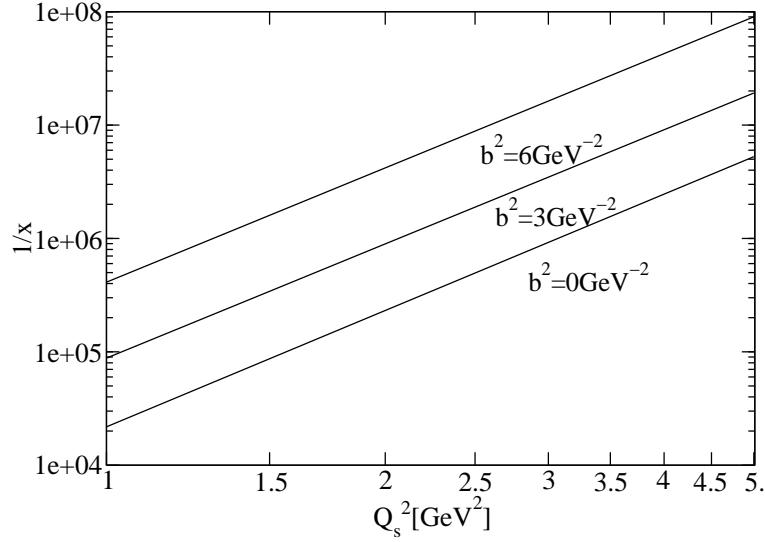}
    }
    \put(100, -480){
      \includegraphics{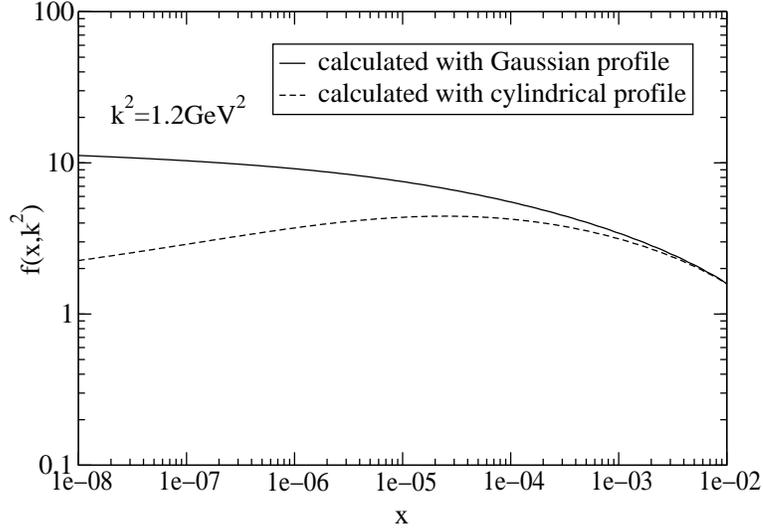}
    }
        \end{picture}
\vspace{16cm}
\caption{\footnotesize{Critical line obtained for three values of the impact parameter
(up). The unintegrated gluon density integrated over impact parameter for different profile
functions. Continuous line corresponds to Gaussian profile while scatter line corresponds to
cylindrical target.}}
\label{fig:satline}
\end{figure}
In fig (\ref{fig:bdepglue}) we plot the unintegrated gluon density as a function of
the impact parameter for fixed values of $\k^2$ and for $x=10^{-8}$.
We observe that at small values of $x$ and central values of $\b$ the
saturation effects are strongest and lead to the depletion of the gluon density.
On the other hand, due to large distance phenomena as for example confinement which we model
via Gaussian input,
there are less gluons for peripheral $\b$. The net result is clear.
The impact parameter dependent gluon density has a maximum as a function of distance from the target.\\
Fig. \ref{fig:funkx}(down) visualizes the unintegrated gluon density as a function
of $x$.
At large $x$ for fixed gluon momentum BFKL/DGLAP effects lead to a
strong growth of the gluon density. At certain value of $x$ the nonlinear
term becomes equal to linear and cancels out with it. This effect leads to
occurrence of a maximum which leads to a
definition of a saturation scale \cite{KuM} i.e. we define the saturation scale
as $Q_s^2$ for which:
$
\frac{\partial f(x,Q_s^2,\b)}{\partial{\ln 1/x}}=0\, .
$
Using that definition of saturation line we plot critical line for three values of impact parameter
(\ref{fig:satline}). The extracted value of parameter $\lambda$ according to prescription
$Q_s^2=\left(\frac{x_0}{x}\right)^{\lambda}$ gives $\lambda=0.293$ for $b^2=0GeV^{-2}$,
$\lambda=0.298$ for $b^2=3GeV^{-2}$ and $\lambda=0.298$ for $b^2=6GeV^{-2}$. Those values agree with
the $\lambda$ parameter obtained in previous section where we studied saturation line
coming from wave front analysis. It was obtained for $\alpha_s=0.1$ and $\lambda=0.29$.
That value is also very close to the one in $GBW$ model which we already referred to. It is worth to
note that the normalization that we 
obtained in case of $b^2=0GeV^{-2}$ was not far from the GBW model i.e. $x_0=0.45\times 10^{-4}$
while that value in GBW was $0.41\times 10^{-4}$.

A similar  maximum is not seen for the gluon
density integrated over the impact parameter Fig. \ref{fig:satline}(down),
it flattens, but does not fall as very small $x$ is approached. This is due
to the large contribution to the integral from the peripheral region where density of gluons has not
saturated yet.
In Fig. \ref{fig:funkx} we observe well known fact that the lower $\k$ is the
earlier saturation effects manifest themselves. This can be understood in our approach from the
structure of the integral in the nonlinear part. 
The lower limit of integration is given by value of $\k^2$
at which the gluon is probed and extends to the infinity.
The lower the momentum is, the longer is the path of integration.
Let us now proceed to calculate the $F_2$ structure function. The method of
calculation is the following. We use the solution of (\ref{eq:eqdorozw2}) to obtain
the unintegrated gluon density which depends on the distance from the center of the
target. Then we integrate it over the distance from the center of the target.
The next step is application of the KMS prescription to calculate $F_2(x,Q^2)$. 
In this approach, i.e. the sea quark contribution is calculated within the $k_T$ factorisation
including NLOln1/x corrections coming from exact kinematics. The valence quarks
are taken from  \cite{GRV}. Within the KMS prescription: $F_2^{sea}=\sum_q S_q$:
where:
\begin{equation}
S_q(x,Q)=\frac{Q^2}{4\pi^2\alpha_{em}}\int{d\k^2\over \k^4}\int_x^{a_q(\k^2)}
{dz\over z}  S_{box}^{q}(z,\k^2,Q^2)f\left( {x\over z},\k^2\right)
\label{eq:ktfac}
\end{equation}
where the impact factors corresponding to the quark box contributions
to gluon-boson fusion
process are the same as those used in ref. \cite{KMS} (see also \cite{BE}), i.e.:
\be
S_q^{box}(z,\k,Q)=\alpha_{em}\int_0^1 d\beta d^2\mbox{\boldmath
{$\kappa^{\prime}$}}
\alpha_s\delta(z-z_0)
\label{eq:ktfac2}
\ee

\begin{equation}
\left\{[\beta^2+(1-\beta)^2]\left({\mbox{\boldmath {$\kappa$}}\over D_{1q}}-
{\mbox{\boldmath {$\kappa$}}-{\bf k}\over D_{2q}}
\right)^2+[m_q^2 +4Q^2\beta^2(1-\beta)^2
\left({1\over D_{1q}}-{1\over D_{2q}}
\right)^2\right\}
\label{eq:impf}
\end{equation}
where $\mbox{\boldmath
{$\kappa^{\prime}$}}
=\mbox{\boldmath {$\kappa$}}-(1-\beta){\bf k}$ and

$$
D_{1q}=\kpp^2+\beta(1-\beta)Q^2 + m_q^2
$$

$$
D_{1q}=(\kpp-\k)^2+\beta(1-\beta)Q^2 + m_q^2
$$
\begin{equation}
z_0=\left[1+{\kpp^{\prime 2} +m_q^2\over \beta(1-\beta)Q^2}+{k^2\over Q^2}\right]^{-1}
\label{eq:ddz}
\end{equation}
\begin{figure}[t!]
\centerline{\epsfig{file=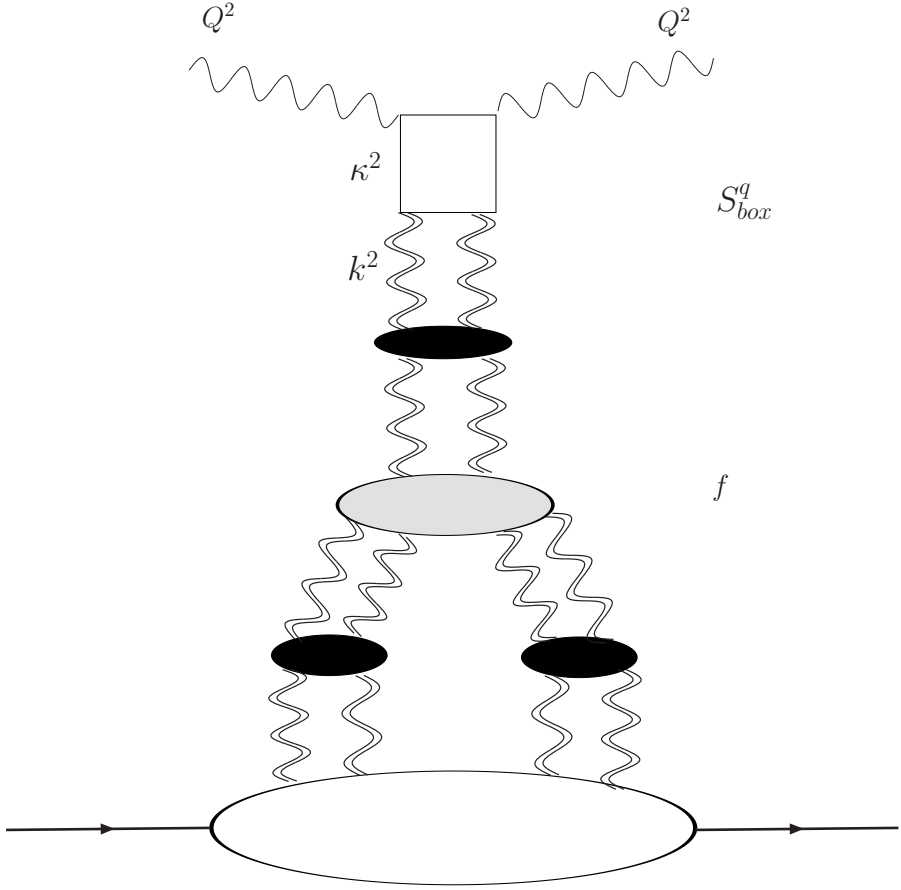,height=12cm}}
\caption{\em Diagrammatic representation of the $F_2$ calculation.}
\label{fig:convbox}
\end{figure}
The effect of kinematical NLOln1/x corrections is included  via argument of unintegrated gluon
density in formula (\ref{eq:ktfac}) which is defined as $x_g=x/z$. This is an
effect coming from exact kinematics of the gluon-boson fusion. At
leading order we would have $z_0=1$ in (\ref{eq:ddz}) and the integral over $z$
would (neglecting quark masses) reduce  (\ref{eq:ddz}) to leading LOln1/x formula (\ref{eq:ktLOx}).
So, at leading order we have $\phi_q(\k,Q)\equiv\int_z^1dz\frac{dz}{z}S_q^{box}(z,\k^2,Q^2)$, where
$\phi(\k,Q)$ was introduced in the first chapter.\\ The full NLOln1/x corrections are still not
known, but there is activity in that direction \cite{BGQ,BGK}. In the calculation of
the (effective) quark distributions we use the impact factors (\ref{eq:impf})
corresponding to the  massless quarks and the (charmed) quark mass  effects are
 included  in the threshold  factor:
\begin{equation}
a_c(\k^2)=\left(1+{\k^2+m_{c}^2\over Q^2}\right)^{-1}
\label{acs}
\end{equation}
In the
impact factors we use (\ref{eq:impf}) with $m_u=m_d=m_s=0$ and $m_c=1.4GeV$.
The various components entering the box in the KMS prescription are treated in
the following way
\begin{itemize}
\item region when $\k^2,\kpp'^2\!\!<\k_0^2$
is assumed to be dominated by the soft pomeron exchange
\begin{equation}
S^{a}=S_u^P+S_d^P+S_s^P \end{equation}
where
\begin{equation}
S_u^P=S_d^P=2S_s^P=0.037\,\,\ln(462/x)\,\,(1-x)^8\,
\end{equation}
where we modified the power like contribution from nonperturbative
region in order to be
consistent with saturation picture
\item
$\k^2<\k_0^2<\kpp'^2$ the strongly ordered approximation is applied, what
results in neglecting $\k^2$ in comparison to $Q^2$ and $\kpp^2$
\item
The region $\k^2>\k_0^2$ is treated perturbatively and perturbative
expression for $S_q$ is used
\end{itemize}
\begin{figure}[th!]   
\begin{picture}(490,10)     \put(40,0){       \includegraphics{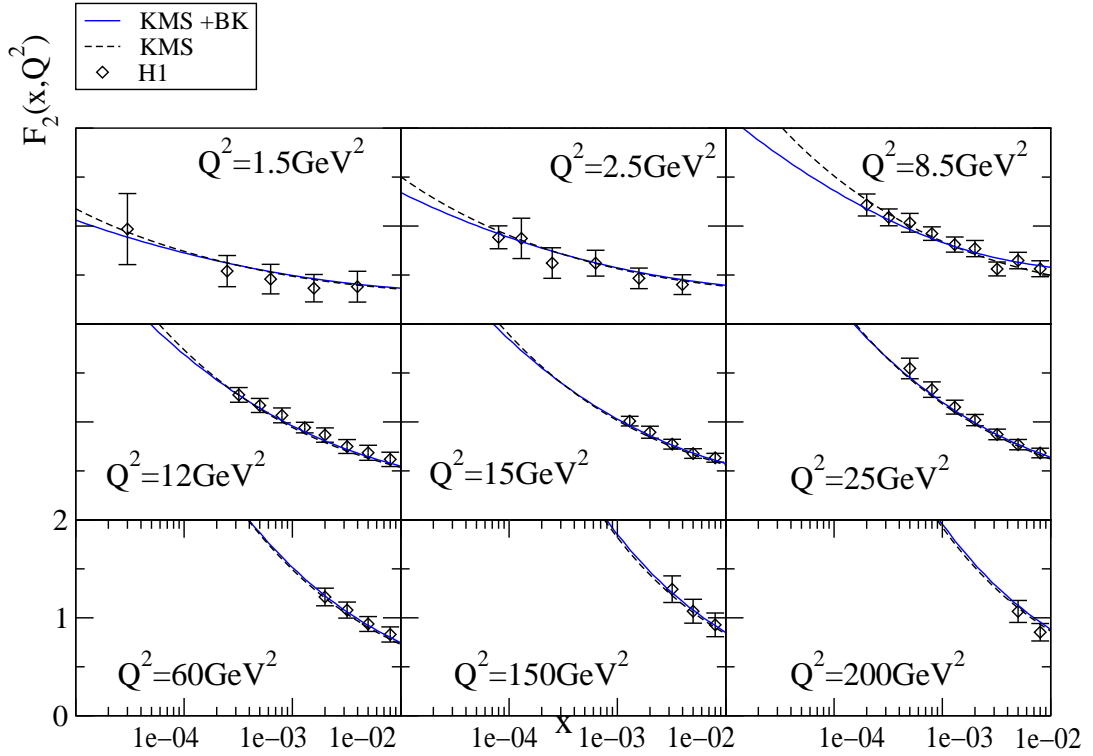}
    }
        \end{picture}
\vspace{12cm}
\caption{\footnotesize{$F_2$ structure function.}}
\label{fig:F22}
\end{figure}
The charm component of $\Sigma$ is generated
perturbatively, i.e. there is no soft pomeron contribution to it and it is given by:
\bea
S_{q=c}(x,Q^2,)=\int^{a}_x\frac{dz}{z}\int_0^{\k_0^2}S_{box}(z,\k^2=0,Q^2;m_c)
\frac{d\k^2}{\k^2}f(\frac{x}{z},\k^2)\nonumber\\
+\int^{a}_x\frac{dz}{z}\int_{\k_0^2}^{\infty}S_{box}(z,\k^2=0,Q^2;m_c)
\frac{d\k^2}{\k^2}f(\frac{x}{z},\k^2)
\eea
With that formalism we obtained good description of data which
made it possible to fix parameter of the input gluon distribution:
\be
xg(x,\k_0)=1.572(1-x)^{2.5}
\ee
The effect of nonlinearity is, however, hardly visible and is just at the edge of HERA
kinematical region Fig. \ref{fig:F22}. We also would like to stress that the effect of 
saturation is better visible and more pronounced in case of the impact parameter dependent
gluon density. This observation suggests that nonlinear effect may easier
easy to observe for quantities which are sensitive to that parameter or its
Fourier conjugate: the momentum transfer.
\section{Rescattering corrections in the diffractive Higgs boson production.}
In this section we will compute the hard rescattering correction to diffractive
central Higgs production at LHC energies \cite{BBKM}.
The diffractive Higgs production in collision of protons is of interest since the
background is much smaller then the expected signal. The other advantage is that
the mass of produced Higgs boson can be measured in two independent ways; first
with the high accuracy by measuring the missing mass to the forward out going
protons and, second by the $H\rightarrow b\overline b$ decay.
\subsubsection{Standard approach}
\begin{figure}[b!]
\centerline{\epsfig{file=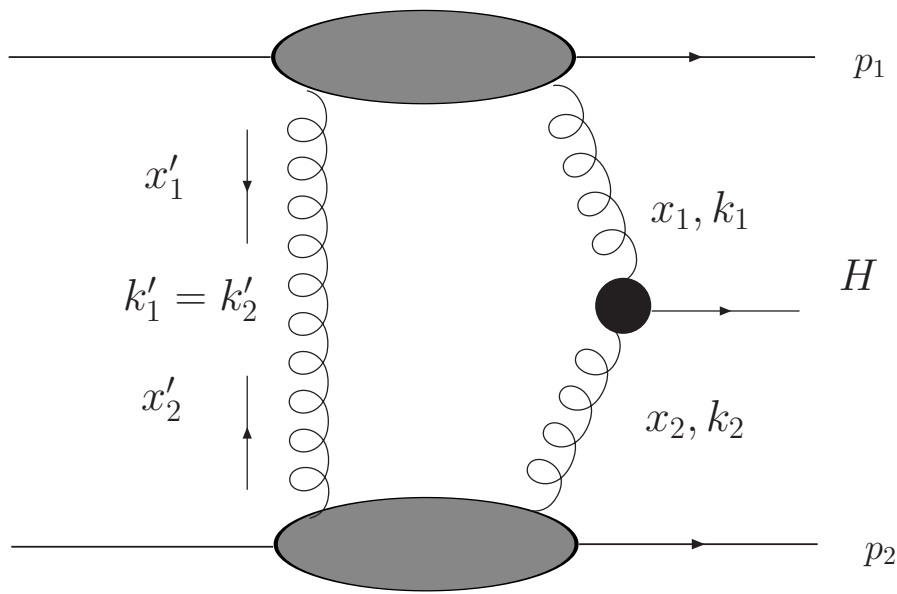,height=8cm}}
\caption{\em Kinematics of the Two Pomeron fusion contribution to the exclusive Higgs boson production.}
\label{fig:dis}
\end{figure}
In the standard pQCD \cite{KMR1}-\cite{KMR7} approach, it is without including the hard rescattering
corrections the amplitude for process under studies is given by:
\be
M_0(y)=i2\pi^3A\int\frac{d^2\k}{2\pi \k^4}f^{off}(x_1,\k^2;\mu)
f^{off}(x_2,\k^2;\mu)
\label{eq:amplhiggs}
\ee
Here, the constant above is given by:
\be
A^2=K\sqrt G_F\alpha_S^2(M_H^2)/9\pi^2
\ee
where $G_F$ is the Fermi coupling constant, and $K\approx 1.5$ is the NLO
$K-$factor.
The quantity $f^{off}(x_1,\k^2;\mu)$ is so called off-diagonal gluon distribution and is
given by:
\be
f(x,\k^2;\mu)=R_{\xi}Q^2\frac{\partial}{\partial Q^2}
\biggl[x g(x,Q^2)\sqrt{T_g(Q,\mu)}\biggr]_{Q^2=\k^2}.
\label{unglue}
\ee
where $T_g(k,\mu)$ is the Sudakov form-factor and it accounts for probability that there will 
be no multiple emission from fusing gluons, 
\be
R_{\xi}=\frac{2^{2\lambda +3}}{\sqrt{\pi}}\frac{\Gamma(\lambda +5/2)}{\Gamma(\lambda+4)} 
\ee
is the correction factor and allows us to obtain off-diagonal gluon distribution if we now
diagonal, the $\mu$ scale is $\mu\approx 0.62M_{H}$ .
The last important point is the so called opacity factor $\Omega(s,b)$ which accounts for 
nonperturbative soft
rescattering. Those effects are factorised from the hard process leading to the Higgs production under
assumption that the hard process leading to the production of Higgs occurs on short time scale.
Assuming this factorisation is fulfilled the effect of soft physics enters just by multiplication of
the hard amplitude by so called gap survival factor which depends on the opacity $S^2=exp(\Omega(s,\b))$.
The differential cross section following from the amplitude (\ref{eq:amplhiggs}) (without including gap
survival factor) is:
\be
\frac{d\sigma_{pp\rightarrow pHp}(y)}{dydt_1dt_2}=\frac{|M_0(y)|^2}{256\pi^3}
\ee
After inclusion of the proton elastic form factors  
one obtains the dependence of the cross section
on momentum transfers $t_1=(p_1-k_1)^2$ and $t_2=(p_2-k_2)^2$ of the form of $\exp(R^2(t_1+t_2)/2)$.
Thus, after the integration over momentum transfers $t_1$ and $t_2$,
the single differential cross section reads:
\be
{d\sigma_{pp\rightarrow pHp}(y) \over dy} =  {|M_0(y)|^2 \over 64\pi^3 R^4}.
\ee
For our future developments it is convenient to have the right-hand side of expression above
in impact parameter representation using the Gaussian profile (\ref{eq:zalodb})
we obtain:
\be
{d\sigma^{(0)}(y)_{pp\rightarrow pHp}  \over dy} = {1\over 16\pi}\,
\int d^2 b \int d^2 \b_1 |S(\b_1)S(\b-\b_1)M_0(y)|^2,
\ee
where $\b_1$ and $\b_2=\b-\b_1$ are transverse positions of the production vertex measured from 
centers
of the protons and $b$ is the impact parameter of the collision.

\subsection{Hard rescattering corrections}
The motivation to include hard rescattering corrections to diffractive Higgs production 
comes from the large rapidity gap between the colliding protons. At rapidity $19 Y$ apart 
from single logarithms of Higgs boson mass small $x$ logarithms should be resumed.
To be specific let us consider process where two protons scatter at large rapidity. 
According to our previous discussion of saturation interacting partons may overlap and
recombine what will have impact on the Higgs production.
That effect may be sizable because of large rapidity available. 
We would like to stress here that we are going to consider only a subclass of the screening
corrections which should be included in BFKL-based description of the Higgs production. In
particular we are going to restrict ourselves to pomeron fan diagrams configurations in the
exchanged system which leads to Balitsky-Kovchegov equation for the gluon density. The considered
configuration is therefore highly asymmetric and should be considered as estimate of importance
of rescattering corrections.  
To be specific the amplitude consists of following elements:
\begin{itemize}
\item The two scale unintegrated gluon distribution, $f^{(a)}(x_1,\k_1;\mu_a)$
describing gluons above the TPV. This gluon distribution is evaluated at relatively
large $x_1$. The scale $\mu_a=\k_1$
\item
The two-gluon Green's function $G_A(\k_2,\k_3;x_1,x_2;\mu_a,\mu_b)$ that 
describes the propagation of the two gluon singlet state between the Higgs production vertex and the TPV. 
The diffusion approximation for the Green's function  
is going to be considered $G_{BFKL}(\k_2,\k_3;x_1,x_2;\mu_b)$.
The Green's functions will have typical $\mu_b\approx m_H/2$, which belongs to the Higgs production vertex. 
In order to account for the scale dependence at the Higgs vertex, the Sudakov form factor, $\sqrt{T_g(k_k;\mu_b)}$ 
will be included into the Green's function:
\be
G_{BFKL}(\k_2,\k_3;x_1,x_2;\mu_b)=\frac{|\k_2| |\k_3|}{2\pi}\frac{\sqrt{T_g(\k_3,\mu_b)}}
{\sqrt{\pi D log(x_1/x_2)}}\exp \left[\frac{\log^2(\k_2^2/\k_3^2)}
{4Dlog(x_1/x_2)}\right]\left(\frac{x_1}{x_2}\right)^{\lambda}
\label{eq:bfklgreen}
\ee
where $D=14\zeta(3)$
\item The Higgs boson production vertex which does not differ from the TPF
\item The off-diaginal unintegrated gluon distribution, $f^{b}(x_3,\k_3,\q-\k_3;\mu_b)$ gluon distribution 
that enters the Higgs boson production vertex from the lower 
side, with the total transverse momentum $\q$. That gluon distribution is taken to be the 
same as in the case of the TPF i.e. it is an off-diagonal gluon distribution given  with $\mu_b=m_H/2$
\item
The off-diagonal unintegrated gluon distribution, $f^{(c)}(x_4,\k_4,-\q-\k_4;\mu_c)$, 
for the screening Pomeron that propagates across the large rapidity distance from the 
lower proton to the TPV with momentum transfer $-\q$. The scale $\mu_c$ is set by the gluon 
virtuality, so one has $f^{(c)}(x_4,\k_4,-\q-\k_4;\k_4)$.
\end{itemize}
\begin{figure}[t!]
\centerline{\epsfig{file=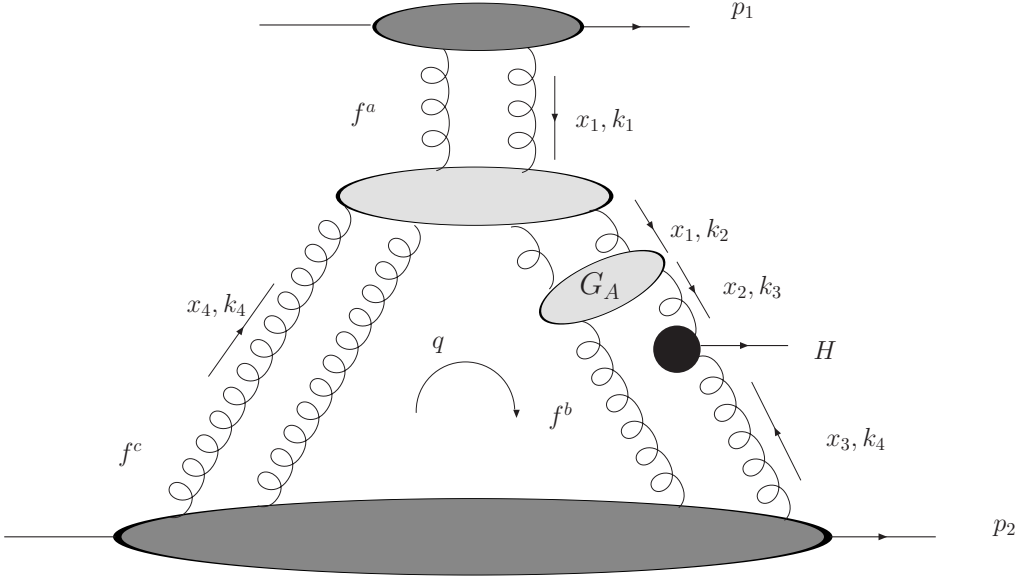,height=8cm}}
\caption{\em The hard rescattering correction to the exclusive Higgs production.}
\label{fig:dis}
\end{figure}
After these definitions the dominant part of the diagram can be written:
\bea
Im M^{(1)}_{corr}(y)=-16\pi^2 2\pi^3 A\int_{x_a}^{x_b}\frac{dx_4}{x_4}
\int\frac{d^2\q}{(2\pi)^2}\int\frac{d^2\k_2}{2\pi\k_2^4}\int\frac{d^2\k_3}{2\pi \k_3^4}
\int\frac{d^2\k_4}{2\pi \k_4^4}\nonumber\\
\int\frac{d^2\k_1}{2\pi \k_1^2}f^{(a)}(x_1,\k_1)V(\k_1,-\k_1,\k_2,-\k_2,\k_4,-\k_4)
G_A(\k_2,\k_3;x_1,x_2;m_H/2)\nonumber\\
f^{(b)}(x_3,\k_3,\q-\k_3;m_H/2)f^{(c)}(x_4,\k_4,-\q-\k_4)
\eea
where:\\
$x_3=\frac{M_H}{\sqrt s}e^{-y}$,  $x_2=\frac{M_H}{\sqrt s}e^{-y}$,
$x_1=\frac{\k_0^2}{x_4s}$ and $x_a\simeq\frac{\k_0^2}{s}$, $x_b\simeq\frac{\k_0^2}{x_2s}$.
The scale $\k_0^2$ that enters the definitions $x_i$ corresponds to the average values of the
virtualities, we set that scale $\k_0^2=1GeV^2$. The scattering is assumed to be forward, however,
due to the fact that below the TPV the loop is formed the momentum transfer ($\q$) dependence 
enters $f^{(b)}$,
$f^{(a)}$. The diagram accounting for replacement of target an projectail is accounted by
replacing $y\rightarrow -y$.
The correction amplitude is then:
\be
M_{corr}(y)=M^{(1)}_{corr}(y)+M_{corr}^{(1)}(-y)
\ee
The integral above is infrared safe due to emergence of the saturation scale
$Q_s^2(x_4)$
in gluon $f^{(c)}$. That scale at interesting us values of $x_4$ as HERA data indicates is
well above the cut-off scale $\k_0^2$.
To implement the soft rescattering correction which will enter through so called 
opacity factor
it is needed to have formula for the differential cross section in impact parameter 
representation. 
It reads:
\be
{d\sigma^{(0+1)}(y)_{pp\to pHp}  \over dy} = {1\over 16\pi}\,
\int d^2 \b \int d^2 \b_1 |S(\b_1)S(\b-\b_1)M_0(y)+
 M_{\mathrm{corr}}(y,\b,\b_1)|^2,
\ee
where the dominant parts of both $M_0(y)$ and
\bea
Im M^{(1)}_{corr}(y,\b,\b_1)=-16\pi^2 2\pi^3 A\int_{x_a}^{x_b}\frac{dx_4}{x_4}
\int\frac{d^2\q}{(2\pi)^2}\int\frac{d^2\k _2}{2\pi\k_2^4}\int\frac{d^2\k_3}{2\pi \k_3^4}
\int\frac{d^2\k_4}{2\pi \k_4^4}\nonumber\\
\int\frac{d^2\k_1}{2\pi \k_1^2}[f^{(a)}(x_1,\b-\b_1)\otimes V](\k_2,k_4)
G_A(\k_2,\k_3;x_1,x_2;m_H/2)\nonumber\\
f^{(b)}(\x_3,\k_3,\b_1;m_H/2)f^{(c)}(x_4,\k_4,\b_1)-\{y\rightarrow-y\}
\eea
The impact parameter dependence of
$ f^{(a)}$ and $ f^{(b)}$ is given by $S(b)$ and may be
factorised out, in analogy with the case of the Two Pomeron Fusion amplitude.
The impact parameter dependence of $ f^{(c)}$ is determined by the
BK~evolution, and it is different from $S(\b)$ because of nonlinearity.
Having defined $M_{corr}(y,\b,\b_1)$, we can write down the cross section for the
exclusive production, taking into account the soft rescattering
\be
{d\sigma^{(0+1),\Omega} _{pp\longrightarrow pHp} (y) \over dy} = {1\over 16\pi}\,
\int d^2 \b \int d^2 \b_1\,
|S(\b_1)S(\b-\b_1)M_0(y)+M_{corr}(y,\b,\b_1)|^2\,
\exp(-\Omega(s,\b)),
\label{final}
\ee
where $\,\Omega(s,\,b)\,$ is the opacity of the $\,pp\,$ scattering
at squared energy $s$ and impact parameter $b$.
Here we adopt the opacity factor of
the two channel eiconal model proposed in \cite{KMR11,KMR12}.
\subsection{Hard rescattering corrections - results and discussion.}
\begin{figure}[th!]
\centerline{\epsfig{file=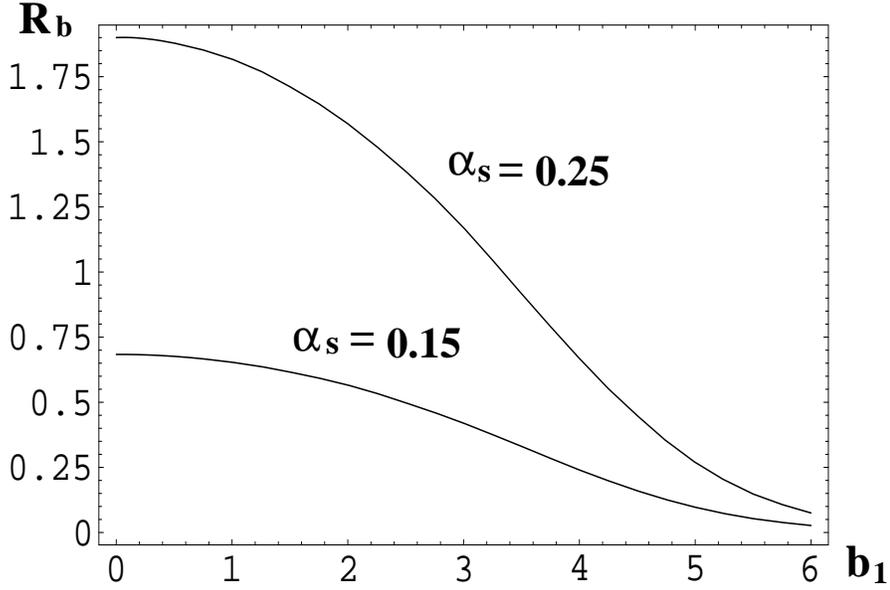,height=8cm}}
\caption{\em The ratio of the hard rescattering correction to the 
Two Pomeron Fusion at b=0, as
a function of the transverse, distance, $b_1$ of the Triple Pomeron Vertex from 
the center of the
target proton, for the two values 
$\alpha_s=0.15$ and $\alpha_s=0.25$ at $\sqrt s=14TeV$.}
\label{fig:Amp16}
\end{figure}

All numerical calculations of the amplitude and of the cross section for the central
exclusive Higgs boson production have been performed choosing $y=0$ and assuming the LHC energy
$\sqrt s=14TeV$. The mass of the Higgs boson was assumed to be $M_H=120GeV$.
In Fig. \ref{fig:Amp16} we show ratio of corrected amplitude to TPF amplitude as function of 
transverse distance $\b_1$ of the Higgs boson production vertex from the center of the target 
proton:
\be
R_b(b_1)=\left[\frac{M_{corr}(y,b,b_1)}{M_0(y,b,b_1)}\right]_{y=0,b=0}
\label{eq:rat}
\ee
\begin{table}
\begin{center}
\begin{tabular}{|c|c|c|c|c|c|}
\hline
\, & \, & \,& \,& \, & \, \\
$d\sigma_{\mathrm{TPF}}/dy$ & $\,d\sigma/dy\,(\alpha_s\,=0.15)\,$ &
$\,d\sigma/dy\,(0.17)\,$ & $\,d\sigma/dy\,(0.2)\,$ &
$\,d\sigma/dy\,(0.22)\,$ & $\,d\sigma/dy\,(0.25)\,$ \\
\, & \,  & \,& \,& \, & \,\\
\hline
\, & \, & \,& \,& \, & \, \\
0.4 & 0.16 & 0.12 & 0.042 & 0.08 & 0.14 \\
\, &  \, & \,& \,& \, & \, \\
\hline
\end{tabular}
\label{Ampl7}
\end{center}
\end{table}
The amplitudes are integrated over $y_1$.
We observe that the ratio (\ref{eq:rat}) decreases with increasing distance, $\b_1$
from the proton center. Thus, the corrections term tends to suppress the Higgs production 
inside the
target proton. The correction is large for both choices of $\alpha_s$.
In Table 1 we show values of differential $d\sigma/dy$ cross section for TPF and TPF with
rescattering corrections included. Results for different values of $\alpha_s$ are presented.
At $\alpha_s=0.2$ the absolute value of the hard rescattering correction exceeds the Two 
Pomeron
Fusion term and this explains why, at this value of $\alpha_s$, the corrected cross section 
takes
its smallest value. For all considered values the cross section with rescattering corrections 
is
smaller then TPF and is sensitive to the triple pomeron coupling.
As we have already said we considered only subclass of diagrams that contribute to the described
process. The more sophisticated considerations would require fan diagrams from both sides 
\cite{MB}
and the inclusion of the pomeron loop or BKP states.
\chapter{Summary and conclusions}
In this thesis we have studied aspects of the high energy limit of QCD. We investigated theoretical
problems related to the physics of saturation and its implications for phenomenology.\\ 
In the second chapter we constructed amplitudes for elastic scattering of hadronic objects with
the exchange of a pomeron loop and also four gluon BKP state. To find preferred by the TPV regions
of phase space we  studied collinear and
anticollinear limits of the Triple Pomeron Vertex. This analysis allowed us to gain some insight
into the complicated structure of the vertex. In particular using this analysis one can see which kind of
scattering proceses are mediated via triple interaction. Our analysis has shown that in the collinear
limit the TPV does not give the expected power of logarithm when convoluted with BFKL ladders and impact
factors. We find this result striking and the possible explanation is that the preferred configuration
for the system is the configuration where there is no triple interaction when the scales are strongly
ordered. The preferred configuration in that situation may be when gluons directly couple to 
the quark loop. We also studied the anticollinear limit of the TPV and we found
that the leading contribution does not give the expected power of logarithms and comes from the
anticollinear pole. The situation changes at
higher orders and also when we allow momentum to be transfered then there
is a contribution with the expected power of logarithm. Similar analysis was carried out also for a finite
number of colors. We show that in this case there is the expected power of logarithms both in the collinear
and the anticollinear limit.\\
The result involving infinite number af colors was also obtained by performing angular angular averaging over 
the azimuthal angles of momenta variables in the vertex. We found that the real emission part consists 
of a term which corresponds to the anticollinear pole accompanied by theta functions indicating  anticollinear
ordering of scales. The virtual term (which is subleading) also includes a theta function indicating
anticollinear ordering of scales . This result is in agreement with the performed
twist analysis. 
In chapter four we applied the Triple Pomeron Vertex together with the BFKL kernel to propose
the Hamiltonian for the evolution of the proton which we assumed to be state of reggeized gluons. 
We derived coupled evolution equations for the system of these gluons. These equations are very similar 
to those studied by Balitsky. We however, propose them in momentum space where the interpretation in terms 
of Feynmann diagrams is much clearer. We show that assuming mean field approximation which
basically means large $N_c$ limit we obtain
nonlinear closed equation which is equivalent to the the BK equation for the unintegrated gluon density.
Finally we discuss the relation of BK for the unintegrated gluon density to other known nonlinear
equations.\\   
The fifth chapter is devoted to phenomenology. We develop a procedure to make the gluon 
density that follows from BK more realistic in order to apply it to phenomenology.
The important point  is to find an efficient way to implement subleading corrections in $\ln1/x$ 
corrections and to choose realistic proton profile function.   
We follow the method of KMS which is easy to implement and is consistent with exact NLOln1/x calculations. 
Included corrections
have a large impact on the behavior of the gluon density. They are especially
important for large value of momentum. The effect of
nonlinearity is not so important at this region. This is due to the fact that the path of
integration in the nonlinear term is short. It however, becomes important for smaller values
of the gluon momentum and small $x$ where it leads to the depletion of the gluon density (for cylindrical target).
Similar behavior is not observed for a gluon density with a Gaussian profile (integrated over impact
parameter) since at largest
distances the effects of saturation are weaker. For that reason we argue that the Gaussian profile
is more physical since it does not lead to over-saturation of the gluon density.
With the introduced setup we perform a fit to $F_2$ data to constrain gluonic input distribution. 
The form of the input gluon distribution  can be still chosen to be more optimal. At present we choose 
it to be power like and all depletion is due to the nonlinearity of the equation. 
We also study the saturation line that follows from BK and we show that the saturation effects on 
level of the gluon density are
almost negligible in the HERA kinematical region. We wish to check in the future analysis another functional
form of it. Finally we develop the method to include hard rescattering corrections to Diffractive
Higgs boson production at the LHC. We argue that at the rapidity available at LHC one can expect hard
rescattering corrections to be important (this follows from studying the saturation line).
Those corrections should be treated separately from soft ones which were already studied. 
The crucial points here are:
\begin{itemize}
\item the presence of the TPV which allow
pomerons to rescatter 
\item constrained by HERA data the gluon density coming from the solution of
(\ref{eq:eqdorozw2}). 
\end{itemize}
The obtained corrections turn out to be rather large and decrease the cross
section for Higgs production. We however, keep in mind that those results are a rather rough
estimate. To be more realistic one should consider more symmetric situation where fan diagrams 
are allowed to propagate from two sources. One should also include in this processes contributions coming
from closed pomeron loops (\ref{eq:loop1}) or even BKP states(\ref{eq:loop2}). The task to include these 
diagrams in their full complexity
is difficult. We can however, try to investigate in a future analysis the importance of the contribution 
from the different parts of phase space and apply our results of collinear analysis to simplify the problem.
    
\chapter*{Appendices}
\setcounter{section}{0}
\setcounter{equation}{0}
\addcontentsline{toc}{chapter}{Appendices}
\chaptermark{Appendices}
\renewcommand{\thesection}{\Alph{section}}
\renewcommand{\theequation}{\Alph{section}.\arabic{equation}}

\setcounter{equation}{0}
\section{Details of the pomeron loop calculation}

In this appendix we will give short outline of sequence of steps that led us to (\ref{eq:loop1},
\ref{eq:loop2}).
The general structure is transparent and reflects the Regge factorization.
To set up the convention for the impact factors let us begin with the quark-quark scattering
due to the color singlet exchange.
The averaged over helicities and colors amplitude in eiconal approximation equals to:
\be
\begin{split}
A_{qq\rightarrow qq}^{(1)}(s,t=0)=2is\pi\int
\frac{d^2\k_1}{(2\pi)^3}\phi_q^{a_1a_2}\frac{1}{\k_1^4}\phi_q^{a_1a_2},
\end{split}
\ee
where as usual $\k_1$ is the momentum of $t$-channel gluon 
and $s$ is the square of
the total energy,
$\phi_q^{a_1a_2}=\delta^{a_1a_2}\frac{g^2}{2N_c}$.
The next step is to calculate the elastic scattering of quarks at three loop level due to the color
singlet exchange, Fig. \ref{fig:diagramy1}(left).
Including $s$ and $u$-channel contributions we arrive at:
\be
\begin{split}
A_{qq\rightarrow qq}^{(1)}(s,t=0)=2is\pi\phi_q^{a_1a_2} 
\frac{1}{2!}\ln\frac{s}{s_0}\int\frac{d^2\k_1}{(2\pi)^3}\frac{d^2\k_2}{(2\pi)^3}
\frac{d^2\k_3}{(2\pi)^3}\frac{1}{\k_1^4}\frac{1}{\k_2^4}
\frac{1}{\k_3^4}\\
N_c^2K_{BFKL}(\k_1,\k_2)K_{BFKL}(\k_2,\k_3)\phi_q^{a_1a_2}
\end{split}
\label{eq:ampli1}
\ee
where $\k_1$, $\k_2$, $\k_3$ are momenta of $t$-channel gluons.

To investigate the formation of the pomeron loop it is enough to calculate two loop correction to the process
above. The correction, that we are interested in, comes as two additional gluons in the $t$-channel
exchanged between gluons in the $s$-channel, Fig. \ref{fig:diagramy1}(right). We are in particular interested in color singlet contribution which
therefore does not contribute to reggezation. In calculation of that contribution we make use of dispersion
technics.
The result is:
\newpage
$$
Disc_{central}A^{(2)}_{qq\rightarrow qq}(s,t=0)= 2s\pi^2 2\pi P^{a_1'a_2'b_1'b_2'}P^{a_3'a_4'b_3'b_4'}
\frac{4}{(2!)^2}\frac{1}{2!} \ln^2\frac{s}{s_0}$$
$$\phi_{a_1a_2} $$
$$
\int\frac{d^2\k_1}{(2\pi)^3}\frac{d^2\k_2}{(2\pi)^3}\frac{d^2\k_3}{(2\pi)^3}
\frac{d^2\r}{(2\pi)^3}\frac{d^2\k_2^{'}}{(2\pi)^3}
\frac{K^{\{a_i\}\rightarrow\{a_i'\}}_{2\rightarrow 4}(\k_2,\k_2-\r,\k_2',\k_2'+\r) 
K^{\{b_i'\}\rightarrow\{a"_i\}}_{2\rightarrow 4}
(\k_2,\k_2\!-\!\r,\k_2',\k_2'+\r) }{\k_1^4\k_2^2(\k_2\!-\!\r)^2(\k_2'+\r)^2
\k_2^{'2}\k_3^{4}}
$$
\be
\phi_{a"_1a"_2}
\label{eq:disc22}
\ee
The combinatorial factor $\left(\frac{1}{2!}\right)^2$ in the formula above reflects symmetry with
respect to interchange of gluons in left and right-hand-side parts of considered diagram. The factor
$\frac{1}{2!}$ comes together with logarithm of the energy and indicates the process of the exponentiation.
Other contributions as peripheral cut may be
obtained by means of the AGK cutting rules. These rules relate combinatorial factors associated with
different cuts of an amplitude.
Applying AGK cutting rules to the system of four reggeized gluons we obtain the combinatorial
factor of single multiplicity diagram(there are two such diagrams).
The factor associated with every of them is $-\frac{1}{3!}$.

Summing up all contributions, using relation of discontinuity in direct channel
to the total amplitude, we obtain :
\begin{figure}[t!]
\centerline{\epsfig{file=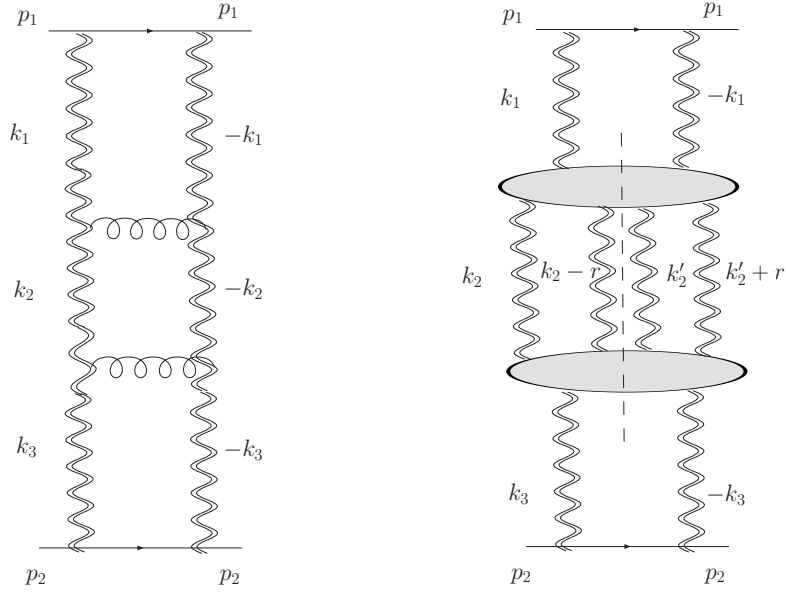,height=8cm}}
\caption{\em Left: amplitude for quark quark scattering due to color singlet exchange in high energy limit
QCD (\ref{eq:ampli1}). Right: central cut given by given by (\ref{eq:disc22}). }
\label{fig:diagramy1}
\end{figure}
\begin{figure}[t!]
\centerline{\epsfig{file=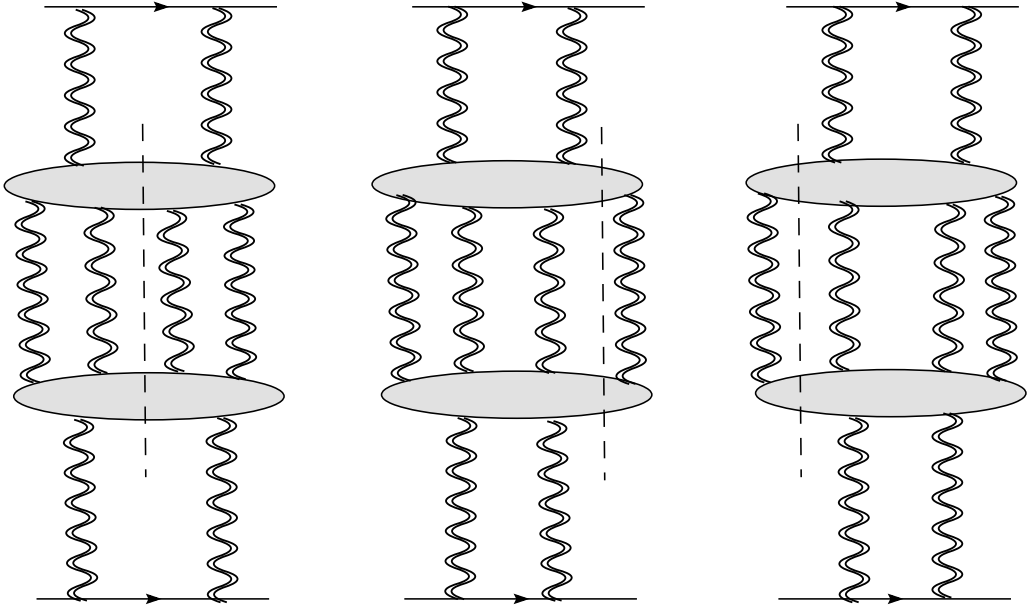,height=8cm}}
\caption{\em Considered cuts.}
\label{fig:diagramy}
\end{figure}
\newpage
$$
A^{(2)}_{qq\rightarrow qq}(s,t=0)= -2s\pi^2 i\pi\frac{8}{4!}
P^{a_1'a_2'b_1'b_2'}P^{a_3'a_4'b_3'b_4'}
\frac{1}{2!} \ln^2\frac{s}{s_0}$$
$$\phi_{a_1a_2} $$
$$
\int\frac{d^2\k_1}{(2\pi)^3}\frac{d^2\k_2}{(2\pi)^3}\frac{d^2\k_3}{(2\pi)^3}
\frac{d^2\r}{(2\pi)^3}\frac{d^2\k_2^{'}}{(2\pi)^3}
\frac{K^{\{a_i\}\rightarrow\{a_i'\}}_{2\rightarrow 4}(\k_2,\k_2-\r,\k_2',\k_2'+\r) 
K^{\{b_i'\}\rightarrow\{a"_i\}}_{2\rightarrow 4}
(\k_2,\k_2\!-\!\r,\k_2',\k_2'+\r) }{\k_1^4\k_2^2(\k_2\!-\!\r)^2(\k_2'+\r)^2
\k_2^{'2}\k_3^{4}}
$$
\be
\phi_{a"_1a"_2}
\label{eq:disc22}
\ee 
The combinatorial factor of system where two pairs of gluons form bound states is easy to
find. We have three possibilities of pairing two gluons to form bound states out of
four gluons. This yields the factor $2/2!$. In this configuration we have a pomeron loop topology.
In order to indicate the reflection symmetry of the loop we decide to keep the combinatorial factor
$1/2!$ and to redefine vertex by distributing the factor $2$ and also the factor of $\pi^2$.
Adding more rungs and including virtual
corrections results in generating more logarithms of energy which after exponentiation can be
replaced by BFKL Green's functions.
Inclusion of other contributing diagrams that mediate transition from two glon state to four gluons results
in replacing the kernel $2\rightarrow 4$ with Triple Pomeron Vertex \cite{BV}. The TPV in our case differs
slightly from the one in \cite{BV} and also in \cite{BLV}. The difference is due to because 
additional color indices for two gluons coming into the vertex (when we consider merging) and also we
included factor of $\pi$ and $\sqrt 2$ in a normalization of the TPV. Our motivation was to treat 
the vertex as locally as it was possible and that caused the redefinition.
\newpage
\setcounter{equation}{0}
\section{Angular averaging of the Triple Pomeron Vertex. }
In the forward direction the TPV takes simple form and is expressed in terms of four $G$ functions.
After averaging over angles, each of them gives the same contribution, so,
to average the vertex over angles it is enough to do that for one $G$ function.
Let us begin with the real part of the $G$ function
\be
G_1(\l,\m)=\frac{\k^2\l^2}{(\k-\l)^2}+\frac{\k^2\m^2}{(\k+\m)^2}
-\frac{(\l+\m)^2}{(\k-\l)^2(\k+\b)^2}
\ee
Let us denote the first term in formula above A, the second B and the third C.\\
An angle between $\l$ and $\m$  is $\alpha$, an angle between $\m$ and $\k$ is $\beta$
and an angle between $\l$ and $\k$ is $\alpha\!+\!\beta$.\\
Using:
\be
\int_0^{2\pi}\!\!d\alpha\frac{1}{a+b\cos\alpha}=\frac{2\pi}{|a^2-b^2|}
\label{eq:formula1}
\ee
we obtain:
\be
I_A=\int_0^{2\pi}d{\alpha}\int_0^{2\pi}d{\beta}\,A=
\frac{4\pi^2\k^2\l^2}{|\l^2-\k^2|}
\ee
and similarly for B:
\be
I_B=\int_0^{2\pi}d{\alpha}\int_0^{2\pi}d{\beta}\,B=\frac{4\pi^2k^2m^2}{|\m^2-\k^2|}
\ee
To compute integral in $C=C_1+C_2+C_3$ we split it into three integrals. 
\be
I_{C_1}=\int_0^{2\pi}d\alpha d\beta \frac{\l^2}{(\k^2-2\l\k\cos(\alpha+\beta)+\l^2)(\k^2+
2\m\k\cos\beta+\m^2)}
\ee
\be
I_{C_2}=\int_0^{2\pi}d\alpha d\beta\frac{\l^2\m^2\cos\alpha}{(\k^2-2\l\k\cos(\alpha+\beta)+\l^2)(\k^2+
2\m\k\cos\beta+\m^2)}
\label{eq:integral2}
\ee
\be
I_{C_3}=\int_0^{2\pi}d\alpha d\beta \frac{\m^2}{(\k^2-2\l\k\cos(\alpha+\beta)+\l^2)(\k^2+
2\m\k\cos\beta+\m^2)}
\ee
Let us begin with (\ref{eq:integral2}). To compute this integral we use:
\be
\int_0^{2\pi}d\alpha\frac{\cos\alpha}{a+b \cos\alpha +c\sin\alpha}=
\frac{2\pi b}{b^2+c^2-a\left(a+(b-a)\sqrt{\frac{a^2-b^2-c^2}{(a-b)^2}}\right)}
\label{eq:calk1}
\ee
In our case $a=1$, $b=-\frac{2|\l||\k|}{\l^2+\k^2}$, $c=\frac{2|\l||\k|}{\l^2+\k^2}sin(\beta)$. 
Inserting this in (\ref{eq:calk1}) and using formula which is written below to compute the 
integral over second angle 
\be
\int_0^{2\pi}d\beta\frac{\cos\beta}{1+a \cos\beta}=-\frac{2\pi a}{(1-a)(1+a+\sqrt{\frac{1+a}{1-a}})}
\ee
we obtain:
\be
I_{C_2}=\frac{32l^2\m^2\k^6}{(\l^2+\k^2)(\m^2+\k^2)(\l^2+\k^2+|\l^2-\k^2|)(\m^2+\k^2)+|\m^2-\k^2|}
\ee
Applying twice (for each angle) \ref{eq:formula1} we obtain for $I_{C_1}$ 
\be
I_{C_1}=\frac{4\pi^2\l^2\k^4}{|\l^2-\k^2||\m^2-\k^2|}
\ee
\be
I_{C_3}=\frac{4\pi^2\m^2\k^4}{|\l^2-\k^2||\m^2-\k^2|}
\ee
The total contribution is given by summing up $I_A$, $I_B$, $I_{C_1}$, $I_{C_2}$ , $I_{C_3}$
The result can by greatly simplified if we notice that we can consider situation when for instance 
$\k^2\!>\l^2,\m^2$.
Using that condition we may drop the absolute value sign. Adding all terms we obtain:
\be
\sum_{A,..,C_{3}} I=\frac{2\l^2\m^2\k^2-2\l^2\k^4-2\b^2\k^4+2\k^6}{(\l^2-\k^2)(\m^2-\k^2)}
\ee   
Expanding the denominator and simplifying
we obtain:
\be
\sum_{A,..,C_{3}}I= 2\k^2
\ee
In case when $\k^2\!<\l^2,\m^2$ we have for $I_{C_3}$
\be
I_{C_3}=\frac{2\l^2\m^2\k^2}{(\k^2-\l^2)(\k^2-\m^2)}
\ee
which has opposite sign to the sum of other contributions so, the total sum gives zero.
The same situation holds in the remaining cases $\m^2\!\!>\!\!\k^2\!\!>\!\!\l^2$, $\l^2\!\!>\k^2\!\!>\!\!\m^2$. This result can be
simply written as:
\be
\frac{1}{(2\pi)^2}\int_0^{2\pi}\!\!d\alpha d\beta\!\!G_1(\l,\m)=2\k^2\theta(\l^2-\k^2)\theta(\m^2-\k^2),
\ee
where the factor $1/(2\pi)^2$ comes from averaging.
Let us now perform angular averaging of disconnected pieces of the $G(\l,\m)$ function.
\be
G_2(\l,\m)=-\k^4\ln\frac{\l^2}{(\l+\m)^2}\delta(\l^2-\k^2)-\k^4\ln\frac{\m^2}{(\l+\m)^2}\delta(\m^2-\k^2)
\ee 
To compute the integral over angles we split the logarithm:
\be
I_D=\int_0^{2\pi}d\alpha d\beta[\ln\l^2-\ln(\l^2+\m^2+2\l \m \cos\alpha)]
\ee 
If we assume that $|\l|>|\m|$ we obtain:
\be
I_D=4\pi^2(\ln\l^2 -\ln\l^2)=0
\ee
In case when $|\m|\!\!>\!\!|\l|$ we obtain:
\be
 I_D=4\pi^2(\ln\l^2 -\ln\m^2)=\ln\frac{\l^2}{\m^2}
\ee
Introducing $\theta(z)$ function we can write this result in a more compact form:
\be
I_D=4\pi^2\ln(\l^2/\m^2)\theta(\m^2-\l^2)
\ee
And similarly in the second term:
\be
I_E=4\pi^2\ln(\m^2/\l^2)\theta(\l^2-\m^2)
\ee
The integrated over angles and averaged $G_2$ function reads:
\be
G_2(\l,\m)=-\k^4\ln\frac{\l^2}{\m^2}\theta(\m^2-\l^2)\delta(\l^2-\k^2)
-\k^4\ln\frac{\m^2}{\l^2}\theta(\l^2-\m^2)\delta(\m^2-\k^2)
\ee

\newpage
\newpage
\chapter*{Acknowledgements}
First of all I would like to thank my supervisor Professor Jochen Bartels for many 
interesting discussions and supervising me during my PhD studies.\\
Special thanks to Krzysztof Golec-Biernat for discussions, many useful comments, and suggestions.\\ 
I would like to thank Jochen Bartels, Sergey Bondarenko, Leszek Motyka, Anna Maria Stasto for
fruitful collaboration.
The interesting and stimulating discussions with Grigorios Chachamis, Sergey Bondarenko, 
Hannes Jung, Carlo Ewerz, Frank Fugel, Martin Hentschinski, Henri Kowalski,
Leszek Motyka, Misha Lublinsky, Krisztian Peters, Gavin Salam, Michele Salvatore, Sakura Schafer-Nameki, 
Florian Schwennsen, 
Anna Maria Sta\'sto, Bowen Xiao and Gian Paolo Vacca are kindly acknowledged.\\
I am grateful to Professor Stan Brodsky, Professor Lev Lipatov and Professor Alfred Mueller for very
instructive discussions.\\
For reading parts of the manuscript and many useful comments I would like to thank Steve Aplin, Krzysztof Golec-Biernat, 
Martin Hentschinski, Krzysztof Komar, Jaroslaw Lukasik, Michele Salvatore, Lydia Shlegova,
Michael Price.\\
The financial support of Graduiertenkolleg "Zuk\"unftige Entwicklungen in der Teilchenphysik" is
kindly acknowledged.\\
I would like to thank my wife Edyta, my familly, and my friends for their interest in my work.

Finally I would like to express my gratitude to Professor Jan Kwieci\'nski who passed away few 
years ago. I am grateful for introducing me to the physics of High Energy QCD  and for first
discussions on it.\\

\end{document}